\documentclass[twocolumn,reprint,aps,prx, superscriptaddress]{revtex4-1}
\usepackage[utf8]{inputenc}
\usepackage{color}
\usepackage{amsmath}
\usepackage{amssymb}
\usepackage{graphicx}
\usepackage{bbold}
\usepackage[unicode=true,
 bookmarks=false,
 breaklinks=false,pdfborder={0 0 1},backref=false,colorlinks=true]
 {hyperref}
\hypersetup{
 linkcolor=blue,citecolor=blue,urlcolor=blue,linkcolor=blue,citecolor=blue,urlcolor=blue}
\makeatletter

\usepackage{braket}

\definecolor{mygreen}{rgb}{0,0.5,0}
\definecolor{myblue}{rgb}{0,0,0.75}
\definecolor{mymagenta}{cmyk}{0,1,0,0.12}

\makeatother

\begin{document}

\title{A Unidirectional On-Chip Photonic Interface for Superconducting Circuits}
\author{P.-O. Guimond}
\thanks{Corresponding author: \href{mailto:pierre-olivier.guimond@uibk.ac.at}{pierre-olivier.guimond@uibk.ac.at}}
\affiliation{Center for Quantum Physics,
University of Innsbruck, Innsbruck A-6020, Austria}
\affiliation{Institute for Quantum Optics and Quantum Information, Austrian Academy of Sciences, Innsbruck A-6020,
      Austria}
      
\author{B. Vermersch}
\affiliation{Center for Quantum Physics,
University of Innsbruck, Innsbruck A-6020, Austria}
\affiliation{Institute for Quantum Optics and Quantum Information, Austrian Academy of Sciences, Innsbruck A-6020,
      Austria}
\affiliation{Univ. Grenoble Alpes, CNRS, LPMMC, 38000 Grenoble, France}
\author{M. L. Juan}
\affiliation{Institute for Experimental Physics, University of Innsbruck, A-6020 Innsbruck, Austria}
\affiliation{Institute for Quantum Optics and Quantum Information, Austrian Academy of Sciences, Innsbruck A-6020,
      Austria}
\author{A. Sharafiev}
\affiliation{Institute for Experimental Physics, University of Innsbruck, A-6020 Innsbruck, Austria}
\affiliation{Institute for Quantum Optics and Quantum Information, Austrian Academy of Sciences, Innsbruck A-6020,
      Austria}
\author{G. Kirchmair}
\affiliation{Institute for Experimental Physics, University of Innsbruck, A-6020 Innsbruck, Austria}
\affiliation{Institute for Quantum Optics and Quantum Information, Austrian Academy of Sciences, Innsbruck A-6020,
      Austria}
\author{P. Zoller}
\affiliation{Center for Quantum Physics,
University of Innsbruck, Innsbruck A-6020, Austria}
\affiliation{Institute for Quantum Optics and Quantum Information, Austrian Academy of Sciences, Innsbruck A-6020,
      Austria}
\date{\today}

\begin{abstract}
We propose and analyze a passive architecture for realizing on-chip, scalable cascaded quantum devices. In contrast to standard approaches, our scheme does not rely on breaking Lorentz reciprocity. Rather, we engineer the interplay between pairs of superconducting transmon qubits and a microwave transmission line, in such a way that two delocalized orthogonal excitations emit (and absorb) photons propagating in opposite directions. We show how such cascaded quantum devices can be exploited to passively probe and measure complex many-body operators on quantum registers of stationary qubits, thus enabling the heralded transfer of quantum states between distant qubits, as well as the generation and manipulation of stabilizer codes for quantum error correction.
\end{abstract}
\maketitle

\section{Introduction}
Over the last two decades, superconducting circuit technologies have emerged among the most promising platforms for realizing quantum processors \cite{Devoret1169,Gambetta2017}. One avenue consists in designing quantum networks in a modular approach, where distant stationary qubits interact by exchanging photons as ``flying qubits'' propagating in waveguides \cite{Kimble2008}. As the size of experiments and number of qubits in quantum networks scale in complexity, controllable routing of quantum information between distinct components becomes a requirement \cite{GU20171}. In most current experiments, this task is taken care of using ferrite junction circulators, which break Lorentz reciprocity via the Faraday effect \cite{RevModPhys.25.253,PhysRevApplied.10.047001}. However, as these devices are bulky, lossy, and use large magnetic fields, they are not suitable for on-chip integration, and new, scalable alternatives must be developed. To address this challenge, several approaches were proposed in recent years. Most strategies require active devices \cite{Kamal2011,PhysRevLetters112.1677012014,Estep2014,PhysRevApplied.4.034002,PhysRevX.5.041020,Roushan2016,Barzanjeh2017,PhysRevX.7.041043,Kamal2017,Metelmann2018}, where reciprocity is broken by the interplay of several pump fields with precise phase relations, at the cost of adding energy to the system. On the other hand, passive devices have also been proposed based on superconducting junction rings, where circulation is obtained using a constant flux bias; these are however highly sensitive to charge noise \cite{PhysRevA.82.043811,Muller2018}. 

\begin{figure*}
\includegraphics[width=\textwidth]{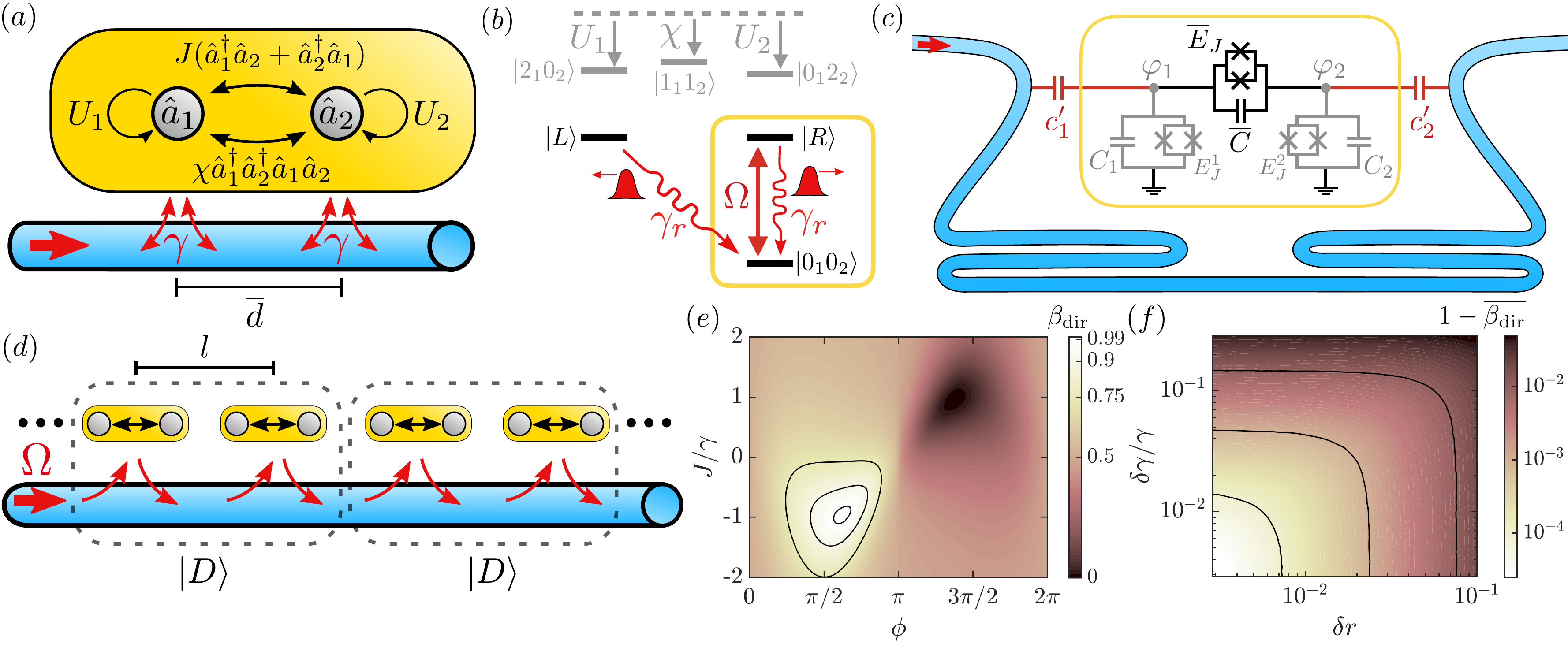}
\caption{\label{fig:fig1}\emph{Unidirectional coupling of quantum emitters to a transmission line.} (a)~Model for realizing a giant unidirectional emitter (GUE) using non-linear coupling between two artificial atoms coupled to a waveguide. (b)~Corresponding level structure obtained with specific parameters (see text). An effective two-level system with states $\ket{0_10_2}$ and $\ket{R}$ is obtained, which couples to right-propagating modes of the transmission line. (c)~Superconducting circuit implementation, where two transmons ($k=1,2$) are coupled at two points to a meandering transmission line, and interact via a SQUID.  (d)~Driven-dissipative cascaded quantum network realized with several GUEs as effective two-level emitters {unidirectionally} coupled to a transmission line. The system dissipates towards a pure steady-state with emitters pairing up in an entangled state $\ket{D}$. (e)~Directionality $\beta_\text{dir}$ of emitted photons, with $\Delta_{k}=0$ and $r_{k}=0.2$, $\gamma_{k}=\gamma$. (f)~Averaged directionality $\overline{\beta_\text{dir}}$ for $J=J_\text{opt}$, $\phi=\phi_\text{opt}$ and $\Delta_k=0$, obtained with uniformly distributed $r_1$, $r_2$, $\gamma_1$ and $\gamma_2$, with means $\overline{r_k}=0.2$, $\overline{\gamma_k}=\gamma$ and standard deviations $\sqrt{\overline{r_k^2}-{\overline r_k}^2}=\delta r$, $\sqrt{\overline{\gamma_k^2}-{\overline \gamma_k}^2}=\delta \gamma$.} 
\end{figure*}

In this work, we tackle the problem of quantum information routing from a different angle; rather than circulators, we design effective integrated qubits as composite objects coupled to a meandering 1D transmission line [see Fig.~\ref{fig:fig1}(a-c)], with the requirement that photons propagating in one direction are absorbed and reemitted along the same direction, without breaking reciprocity. Coherently driving several such \emph{unidirectional} quantum emitters through the transmission line gives rise to an effective cascaded driven-dissipative dynamics, as represented in Fig.~\ref{fig:fig1}(d), where photons radiated by each emitter coherently drives other emitters downstream; in the literature, this paradigm is sometimes referred to as ``chiral quantum optics'' \cite{Lodahl2017}, and features interesting steady-state properties, as will be discussed below.

In analogy to ``giant'' artificial atoms \cite{Kockum2014,PhysRevA.95.053821,PhysRevLetters120.140404,Andersson2019}, which couple to a photonic or phononic waveguide at several points separated by distances comparable to the wavelength, our approach consists in designing a \emph{giant unidirectional emitter} (GUE), here realized using two artificial atoms as anharmonic oscillators, as represented in Fig.~\ref{fig:fig1}(a). These atoms are coupled to a waveguide, at two points separated by a distance $\overline d\sim\lambda_0/4$, with $\lambda_0$ the photon wavelength. By designing the interaction between artificial atoms, our composite object effectively admits a $V$-level structure with two delocalized excited states $\ket{L}\sim(i\ket{1_10_2}+\ket{0_11_2})/\sqrt{2}$ and $\ket{R}\sim(\ket{1_10_2}+i\ket{0_11_2})/\sqrt{2}$ (with $\ket{n_{k}}$ denoting Fock state $n=0,1,\ldots$ of atom $k=1,2$), with the remarkable property that their transitions to the ground state $\ket{0_10_2}$ couple respectively only to left- and right-propagating modes of the waveguide [see Fig.~\ref{fig:fig1}(b)], which is due to a destructive interference in the photon emission (and absorption).  Below we will analyze an implementation of this model with superconducting transmon qubits coupled via a superconducting quantum interference device (SQUID) [see Fig.~\ref{fig:fig1}(c)]. 

As we will show later on, these composite emitters can be used as unidirectional photonic interfaces for additional long-lived stationary qubits (represented below in Fig.~\ref{fig:fig6}), which has immediate applications for quantum information processing and quantum computing. In our approach, quantum information is manipulated and directed passively, using an itinerant probe field as ``flying qubit'' propagating in the waveguide. This forms a naturally scalable architecture for quantum networking, which we will illustrate in particular with the realization of quantum state transfer between distant stationary qubits, and with the generation and manipulation of stabilizer codes for quantum error correction~\cite{Gottesman1997}. Our architecture is passive and tunable \emph{in situ}, and, as we will show, the required experimental parameters and imperfections are achievable with current technology.



Our results presented below are organized as follows. In Sec.~\ref{sec:1TLS} we describe and analyse the design of giant unidirectional emitters (GUEs) as composite artificial atoms with an effective V-level structure, with each transition absorbing and emitting photons along a single direction in a waveguide, and present in Sec.~\ref{sec:impl} a possible implementation with superconducting transmon qubits. Next, we study in Sec.~\ref{sec:NTLS} the cascaded driven-dissipative dynamics arising when several such unidirectional emitters are driven via the waveguide. Finally, in Sec.~\ref{sec:qinforouting} we describe how these emitters can act as unidirectional photonic interfaces for additional long-lived stationary qubits, which enables applications for quantum networking such as quantum state transfer between distant stationary qubits, and the generation and manipulation of stabilizer codes for quantum error correction.

\section{Model of unidirectional quantum emitters}\label{sec:1TLS}


Our model for designing unidirectional quantum emitters is represented in Fig.~\ref{fig:fig1}(a), and consists of two interacting artificial atoms as anharmonic oscillators coupled at two distant points to a waveguide. The dynamics of these two atoms, within the rotating wave approximation, is described by the Hamiltonian (with $\hbar=1$)
\begin{equation}\label{eq:Hadef}
\begin{aligned}
\hat H_a=&\sum_{k=1}^2 \omega_k  \hat a^\dagger_k \hat a_k-\left(U_k/2\right) \hat a^\dagger_k\hat a^\dagger_k\hat a_k\hat a_k
\\  &+ J\big(\hat a^\dagger_1\hat a_2+\hat a^\dagger_2\hat a_1\big)-\chi \hat a^\dagger_1\hat a_1\hat a^\dagger_2\hat a_2.
\end{aligned}
\end{equation}
Here $\omega_k$ is the transition frequency of each atom $k$, $U_k$ denotes their anharmonicity, and $\hat a_k$ is their annihilation operator, which satisfies $[\hat a_k,\hat a^\dagger_l]=\delta_{k,l}$. The second line in Eq.~\eqref{eq:Hadef} describes the interaction between atoms, with linear exchange interaction rate $J$, and non-linear cross-Kerr frequency $\chi$, which can be implemented with two superconducting transmon qubits coupled via a SQUID (see Fig.~\ref{fig:fig1}(c) and discussion below).  

The waveguide has a continuous spectrum of modes described over the relevant bandwidth by the bare Hamiltonian $\hat H_\text{ph}=\int d\omega \omega [\hat b_R^\dagger(\omega)\hat b_R(\omega)+\hat b_L^\dagger(\omega)\hat b_L(\omega)]$, where $\hat b_{d}(\omega)$ is the annihilation operator for photons with frequency $\omega$ propagating to the right (with $d=R$) or to the left (with $d=L$), and satisfies $[\hat b_{d}(\omega),\hat b^\dagger_{d'}(\omega')]=\delta(\omega-\omega')\delta_{d,d'}$. Finally, the coupling between the atoms and the waveguide yields, within the rotating wave approximation, the Hamiltonian
\begin{equation}
\begin{aligned}\label{eq:Hintdef}
\hat H_\text{int}=\frac{1}{\sqrt{2\pi}}\int d\omega \Big[&\hat b_R^\dagger(\omega)\big(e^{i\omega \overline d/v_g} \hat L_1+ \hat L_2\big) 
\\+& \hat b_L^\dagger(\omega)\big(\hat L_1+e^{i\omega \overline d/v_g} \hat L_2\big) +\text{h.c.}\Big].
\end{aligned}
\end{equation}
Here {$\hat L_1 =  \sqrt{\gamma_1}(\hat a_1+r_2\hat a_2)$} and $\hat L_2=  \sqrt{\gamma_2}(\hat a_2+r_1 \hat a_1)$ are the coupling operators associated to each coupling point, with coupling rates $\gamma_k$ (which we assume constant over the relevant bandwidth) and small cross-coupling coefficients $r_k$ (see implementation below), $\overline d$ is the distance of separation between the two coupling points along the waveguide, and $v_g$ is the group velocity of photons in the waveguide. 

Within a markovian approximation (i.e., assuming $\gamma_k \overline d/v_g\ll 1$), the dynamics of the field can be integrated and treated as a reservoir for the atoms, and we obtain for the Heisenberg equation of motion for an arbitrary atomic operator $\hat O(t)$ the quantum Langevin equation (see details in Supplementary Section \ref{secSM:directionality})
\begin{equation}\label{eq:Otheisenberg}
\begin{aligned}
\frac{d}{dt}\hat O(t)=&-i\left[\hat O,\hat H_\text{eff}\right]+\sum_{d=R,L}\hat L_d^\dagger \hat O \hat L_d -\frac{1}{2}\left\{\hat L_{d}^\dagger \hat L_d, \hat O\right\}
\\ &+\sum_{d=R,L}[\hat b^\text{in}_{d}(t)]^\dagger [\hat O,\hat L_d]+[\hat L_d^\dagger,\hat O]\hat b_{d}^\text{in}(t),
\end{aligned}
\end{equation}
expressed in a rotating frame with respect to a central frequency $\omega_0$, and in an interaction picture with respect to the waveguide Hamiltonian $\hat H_\text{ph}$. Here the effective Hamiltonian reads
\begin{equation}
\begin{aligned}
\hat H_\text{eff}=&-\sum_{k=1}^2 \Delta_k  \hat a^\dagger_k \hat a_k-\big(U_k/2\big) \hat a^\dagger_k\hat a^\dagger_k\hat a_k\hat a_k-\chi \hat a^\dagger_1\hat a_1\hat a^\dagger_2\hat a_2
\\  &+J\big(\hat a^\dagger_1\hat a_2+\hat a^\dagger_2\hat a_1\big)+\sin(\phi) \big( \hat L_2^\dagger \hat L_1+\hat L_1^\dagger \hat L_2\big),
\end{aligned}
\end{equation}
with $\Delta_k=\omega_0-\omega_k$, where the last term emerges from a coherent exchange of photons propagating in the waveguide between the two coupling points, with $\phi=\omega_0 \overline d/v_g$ the phase acquired by a photon in the propagation. On the other hand, the \emph{collective coupling} operators in Eq.~\eqref{eq:Otheisenberg} represent the collective couplings of the atoms to right- and left-propagating photons due to interference of photon emission and absorption in the reservoir, and are defined respectively as $\hat L_R(t)=e^{i\phi}\hat L_1(t)+\hat L_2(t)$ and $\hat L_L(t)=\hat L_1(t)+e^{i\phi}\hat L_2(t)$.
Finally, $\hat b_{d}^\text{in}(t)$ represents the input fields of the waveguide propagating along direction $d$, and is related to the output fields via \cite{gardiner2004quantum}
\begin{equation}\label{eq:inputoutput}
\hat b_{d}^\text{out}(t)=\hat b_{d}^\text{in}(t)+\hat L_d(t),
\end{equation} 
with $[\hat b_{d}^\text{in/out}(t),(\hat b_{d'}^\text{in/out}(t'))^\dagger]=\delta(t-t')\delta_{d,d'}$. The emergence of unidirectional coupling between propagating photons and the composite two-atom system, from Eqs.~\eqref{eq:Otheisenberg} and~\eqref{eq:inputoutput}, occurs under the following two conditions.

 (I) First, the two collective coupling operators $\hat L_{R}$ and $\hat L_L$ must be orthogonal, i.e., $[\hat L_L^\dagger,\hat L_R]=0$, such that each operator $\hat L_d$ couples only to the corresponding input fields $\hat b^\text{in}_d(t)$ in Eq.~\eqref{eq:Otheisenberg}. Here, this condition requires the system parameters to be symmetric, i.e., $r_1=r_2= r$ and $\gamma_1=\gamma_2= \gamma$, while the propagation phase must be set to $\phi=\phi_\text{opt}$, with the optimal propagation phase $\phi_\text{opt}=\pi/2+2\arctan(r)$.
With these parameters, the collective coupling operators reduce to $\hat L_{R/L}=\sqrt{\gamma_r} \hat a_{R/L}$, up to an irrelevant phase factor, with the definition of two orthogonal delocalized atomic modes $\hat a_R=(i\hat a_1+\hat a_2)/\sqrt{2}$ and $\hat a_L=(\hat a_1+i\hat a_2)/\sqrt{2}$, and where the effective coupling strength of the system to the waveguide is given by
$\gamma_r=2\gamma\left(1+2r\cos[\phi_\text{opt}]+r^2\right)$.

(II) Second, the excitations associated to these two modes $\hat a_R$ and $\hat a_L$ must be eigenstates of the effective Hamiltonian $\hat H_\text{eff}$. For states with a single atomic excitation, i.e., $\ket{R}= \hat a^\dagger_R\ket{G}$ and $\ket{L}= \hat a_L\ket{G}$ with $\ket{G}=\ket{0_10_2}$ the ground state of both atoms, this is achieved by taking symmetric detunings $\Delta_1=\Delta_2\equiv\Delta+2r\gamma\sin(\phi_\text{opt})$ and $J= J_\text{opt}$, with the optimal hopping rate given by $J_\text{opt}=-\gamma(1+r^2)\sin(\phi_\text{opt})$. The two excited states $\ket{R}$ and $\ket{L}$ are then eigenstate of $\hat H_\text{eff}$ with eigenenergies $-\Delta$. The non-linear cross-Kerr interaction with frequency $\chi$, on the other hand, is introduced in the model in order to prevent the excitation of the doubly-excited state $\ket{1_11_2}$ when driving the system via the input fields, as we will consider below. 


When these two conditions are fulfilled, the composite emitter will absorb and reemit propagating photons along the same direction. In order to assess this directionality in a more general case, we assume the emitter is prepared in state $\ket{R}$ at time $t=0$ with the waveguide in the vacuum state, and solve the dynamics of the system, which yields the emission of a photon in the waveguide, with the emitter returning to its ground state $\ket{G}$. The temporal shapes of the wavepacket amplitudes of the emitted photon propagating to the right/left are then obtained using a Wigner-Weisskopf ansatz (see details in Supplementary Section \ref{secSM:directionality}) as $f_{R/L}(t)\equiv\bra{G}\hat b_{R/L}^\text{out}(t)\ket{R}=\langle G|\hat L_{R/L}\mathcal L^{-1}[\hat F^{-1}(s)\ket{R}]$, where $\mathcal L[\cdot](s)$ denotes the Laplace transform, and the evolution of the atomic excitation amplitudes is governed by the operator 
\begin{equation}\label{eq:Gsdef}
\hat F(s)=s+i\hat H_\text{eff}+\frac12 \left(\hat L_R^\dagger \hat L_R+\hat L_L^\dagger \hat L_L\right).
\end{equation}

We then define the directionality of photon emission as $\beta_\text{dir}=\int_0^\infty|f_R(t)|^2dt$.
This directionality of emitted photons is represented in Fig.~\ref{fig:fig1}(e,f). Fig.~\ref{fig:fig1}(e) shows that very good directionalities can be achieved even with relatively large imprecisions on $J$ and $\phi$ around their optimal values, e.g. due to fabrication imperfections. Here we obtain $\beta_\text{dir}>99\%$ for $|J-J_\text{opt}|\lesssim\gamma/10$ and  $|\phi-\phi_\text{opt}|\lesssim\pi/10$. This robustness to imperfections is also observable in Fig.~\ref{fig:fig1}(f), where we show the average directionality $\overline{\beta_\text{dir}}$ obtained with random static deviations of $r_k$ and $\gamma_k$. We obtain $\overline{\beta_\text{dir}}>99\%$ as long as the fluctuation in the coupling parameters are below $\delta\gamma\lesssim0.1\gamma$ and $\delta r \lesssim  0.05$. 

\section{Implementation with superconducting circuits}\label{sec:impl}

Our model can be implemented with the circuit represented in Fig.~\ref{fig:fig1}(c), which consists of two superconducting transmon qubits ($k=1,2$) with flux-tunable Josephson energies $E_J^k$ and charging energies {$E_C^k=e^2/(2C^\text{eff}_k)$} \cite{PhysRevA.76.042319}, where $e$ is the elementary charge and $C^\text{eff}_k$ are the effective transmon capacitances (see details in Supplementary Section \ref{secSM:simulations}). The interaction between transmons is mediated by a SQUID, acting as a non-linear element with flux-tunable Josephson energy $\overline E_J$ and with capacitance $\overline C$. We note that such tunable non-linear couplings mediated by Josephson junctions were demonstrated in recent experiments~\cite{PhysRevLetters113.220502,Kounalakis2018,PhysRevLetters122.183601}, and find applications for quantum simulation \cite{PhysRevLetters110.163605,PhysRevA.95.042330,Marcos2014} and quantum information processing~\cite{PhysRevLetters111.063601}. 

Following standard quantization procedures, the Hamiltonian for the circuit can be expressed as in Eq.~\eqref{eq:Hadef} (see details in Supplementary Section \ref{secSM:simulations}). In particular, analytical insight on the resulting system parameters can be gained in the regime of weakly coupled transmons, with $E_C^k\ll  E_J^k$, $\overline E_J\ll E_J^k$ and $\overline C\ll C_k$. In this limit, an estimation of the various parameters of the model can be made in terms of the circuit parameters, with the atomic transition frequencies taking the expression $\omega_k\approx\sqrt{8E_J^kE_C^k}$, while the atomic anharmonicities read
$U_k \approx E_C^k$. The interaction between atoms contains a linear hopping term $J=J_C-J_I$, with a capacitive ($J_C$) and an inductive ($J_I$) contribution reading
\begin{equation}\label{eq:Jexpr}
J_C\approx \omega_0\frac{\overline C}{2\sqrt{C_1^\text{eff}C_2^\text{eff}}}, \ \ \ J_I\approx \omega_0\frac{\overline E_J}{2\sqrt{E_J^1 E_J^2}},
\end{equation}
while the cross-Kerr interaction term reads
\begin{equation}\label{eq:Uexpr}
\chi= 2 \overline E_J\sqrt{\frac{E_C^1E_C^2}{E_J^1E_J^2}}.
\end{equation}
We note that the three Josephson energies in Fig.~\ref{fig:fig1}(c) can be independently controlled via flux biases, allowing for an independent \emph{in situ} fine-tuning of the detunings $\Delta_k$ and the hopping rate $J$. The couplings to the waveguide on the other hand are given by $\gamma_k=(c'_k/C^\text{eff}_k)^2 \omega_0 e^2 Z_0\sqrt{E_J^k/(8E_C^k)}$, with $c'_k$ the coupling capacitances and $Z_0$ the transmission line impedance~\cite{Johansson_2006,Lalumiere:2013io}. The capacitance $\overline C$ introduces as well small cross-coupling coefficients $r_k=\overline C/C_k^\text{eff}$, resulting in photon emission from each artificial atom via both coupling points. 

\section{Driven-dissipative dynamics of cascaded quantum networks}\label{sec:NTLS}

 Although the properties of unidirectional emission of our GUE studied above preserve Lorentz reciprocity, i.e., they are invariant under the exchange of left- and right-propagating modes, driving the system through the waveguide allows one to effectively achieve non-reciprocal interactions \emph{between artificial atoms}. A paradigmatic example of such a situation is represented in Fig.~\ref{fig:fig1}(d), where several GUEs are coherently driven via right-propagating modes, thus driving the $\hat a_R$ transition as represented in Fig.~\ref{fig:fig1}(b). Photons emitted by each emitter will then also propagate to the right, leading to an effective \emph{cascaded} quantum dynamics, where each GUE drives the other ones downstream, without any back-action \cite{Gardiner1993,Carmichael1993,Metelmann2015}.

This scenario has been studied in recent years in a different context, in a field known in the literature as ``chiral quantum optics" \cite{Lodahl2017}, which originated from experiments with quantum emitters in the optical domain, such as atoms \cite{PhysRevLetters110.213604,Mitsch2014,Shomroni2014,Bechler2018a} or quantum dots \cite{Sollner2015a,Coles2016,Barik666,Barik}, coupled to photonic 1D nanostructures. The strong confinement of light in these structures gives rise to a so-called ``spin-momentum locking" effect \cite{Bliokh2015}, allowing for unidirectional couplings between photons and emitters which, in an analogous way to our GUE, does not by itself break Lorentz reciprocity. Besides, building on non-local couplings of quantum emitters to 1D reservoirs, chiral quantum optical systems could also be realized in AMO platforms with broken reciprocity \cite{Ramos2016,Vermersch2016,Grankin2018}. While photon losses inherent to optical platforms form experimental challenges, the near-ideal mode matching of artificial atoms coupled to 1D transmission lines presents new opportunities to realize this paradigm, in the microwave domain \cite{Hoi2011,Hoi2013}. 
Interestingly, it has been predicted that, for several quantum emitters, the ensuing cascaded dynamics in the presence of a coherent drive results in the dissipative preparation of quantum dimers, with quantum emitters pairing up in a dark, entangled state \cite{Stannigel:2012jk,Ramos2014,Pichler2015}, as we will show below.  

 In order to study the dynamics of an ensemble of $N$ GUEs (labeled $n=1,\ldots,N$) interacting via a common waveguide, we employ the \emph{SLH input-output formalism} \cite{Gough2009,Gough2009b,doi:10.1080/23746149.2017.1343097}. The SLH framework provides a methodical approach for modeling such composite quantum systems interacting via the exchange of propagating photons, where we assume that non-Markovian effects, due e.g. to the finite propagation time of photons exchanged by the emitters \cite{Pichler2016}, can be neglected. 
 As detailed in the Supplementary Section \ref{secSM:SLH}, the dynamics of the network of $N$ GUEs can then be obtained from the input-output properties of each individual GUE, by recursively applying composition rules of the SLH formalism in a ``bottom-up'' fashion. The evolution of an arbitrary atomic operator $\hat O(t)$ in the rotating frame then obeys a quantum Langevin equation as expressed in Eq.~\eqref{eq:Otheisenberg}, with a redefinition of the effective Hamiltonian and of the coupling operators. Denoting the various parameters and operators associated with each GUE with a corresponding superscript $n$, we obtain for the effective Hamiltonian
 \begin{equation}\label{eq:HeffNGUEs}
\begin{aligned}
\hat H_\text{eff}=&\sum_n \hat H_\text{eff}^n -\frac{i}2\sum_{n,m<n}\left[(\hat L_R^n)^\dagger \hat L_R^m e^{i\tilde\phi(n-m)}-\text{h.c.}\right]
\\ &-\frac{i}2\sum_{n,m>n}\left[(\hat L_L^n)^\dagger \hat L_L^m e^{i\tilde\phi(m-n)}-\text{h.c.}\right],
\end{aligned}
\end{equation}
with the photon propagation phase $\tilde\phi=\omega_0 l/v_g$ where $l$ is the distance between two neighbouring composite emitters along the waveguide. We note that the two new terms in Eq.~\eqref{eq:HeffNGUEs} correspond to excitation exchange interactions between different GUEs, mediated respectively by right- and left-propagating photons. For the coupling operators on the other hand, we obtain $\hat L_R=\sum_n e^{i\tilde\phi (N-n)} \hat L_R^n$ and $\hat L_L=\sum_n e^{i\tilde\phi (n-1)} \hat L_L^n$, which represent interference in the atom-field coupling between the emitters. 

The presence of a coherent drive via the right-propagating waveguide modes, with amplitude $\alpha(t)$ [and corresponding Rabi frequency $\Omega(t)=\sqrt{\gamma_r}\alpha(t)$], can be accounted for by assuming the initial state of the waveguide $\ket{\alpha_R}$ satisfies $\hat b_d^\text{in}(t)\ket{\alpha_R}=\alpha(t)\delta_{d,R}\ket{\alpha_R}$. Writing $\langle \hat O(t)\rangle=\text{Tr}\big[\hat O\hat\rho(t)\big]$, with $\hat \rho$ the atomic density matrix, the temporal evolution from Eq.~\eqref{eq:Otheisenberg} then yields the master equation 
 \begin{equation}\label{eq:MEsinglechiralqubit}
 \begin{aligned}
\frac{d}{dt}\hat\rho=-&i\left[\hat H_\text{eff}-i\alpha(t) \hat L_R^\dagger+i\alpha^*(t) \hat L_R,\hat\rho\right]
\\+&\mathcal D\big[\hat L_R\big]\hat\rho+\mathcal D\big[\hat L_L\big]\hat\rho,
\end{aligned}
\end{equation}
where $\mathcal D[\hat a]\hat\rho= \hat a\hat\rho \hat a^\dagger-\frac12\{\hat a^\dagger \hat a,\hat \rho\}$. Eq.~\eqref{eq:MEsinglechiralqubit} allows to access the evolution and steady-state values of observables with a finite drive amplitude $\alpha$. In order to account for additional imperfections, we also add in Eq.~\eqref{eq:MEsinglechiralqubit} dephasing terms $2\gamma_{\varphi}\sum_{n,k}\mathcal D[(\hat a^n_{k})^\dagger \hat a_k^n]$ and non-radiative decay terms $\gamma_\text{nr}\sum_{n,k}\mathcal D[\hat a_{k}^n]$.

\begin{figure}
\includegraphics[width=0.5\textwidth]{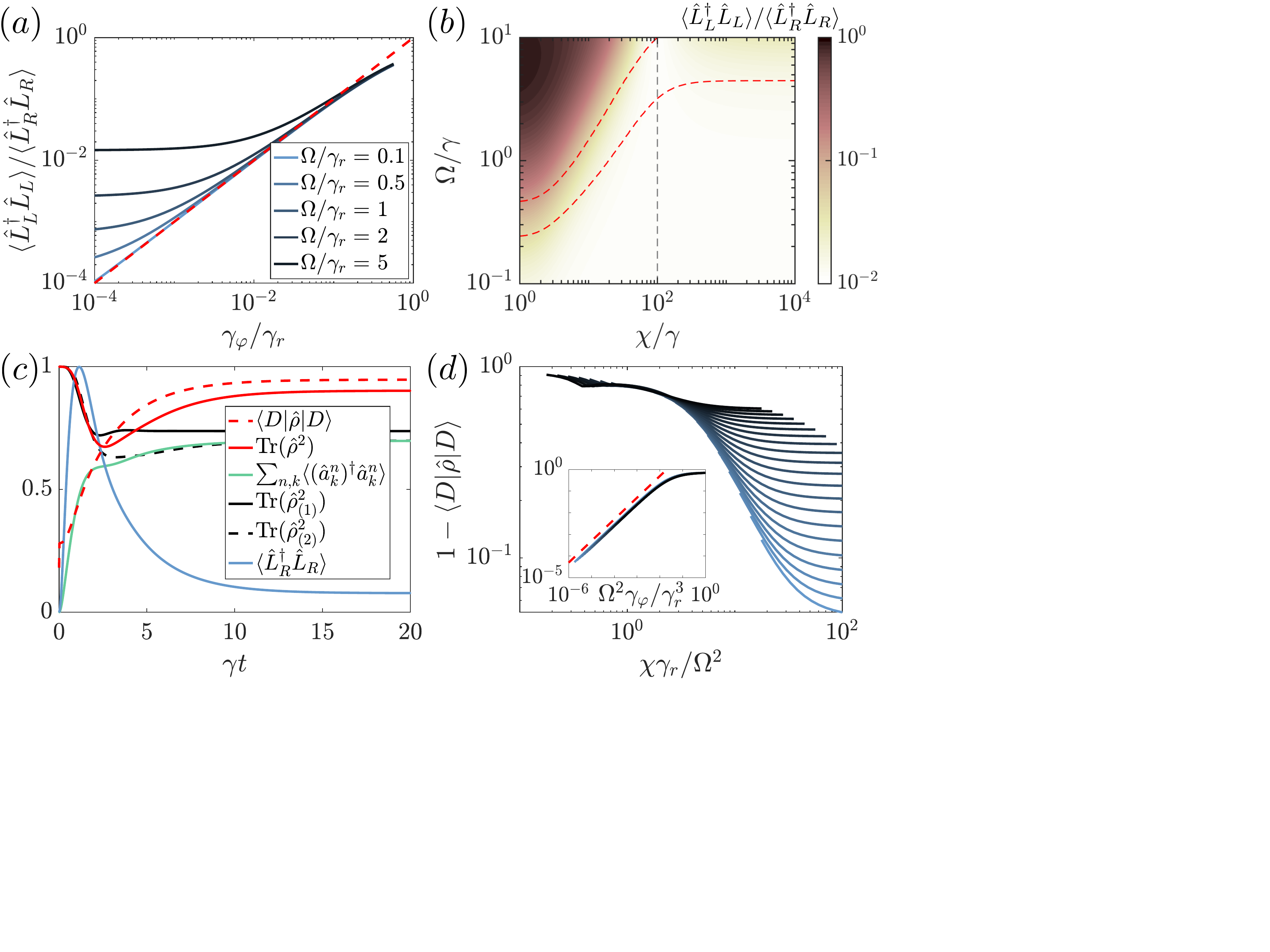}
\caption{\label{fig:fig4} \emph{Driven-dissipative dynamics.} (a,b)~Ratio of left- and right-propagating photon intensities, in the steady-state, emitted by the artificial atoms when coherently driven through the waveguide, with $\gamma_{k}=\gamma$, $\Delta_{k}=0$, $r_{k}=0.2$, $U_{k}=100\gamma$, $\gamma_\text{nr}=0.01\gamma$, $J=J_\text{opt}$, $\phi=\phi_\text{opt}$. (a)~$\chi=50\gamma$. Dashed red: $\gamma_\varphi/\gamma_r$. (b)~$\gamma_{\varphi}=0.01\gamma$. Dashed red: probability of $10^{-2}$ and $10^{-3}$ (resp. top and bottom) of having two or more excitations in the atoms. Dashed grey: $\chi=U_1=U_2$. (c,d)~Cascaded dynamics with $N=2$ GUEs, with $r_k=0.2$, $U_k=500\gamma$, $\tilde\phi=0$, and $\gamma_\text{nr}=\gamma_{\varphi}$. (c)~$\Omega=\gamma$, $\chi=50\gamma$, $\gamma_{\varphi}=0.01\gamma$. (d)~Steady-state overlap $\bra{D}\hat \rho\ket{D}$, with $\Omega\in[1,10]\gamma$ (light to dark blue), and $\gamma_{\varphi}=0.01\gamma$. Inset: $\chi\to\infty$, red dashed curve $\propto \Omega^2\gamma_{\varphi}/\gamma_r^3$. }
\end{figure}

 In Fig.~\ref{fig:fig4}(a,b) we represent the ratio of left- and right-propagating emitted photons obtained in the steady-state of the dynamics for $N=1$, with $J$ and $\phi$ set to their optimal values, and a constant real Rabi frequency $\Omega$ (i.e., the drive frequency is $\omega_0$). Fig.~\ref{fig:fig4}(a) shows that, since directionality arises in our setup as interference of emission of the two atoms, the dephasing rate $\gamma_{\varphi}$ spoils the interference and induces some emission to the left with an intensity scaling linearly for low Rabi frequency $\Omega$. As $\Omega$ increases with respect to the effective anharmonicities $\chi$ and $U_k$, the intensity of left-propagating photons increases, as states with more than a single excitation get populated.  This population increase can also be observed as the dashed red curves in Fig.~\ref{fig:fig4}(b), and we thus require $\Omega\ll \chi$ in order to retain a two-level dynamics. We also note that when $\chi=U_{1}=U_2$ in Fig.~\ref{fig:fig4}(b), the emission to the left vanishes even when states with several excitations are populated, as for these parameters states with several excitations $(\hat a_R^\dagger)^{n_R}(\hat a_L^\dagger)^{n_L}\ket{G}$ become eigenstates of $\hat H_\text{eff}$ for all $n_{R/L}\geq0$, thus preserving the property of unidirectional emission. Note that in the regime of weakly coupled transmons ($\overline C\ll C_k$ and $\overline E_J\ll E_J^k$) considered above, the value of $U$ is limited by the fact that, from Eq.~\eqref{eq:Uexpr} and $U_k\approx E_C^k$, we have $\chi\ll2\sqrt{U_1U_2}$. Achieving larger values thus requires going beyond the weak coupling regime. This is discussed in the Supplementary Section \ref{secSM:simulations}, where we also study the validity of the analytical expressions for the effective model in Eqs.~\eqref{eq:Jexpr} and \eqref{eq:Uexpr}. Typical achievable values for $\chi$ range from $0$ to $\sim 2\pi \times50$ MHz with $U_k=2\pi\times300$ MHz.
 
In the ideal case where the parameters satisfy the properties of unidirectional coupling and the anharmonicities $\chi$ and $U_k$ are large enough with respect to the Rabi frequency $\Omega$ of the drive, the state of the emitters will thus remain within the two-level manifold $\bigotimes_n\{\ket{G}_n, \ket{R}_n\}$. Denoting here $\hat \sigma_+^n=e^{i\tilde\phi n}\ket{R}_n\!\bra{G}$, the dynamics of Eq.~\eqref{eq:MEsinglechiralqubit} then reduces to a \emph{cascaded} master equation~\cite{Gardiner1993,Carmichael1993}
\begin{equation}\label{eq:MEcascade}
\frac{d}{dt}\hat \rho_\text{eff}=-i\left(\hat H_\text{nh}\hat \rho_\text{eff}-\hat \rho_\text{eff} \hat H^\dagger_\text{nh}\right) + \hat L_R\hat \rho_\text{eff} \hat L_R^\dagger,
\end{equation}
where $\hat \rho_\text{eff}$ denotes the density matrix of the system expressed in the reduced $2^N$-dimensional manifold, and where the effective non-Hermitian Hamiltonian reads, assuming $\Omega$ real,
\begin{equation}\label{eq:Heffdef}
\begin{aligned}
\hat H_\text{nh}=&-\Delta \sum_n \hat \sigma_+^n\hat \sigma_-^n-i \Omega(\hat \sigma_+^n-\hat \sigma_-^n)
\\ &-i\frac{\gamma_r}2\sum_n\hat \sigma_+^n\sigma_-^n-i \gamma_r \sum_{n>m}\hat \sigma_+^n\hat \sigma_-^m.
\end{aligned}
\end{equation}
The dynamics generated by Eq.~\eqref{eq:MEcascade} induces an effective non-reciprocal interaction between the qubits: as seen from the expression of Eq.~\eqref{eq:Heffdef}, an excitation in each qubit $m$ can be coherently transferred only to qubits $n>m$ located to its right. While the reduced density matrix of any single GUE is in general mixed, for even $N$ the state of the whole system dissipates towards a pure steady-state $\ket{\Psi}=\bigotimes_{n=1}^{N/2}\ket{D}_{2n-1,2n}$, where, as represented in Fig.~\ref{fig:fig1}(d), neighbouring qubits pair up as dimers in a two-qubit entangled state \cite{Stannigel:2012jk,Ramos2014,Pichler2015}
\begin{equation}
\ket{D}_{2n-1,2n}\propto\ket{G}_{2n-1}\ket{G}_{2n}-2\sqrt{2}\frac{\Omega}{\gamma_r}\ket{S}_{2n-1,2n},
\end{equation}
with {$\ket{S}_{2n-1,2n}=(\ket{R}_{2n-1}\ket{G}_{2n}-\ket{G}_{2n-1}\ket{R}_{2n})/\sqrt{2}$}. Remarkably, once the system has reached this \emph{dark} state $\ket{D}$, all photons emitted by qubit $2n-1$ are coherently absorbed by qubit $2n$, such that each dimer effectively decouples from the waveguide radiation field.

The dynamics obtained for a pair of $N=2$ GUEs is represented in Fig.~\ref{fig:fig4}(c,d). In Fig.~\ref{fig:fig4}(c) we observe the purification process described above where, in the steady-state, the system dissipates towards the pure state $\ket{D}$, as represented in the red curves. Strikingly, although the atoms are excited (see green curve), the amount of scattered photons, represented in blue, vanishes in the steady-state, i.e., the system becomes dark and decouples from the waveguide. We note that in the transient dynamics, i.e., before reaching the steady-state, photons are scattered unidirectionally by the emitters, which leads to a decrease of the purity Tr$(\hat \rho^2)$. Moreover, the purity of the reduced density matrix $\hat \rho_{(n)}$ for each GUE $n$ remains low in the steady-state (see black curves), as they become entangled. The steady-state overlap $\bra{D}\hat\rho\ket{D}$ is represented in Fig.~\ref{fig:fig4}(d), which shows a requirement for a large $\chi$ with respect to the drive intensity $|\Omega|^2/\gamma_r$. The effect of imperfections due to dephasing and finite excitation lifetimes is represented in the inset, which shows that the steady-state overlap with the dark state becomes unity in the limit $\chi\to\infty$ and $\gamma_\text{nr}=\gamma_\phi=0$.

\section{Quantum information routing for quantum networking and computing}\label{sec:qinforouting}

\begin{figure}
\includegraphics[width=0.5\textwidth]{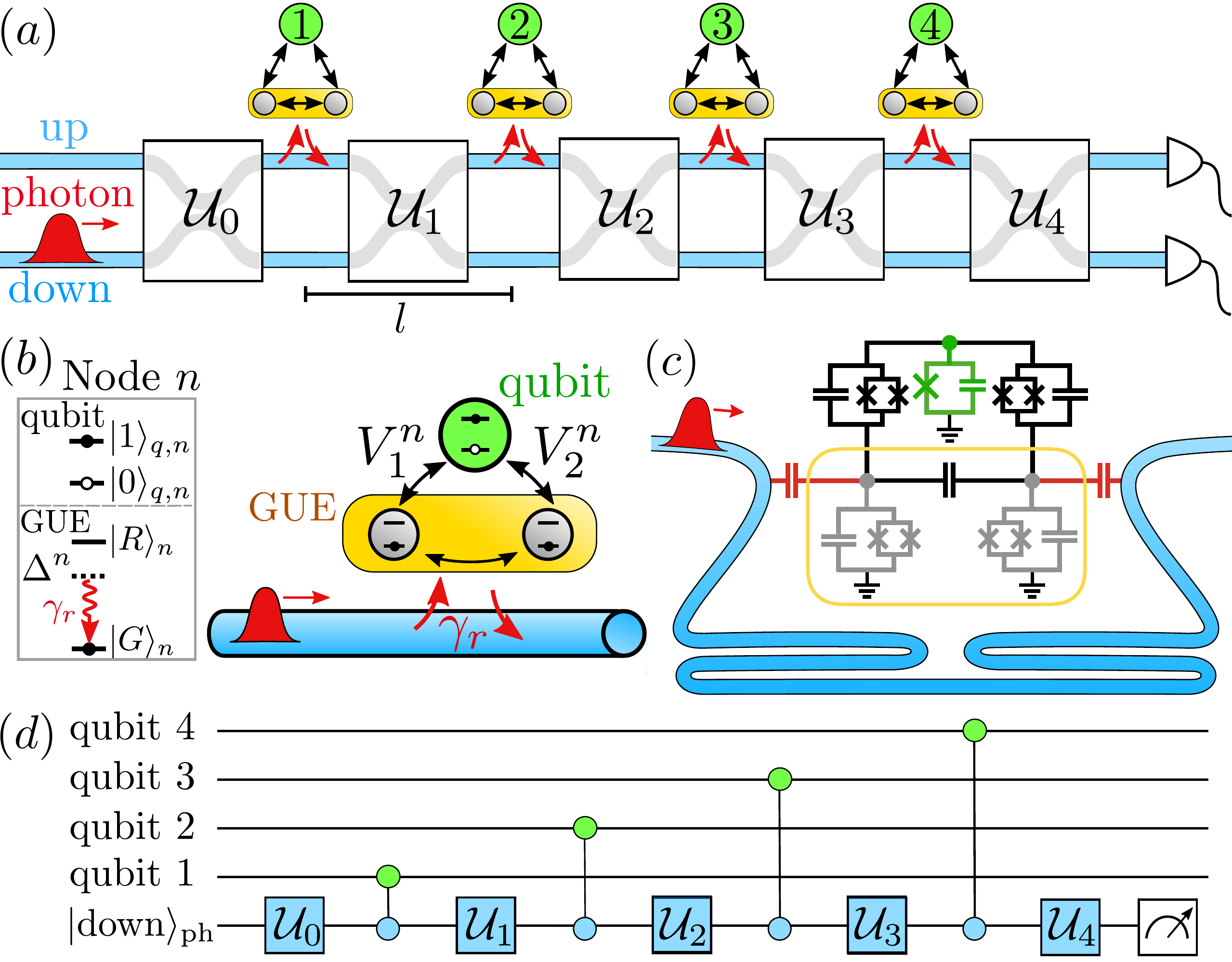}\caption{\label{fig:fig6}\emph{Architecture for quantum information routing.} (a)~An array of qubits ($n=1,\ldots,N$) is coupled to one of two transmission lines, labeled ``up'' and ``down'', via GUEs. A propagating photon scatters sequentially on the qubits, while linear optical elements performing unitary transformations $\mathcal U_n$ couple the transmission lines. A projective measurement of the qubits is performed upon detecting the photon at the output. (b)~Model for the qubit-GUE interaction in each node $n$ with cross-Kerr frequencies $V_1^n$ and $V_2^n$, and (c)~corresponding superconducting implementation adapted from Fig.~\ref{fig:fig1}(c). (d)~Quantum circuit realized using the setup in (a), where the double circles represent controlled-Z gates between the qubits and the photon as virtual ``flying qubit'' with states  $\ket{\text{up}}_\text{ph}$ and $\ket{\text{down}}_\text{ph}$, corresponding to the photon propagating in transmission line ``up'' and ``down'', respectively.}
\end{figure}

Our approach enables the realization of large scale quantum processing units, where quantum information is processed in local nodes, and routed using unidirectional emitters. The setup we have in mind is represented in Fig.~\ref{fig:fig6}(a), where we represent a possible such architecture, with a set of stationary atomic qubits acting as quantum register, and GUEs acting as an interface between a waveguide and the stationary qubits. The idea is to mediate effective long-range multi-qubit interactions by using (i) sequences of scattering events induced by unidirectional couplings between a single photon as ``flying qubit'' and each stationary qubit, (ii) local single-qubit operations, and (iii) linear optics represented by unitary operations $\mathcal U_n$ acting on two waveguides, including in particular 50/50 beam-splitter operations. The applicability of this architecture is illustrated below for quantum state transfer between distant stationary qubits, as well as the generation and manipulation of stabilizer codes. 

The scattering events are designed as follows [see Fig.~\ref{fig:fig6}(b)]. Denoting the parameters and operators associated with node $n=1,\ldots N$ with an index $n$, each GUE is initially prepared in its ground state $\ket{G}_n$, and returns to this state after the photon scattering. The coupling between each stationary qubit (with states $\{\ket{0}_{q,n},\ket{1}_{q,n}\}$) and its GUE consists of a purely non-linear cross-Kerr interaction, which can be described the Hamiltonian $\hat H_V=\sum_n\hat H^n_V$, where (see details in Supplementary Section \ref{secSM:implementation})
\begin{equation}\label{eq:HnVdef}
\hat H^n_V=-\ket{1}_{q,n}\!\bra{1} \left[V_1^n(\hat a_1^n)^\dagger \hat a^n_1+V_2^n(\hat a_2^n)^\dagger \hat a^n_2\right],
\end{equation}
ideally with identical frequencies $V_1^n=V_2^n\equiv V$. The effect of this interaction is then to shift the frequency of the excited states of the GUEs by $V$, conditional on qubit atom $n$ being in state $\ket{1}_{q,n}$, without breaking the properties of unidirectional coupling discussed above. A possible implementation of this interaction term with superconducting circuits, adapted from Fig.~\ref{fig:fig1}(c), is represented in Fig.~\ref{fig:fig6}(c), where the qubit atom is coupled via two SQUIDs to the GUE atoms. We note that (i)~the anharmonicity of the GUEs is inconsequential for the applications considered in this section as we consider the scattering of single photons, hence for simplicity the coupling between the artificial atoms of the GUEs are taken purely capacitive, and (ii)~the presence of capacitances in the coupling SQUIDs between the stationary qubit and the GUE induces a small direct coupling between the qubit and the waveguide modes, which could deteriorate the qubit lifetime; however, this coupling can be cancelled by subradiance due to interference in the photon emission from both coupling points, by taking the qubit transition frequency $\omega_q$ such that $\omega_q\overline d/v_g$ is an odd multiple of $\pi$ (see details in Supplementary Section \ref{secSM:implementation}).

The scattering of a photon on a single node $n$, represented in Fig.~\ref{fig:fig6}(b), is described within the input-output formalism by a \emph{single-photon scattering operator} 
\begin{equation}\label{eq:Sndpdnudeltadef}
\hat {\mathcal S}^n_{d',d}(\nu_p,\delta_p)=\bra{\text{vac},G_n}\hat b_{d'}^\text{out}(\nu_p)[\hat b_d^\text{in}(\delta_p)]^\dagger\ket{\text{vac},G_n}, 
\end{equation}
where $\ket{\text{vac},G_n}$ denotes the vacuum state of the waveguide, with the GUE in its ground state $\ket{G}_n$, and the input and output field operators in the frequency domain are defined via $\hat b_d^\text{in/out}(\delta_p)=(-i/\sqrt{2\pi})\int dt \hat b_d^\text{in/out}(t)e^{i\delta_p t}$. 
The single-photon scattering operator represents the action of the temporal evolution operator on qubit $n$, conditional on having an input photon with detuning $\delta_p$ (with respect to $\omega_0$), propagating in direction $d$ [either right (R) or left (L)] be scattered in direction $d'$ with detuning $\nu_p$. We consider a right-propagating input photon with frequency distribution given by some function $f(\delta_p)$ with qubit atom $n$ in some state $\ket{\psi}_{q,n}$, and write the state of the system before the scattering as {$\ket{\text{in}}=\int d\delta_p f(\delta_p)[\hat b_R^\text{in}(\delta_p)]^\dagger\ket{\text{vac},G_n}\ket{\psi}_{q,n}$}. The state after the scattering can then be expressed from Eq.~\eqref{eq:Sndpdnudeltadef} as {$\ket{\text{out}}=\sum_{d'}\int d\delta_p d\nu_p f(\delta_p) \hat{\mathcal S}^n_{d',R}(\nu_p,\delta_p)[\hat b^\text{out}_{d'}(\nu_p)]^\dagger\ket{\text{vac},G_n}\ket{\psi}_{q,n}$}.

The single-photon scattering operator in Eq.~\eqref{eq:Sndpdnudeltadef} can be obtained by using the quantum Langevin equation \eqref{eq:Otheisenberg} and the input-output relation \eqref{eq:inputoutput} (see details in Supplementary Section \ref{secSM:SLH}). In particular, under the conditions for unidirectional coupling of the GUEs to the waveguide as discussed above, we find $\hat {\mathcal S}^n_{L,R}(\nu_p,\delta_p)=0$ and $\hat {\mathcal S}^n_{R,R}(\nu_p,\delta_p)=\delta(\nu_p-\delta_p)\hat \sigma^n(\delta_p)$, with the Dirac $\delta$-function representing the conservation of the photon frequency in the scattering, and where
\begin{equation}\label{eq:mathcalSnRR}
\hat \sigma^n(\delta_p)=t\big(\Delta^n+\delta_p\big)\ket{0}_{q,n}\!\bra{0}+t\big(\Delta^n+\delta_p+V\big)\ket{1}_{q,n}\!\bra{1},
\end{equation} 
with the phase shift $t(\delta_p)=(2i\delta_p+\gamma_r)/(2i\delta_p-\gamma_r)$. The operator $\hat \sigma^n(\delta_p)$ realizes a generic {phase} gate on qubit $n$. Assuming the photon has a sharp frequency distribution $f(\delta_p)$ around $\delta_p=0$ relative to $\gamma_r$, by taking $V=\gamma_r$ this phase gate can be parametrized by the value of the tunable detuning $\Delta^n$ from GUE $n$. When $\Delta^n=-\gamma_r/2$, the two terms in Eq.~\eqref{eq:mathcalSnRR} acquire an opposite $\pi/2$ phase, and the phase gate becomes the Pauli operator $\hat \sigma_z^n=\ket{0}_{q,n}\!\bra{0}-\ket{1}_{q,n}\!\bra{1}$, up to an irrelevant global phase which can be absorbed in a redefinition of the phase of the output field operator $\hat b^\text{out}_{R}(\delta_p)$. When $\Delta^n\gg\gamma_r$ on the other hand, these two terms become identical, and the phase gate reduces to the identity operator $\mathbb 1$. 

This effective unidirectional photon -- qubit interaction finds immediate applications for the detection of individual itinerant microwave photons, which is a current technological challenge \cite{Gleyzes2007,Hadfield2009,Johnson2010,Inomata2016a,PhysRevX.6.031036,Kono2018,Besse2018}. This can be realized here with a Ramsey sequence, by preparing the atomic qubit in state $\ket{+}_{q,n}$, with $\ket{\pm}\equiv(\pm\ket{0}+\ket{1})/\sqrt{2}$. With $\Delta^n=-\gamma_r/2$, a resonant photon will be scattered unidirectionally by the GUE, while qubit atom $n$ will be left in state $\ket{-}_{q,n}$. The photon can then be detected by measuring the qubit state after applying a Ramsey $\pi/2$-pulse, which realizes a quantum non-demolition measurement of the itinerant photon, in analogy to the cavity-QED experiments in Refs.~\cite{Gleyzes2007,Johnson2010,PhysRevX.6.031036,Kono2018,Besse2018}. The resonance frequency $\omega_0$ of this detector can be tuned, while the detection bandwidth is given by $\gamma_r$ (see details in Supplementary Section~\ref{secSM:detector}).

In order to describe the more generic setup in Fig.~\ref{fig:fig6}(a), which now includes two waveguides as well as $N$ nodes, we make use of the SLH input-output formalism as discussed above (see details in Supplementary Section \ref{secSM:SLH}). We write the input and output field operators in the frequency domain as $\hat b_{d,j}^{\text{in/out}}(\delta)$, which now contains an additional index $j\in\{\text{up}, \text{down}\}$ labelling the two waveguides. The {single-photon scattering operator} for the whole system 
\begin{equation}\label{eq:Sgendef}
\hat {\mathcal S}_{d',d}^{j,i}(\nu_p,\delta_p) =\bra{\text{vac},\mathcal G}\hat b^{\text{out}}_{d',j}(\nu_p)[\hat b^{\text{in}}_{d,i}(\delta_p)]^\dagger\ket{\text{vac},\mathcal G},
\end{equation}
where $\ket{\text{vac},\mathcal G}=\ket{\text{vac}}\bigotimes_{n=1}^N\ket{G}_n$, then contains two additional indices representing the input line $i$ and the output line $j$ of the scattered photon. The derivation and general expression of this operator are provided in the Supplementary Section \ref{secSM:SLH}. 

In the ideal case where each GUE scatters photons unidirectionally, the scattering operator factorizes as $\hat {\mathcal S}_{L,R}^{j,i}(\nu,\delta)=0$ and we obtain
\begin{equation}\label{eq:SiRjRdef}
\hat{\mathcal S}_{R,R}^{j,i}(\nu_p,\delta_p)=\delta(\nu_p-\delta_p)e^{i\tilde\phi N}\left[\mathcal U_N\prod_{n=1}^N \left( \hat S_n(\delta_p) \mathcal U_{n-1}\right)\right]_{j,i}, 
\end{equation}
with the convention $\prod_{n=1}^NA_n=A_N\ldots A_1$, where the propagation phase $\tilde\phi=\omega_0 l/v_g$ (with $l$ the distance along the waveguide between two neighbouring nodes [see Fig.~\ref{fig:fig6}(a)]) enters only as a trivial global phase. 
Here $\mathcal U_n$ denote the linear optical elements acting on the photonic channels, as shown in Fig.~\ref{fig:fig6}(a). They can be represented as $2$-dimensional unitary matrices acting on a vectorial space which we denote as $\{\ket{\text{up}}_\text{ph}, \ket{\text{down}}_\text{ph}\}$, where the basis vectors $\ket{\text{up}/\text{down}}_\text{ph}$, correspond to the transmission line (either ``up'' or ``down'') in which the photon propagates. On this vectorial space the objects $\hat S_n(\delta)$ are diagonal matrices of qubit operators, which represent the photon scattering on each node. They are defined as
$\hat S_n(\delta_p)\ket{\text{down}}_\text{ph}=\ket{\text{down}}_\text{ph}$ and 
$\hat S_n(\delta_p)\ket{\text{up}}_\text{ph}=\ket{\text{up}}_\text{ph}\hat \sigma^n(\delta_p)$ as expressed in Eq.~\eqref{eq:mathcalSnRR}. 

The operator $\hat S_n(\delta_p)$ thus realizes a frequency-dependent {controlled-phase} gate between the propagating photon as a ``flying'' control qubit with states $\ket{\text{down}}_\text{ph}$ and $\ket{\text{up}}_\text{ph}$, and qubit atom $n$. For the applications discussed in the following the parameter  $\Delta^n$ will always be chosen such that the effective interaction in $\hat S_n(\delta_p=0)$, between a resonant photon and qubit atom $n$, is either trivial (with $\Delta^n\gg\gamma_r$), or realizes a \emph{controlled-Z} gate $\ket{\text{down}}_\text{ph}\!\bra{\text{down}}+\ket{\text{up}}_\text{ph}\!\bra{\text{up}}\hat\sigma^n_z$ (with $\Delta^n=-\gamma_r/2$) as represented in Fig.~\ref{fig:fig6}(d). 

\begin{figure}
\includegraphics[width=0.5\textwidth]{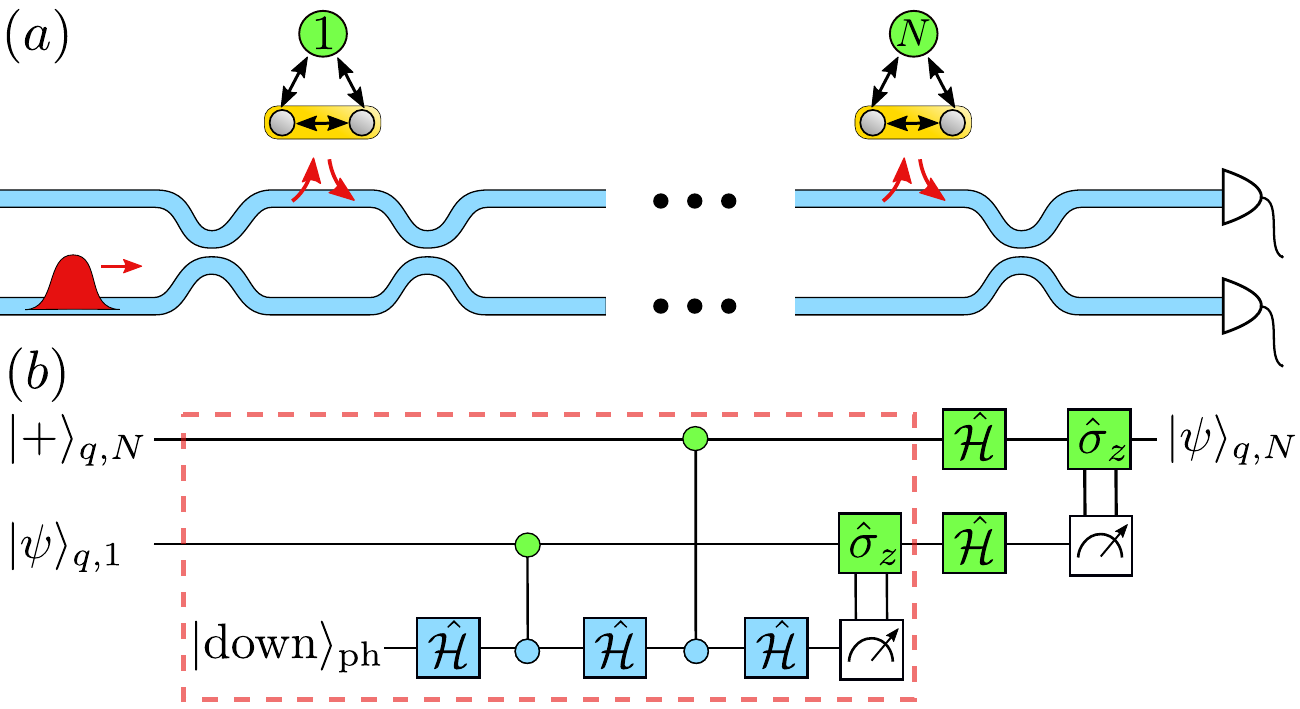}
\caption{\label{fig:fig7}\emph{Protocol for quantum state transfer.} (a)~Setup and (b)~corresponding quantum circuit realizing quantum state transfer from qubit atoms $1$ to $N$. Hadamard photonic gates $\hat {\mathcal H}$ are realized as 50/50 beam-splitters. The dashed red frame represents the action of the scattered photon, with the corresponding quantum circuit realizing a \emph{controlled-Z} gate between the two qubits. Upon detection of the photon at the output and reading out the final state of qubit $1$, the initial superposition state $\ket{\psi}_{q,1}$ is transferred to $\ket{\psi}_{q,N}$.}
\end{figure}

The entanglement structure of the scattering operator $\hat{\mathcal S}_{R,R}^{j,i}(\nu_p,\delta_p)$ in Eq.~\eqref{eq:SiRjRdef} is that of a \emph{matrix product operator}~\cite{Schollwock2011} with bond dimension $2$, which is a consequence of quantum information being carried in the network by a propagating photonic qubit. The photon scattering will thus generate entanglement in the qubit array, which can be used e.g. to prepare it in a matrix product state \cite{Schollwock2011} such as a GHZ state or 1D cluster state \cite{PhysRevLetters86.910} (see details in Supplementary Section \ref{secSM:MPS}). We note that this bond dimension, i.e., the amount of entanglement generatable by scattering a photon in the system, can in principle be increased by expanding the dimensionality of the photonic Hilbert space, e.g. by adding more waveguides.

As a first illustration of the working principles of this passive architecture, we consider one of the most basic protocol requiring quantum information routing, namely quantum state transfer between two stationary qubits. Here, the goal is to transfer a superposition state from one qubit atom, e.g. with $n=1$, to another (possibly distant) one, e.g. with $n=N$, as represented in Fig.~\ref{fig:fig7}(a). This is achieved by engineering the effective photon -- qubit interaction in such a way that the scattering operator in Eq.~\eqref{eq:SiRjRdef} realizes an effective controlled-Z gate \emph{between the distant qubits}, thereby enabling universal quantum computation in our architecture. The corresponding protocol circuit is represented in Fig.~\ref{fig:fig7}(b), which shows how the initial state of qubit $1$ $\ket{\psi}_{q,1}=c_0\ket{0}_{q,1}+c_1\ket{1}_{q,1}$ (with $|c_0|^2+|c_1|^2=1$) is transferred as $\ket{\psi}_{q,N}$ upon detection of the photon at the output, while quantum information is erased from qubit $1$. Here $\hat \sigma_z$ gates are applied conditional on the measurement of the photonic qubit in state $\ket{\text{up}}_\text{ph}$, and of qubit $1$ in state $\ket{1}_{q,1}$. The Hadamard gates are defined for the atomic qubits as $\hat{\mathcal H}=\ket{+}_{q,n}\!\bra{0}+\ket{-}_{q,n}\!\bra{1}$, and are similarly defined for the photonic qubit by replacing $\ket{0/1}_{q,n}$ with $\ket{\text{down}/\text{up}}_\text{ph}$. 

Assuming perfect control over the other parameters of the system, the average fidelity for the quantum state transfer protocol, as defined in the Supplementary Section \ref{sec:QST}, will depend on the photon frequency distribution $f(\delta_p)$ as $\overline{\mathcal F_\text{QST}}=\int d\delta_p |f(\delta_p)|^2\mathcal F_\text{QST}(\delta_p)$, where $\mathcal F_\text{QST}(\delta_p)=1-2(\delta_p/\gamma_r)^2+\mathcal O(\delta_p/\gamma_r)^3$. This sets a bound to the bandwidth $\Delta\omega$ of $f(\delta_p)$ as $\Delta\omega \ll \gamma_r$, and thus to the duration $T$ of the protocol as $T\geq 1/\Delta\omega$ (see below). Standard strategies for heralded quantum communication \cite{PhysRevLetters78.4293} can be translated to our protocol in Fig.~\ref{fig:fig7}(a), by adding ancillary stationary qubits to each node as quantum state ``backups'', thus enabling quantum communication with high fidelity, even with photon losses due for instance to amplitude attenuation in the waveguides or imperfect photon detection (see Supplementary Section \ref{sec:QST}). We note that, as discussed above, the photon detection can also be realized using additional nodes as detectors.

As a second application of our architecture for quantum networking, we now show that the setup of Fig.~\ref{fig:fig6}(a) allows to perform entangling operations on many stationary qubits, and can be used to passively probe and measure many-body operators, such as stabilizers of stabilizer codes for quantum error correction \cite{Gottesman1997}. A standard approach for measuring such stabilizer operators consists in entangling the qubits with an ancilla using two-body interactions; the stabilizers can then be accessed by measuring the ancilla \cite{Aguado2008, Jiang2008a,Muller2011}. Building on a previous protocol for measuring the parity of a pair of quantum dots as unidirectional emitters \cite{PhysRevLetters117.240501}, the measurement of stabilizers is achieved here using an interferometric setup with photonic qubits as ancillas, where the only non-trivial operations on the photons are $\mathcal U_0=\mathcal U_N=\hat{\mathcal H}$, and one obtains for the scattering operator of Eq.~\eqref{eq:SiRjRdef}
\begin{equation}\label{eq:Sstabilizer}
\hat{\mathcal S}_{R,R}^{j,\text{down}}(\nu_p,\delta_p)=\delta(\nu_p-\delta_p)e^{i\tilde
\phi N}\frac{\mathbb 1 +(-1)^{\delta_{j,\text{down}}} \prod_n\hat\sigma^n(\delta_p)}{2}.
\end{equation}

We recall that, with the parameters discussed above, for each stationary qubit $n$ we chose the parameters of the system such that the operator $\hat\sigma^n(\delta_p)$ is either the identity operator $\mathbb1$ or the Pauli operator $\hat \sigma^n_z$ when $\delta_p=0$. Defining an arbitrary subset $\mathcal I$ of the qubit array, the operator in Eq.~\eqref{eq:Sstabilizer} can thus be applied to entangle the state of the output photonic qubit (given by the index $j$) with the parity $\hat{\mathcal P}_{\mathcal I}=\prod_{n\in\mathcal I}\hat \sigma_z^n$ of the interacting qubits, which can then be measured by detecting the photon. More generally, allowing local unitary operations to be performed on the stationary qubits before and after the scattering enables the measurement of any operator of the form $\prod_{n\in \mathcal I} \hat {\boldsymbol \sigma}^n$, where $\hat{\boldsymbol \sigma}^n$ is an arbitrary rotation of $\hat \sigma_z^n$ on the Bloch sphere. Examples of such operators are the stabilizers of cluster states, which are universal resources for quantum computation \cite{Briegel2009}, and of stabilizer codes, where logical qubits are redundantly encoded in many physical qubits and protected by topology \cite{Gottesman1997}. 

\begin{figure}
\includegraphics[width=0.5\textwidth]{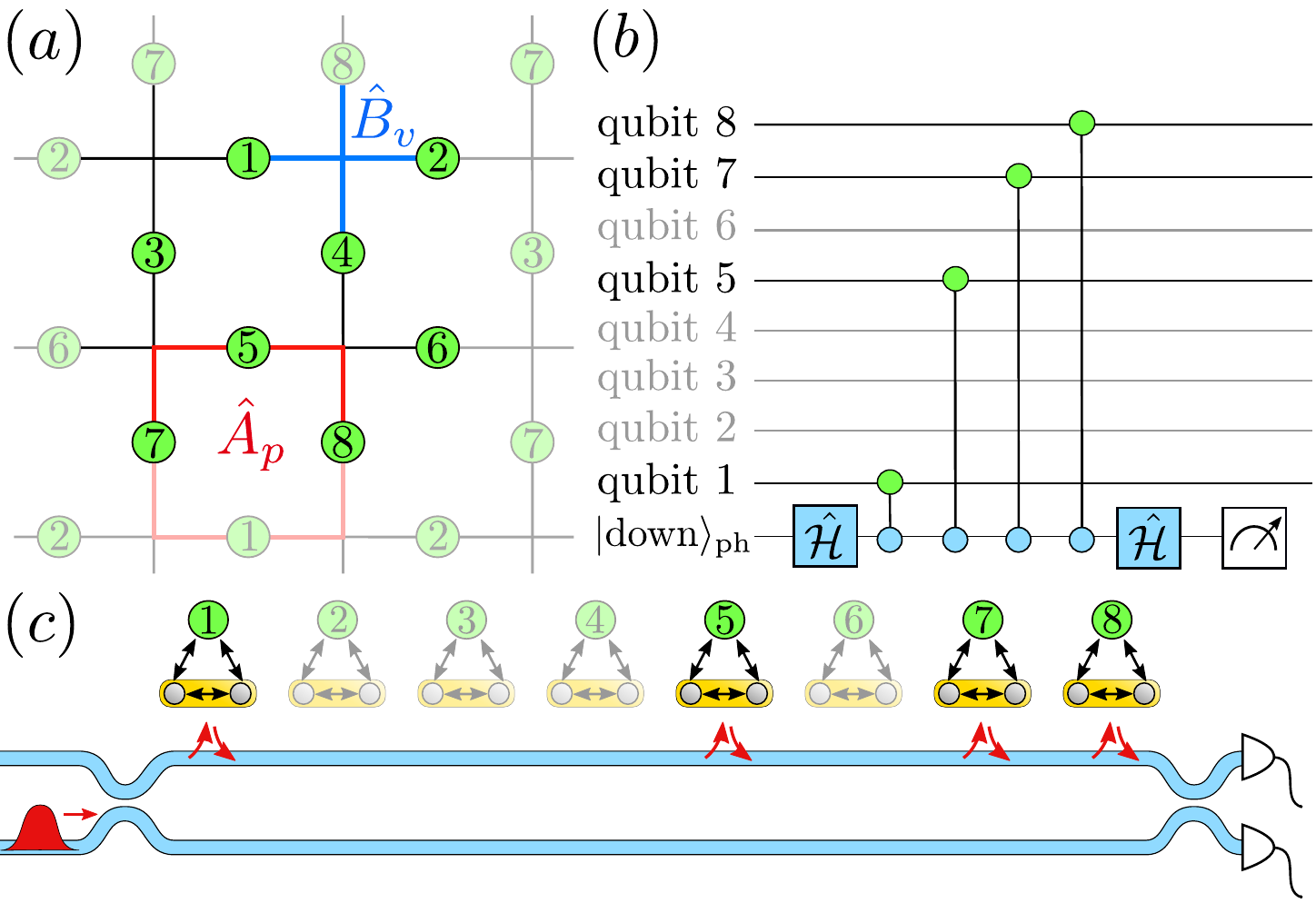}
\caption{\label{fig:fig8}\emph{Toric code generation and manipulation.} (a)~Abstract representation of a toric code, where qubits are located on the edges of a 2D lattice with periodic boundary conditions, with here $N=8$ qubits. The two types of stabilizers $\hat A_p$ and $\hat B_v$ are represented. (b)~Quantum circuit realizing a measurement of the stabilizer $\hat A_p$ represented in (a), and (c)~corresponding interferometric setup, with the detunings $\Delta^n$ of the GUEs chosen such that only nodes $1,5,7$ and $8$ are resonant with the photon.}
\end{figure}

Despite tremendous recent experimental progress towards the realization of stabilizer codes in superconducting platforms \cite{Riste2015,Corcoles2015,Kelly2015,PhysRevX.6.031041,PhysRevLetters119.180501,Gong2019}, scaling up the code distance (i.e., the number of physical qubits) beyond a few qubits remains a great challenge. As we show in the following, our architecture offers a naturally scalable approach to passively probe stabilizers, and thus generate and manipulate stabilizer codes.
As an example of stabilizer code, we consider the toric code \cite{Kitaev:2003jw}, where qubits are located on edges of a lattice with periodic boundary conditions. A minimal instance with $N=8$ qubits is represented in Fig.~\ref{fig:fig8}(a). The toric code has two types of stabilizers: for each plaquette $p$ and each vertex $v$ of the lattice we associate the stabilizers $\hat A_p=\prod_{n\in p}\hat \sigma_z^n$ and $\hat B_v=\prod_{n\in v}\hat \sigma_x^n$, with $\hat\sigma_x^n=\ket{0}_{q,n}\!\bra{1}+\ket{1}_{q,n}\!\bra{0}$. The logical subspace for encoding quantum information then consists of the four states which are eigenstates of all these stabilizers, with eigenvalue $+1$. A protocol for preparing the system in one of these four states consists in initializing all qubits in state $\bigotimes_n\ket{+}_{q,n}$. The plaquette operators $\hat A_p$ are then sequentially measured, and the system can be brought to the desired state by applying single-qubit $\hat \sigma_x^n$ gates afterwards, conditioned on the measurement outcomes (see Supplementary Section \ref{secSM:protocols}). 

In Fig.~\ref{fig:fig8}(b,c) we represent the quantum circuit and the setup realizing the measurement of the operator $\hat A_p$ shown in Fig.~\ref{fig:fig8}(a). Similar protocols, realized by scattering single photons, can be devised for (i) transferring a superposition state from a single additional stationary qubit to a logical quantum superposition state of the stabilizer code, as well as the reverse process, and (ii) realizing arbitrary logical qubit gates on the code subspace, as well as exponentiated string operators for quantum simulation of anyonic \cite{Jiang2008a} and fermionic models \cite{Zhu2018a} (see details in Supplementary Section \ref{secSM:protocols}).

\begin{figure}
\includegraphics[width=0.5\textwidth]{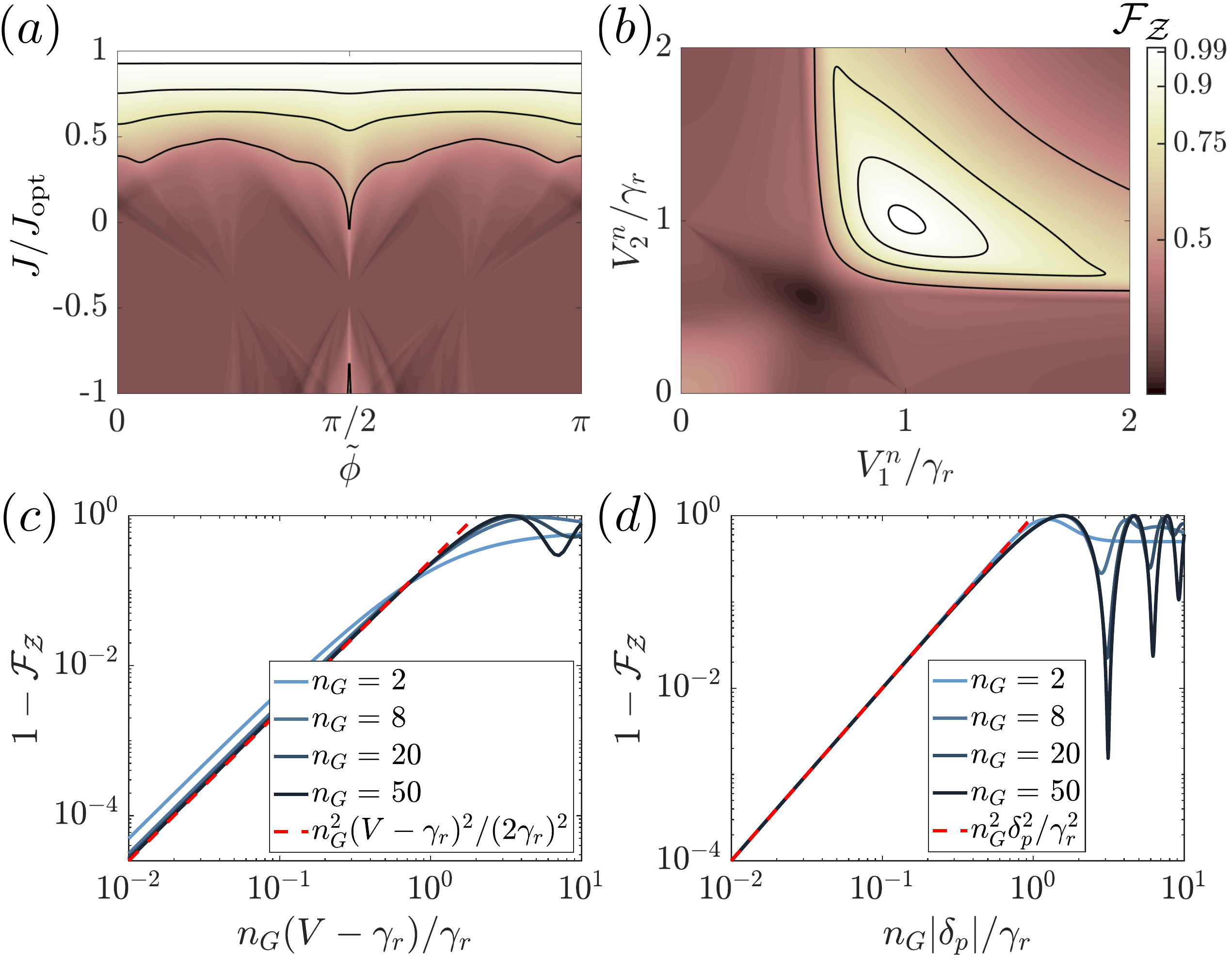}
\caption{\label{fig:fig9}\emph{Stabilizer measurements.} Fidelity $\mathcal F_{\mathcal Z}(\delta_p)$ for the measurement of the parity operator $\hat{\mathcal P}_{\mathcal I}$ on $n_G$ qubits prepared in state $\ket{\Psi_+}$, with $\phi=\phi_\text{opt}$, $\gamma_1=\gamma_2$, $r_k=0$ and $\Delta^n=-\gamma_r/2$. (a)~$V_1^n=V_2^n=\gamma_r$, $\delta_p=0$, $n_G=4$. (b-d)~$J=J_\text{opt}$, $\tilde\phi=0$. (b)~$\delta_p=0$, $n_G=4$. (c)~$\delta_p=0$, $V_1^n=V_2^n=V$. (d)~$V_1^n=V_2^n=\gamma_r$. }
\end{figure}

In order to quantify the efficiency of our scheme, we consider the task of performing a measurement of the parity operator $\hat{\mathcal P}_{\mathcal I}$ on $n_G=|\mathcal I|$ qubits, with the qubits initially prepared in state $\ket{\Psi_+}=\bigotimes_n \ket{+}_{q,n}$. Ideally, detecting the photon at the output of waveguide ``up'' or ``down'' projects this state to state $\ket{\Psi_\text{up}^\text{ideal}}=\frac{1}{\sqrt{2}}(\mathbb 1+\hat{\mathcal P}_{\mathcal I})\ket{\Psi_+}$ or $\ket{\Psi_\text{down}^\text{ideal}}=\frac{1}{\sqrt{2}}(\mathbb 1-\hat{\mathcal P}_{\mathcal I})\ket{\Psi_+}$, respectively. The average fidelity of this process, defined in the Supplementary Section \ref{secSM:protocols}, takes here the expression $\overline{\mathcal F_{\mathcal Z}}=\int d\delta_p|f(\delta_p)|^2\mathcal F_{\mathcal Z}(\delta_p)$, with
\begin{equation}\label{eq:FmathcalZdeltapdef}
\mathcal F_{\mathcal Z}(\delta_p) = \sum_j \left|\bra{\Psi_j^\text{ideal}}\smallint d\nu_p\hat{\mathcal S}^{j,\text{down}}_{R,R}(\nu_p,\delta_p)\ket{\Psi_+}\right|^2,
\end{equation}
which we represent in Fig.~\ref{fig:fig9}. 
In Fig.~\ref{fig:fig9}(a,b) we show this fidelity in situations where the photon scattering is not perfectly unidirectional, with the explicit expression of the scattering operator $\hat{\mathcal S}^{j,\text{down}}_{d',R}(\nu_p,\delta_p)$ from Eq.~\eqref{eq:Sgendef} provided in the Supplementary Section \ref{secSM:SLH}. In these cases where the dynamics is not purely cascaded, the fidelity also depends on the propagation phase $\tilde\phi$, in contrast to Eq.~\eqref{eq:Sstabilizer}. We observe robust fidelities of $\mathcal F_{\mathcal Z}(\delta_p)\gtrsim 99\%$ for small fluctuations of $V^n_{1,2}$ below $\sim 2\%$ and $J$ below $\sim 5\%$ around their optimal values. Fig.~\ref{fig:fig9}(c,d) represents situations where the photon scatters unidirectionally on each node, and shows that the infidelity $1-\mathcal{F_{\mathcal Z}}(\delta_p)$ scales quadratically with the deviation of $V$ around $\gamma_r$, with the number of interacting qubits $n_G$, and with the detuning of the photon $\delta_p$.

As an estimation of experimentally achievable performances, we consider $V=\gamma_r=2\pi \times 50$ MHz. From Fig.~\ref{fig:fig9}(d), the gate infidelity intrinsic to our protocol remains below $1\%$ as long as the photon detuning is below $|\delta_p|\lesssim 0.1 \gamma_r/n_G$. This sets a bound to the duration $T$ of a stabilizer measurement, as the bandwidth $\Delta\omega$ of the photon frequency distribution $f(\delta_p)$ must satisfy $T\Delta\omega \geq 1$. For instance, assuming the photon wavepacket has a truncated gaussian temporal distribution, we obtain an average fidelity $\overline{\mathcal F_{\mathcal Z}}$ above $99\%$ with $T=400$ ns for $n_G=4$ (see Supplementary Section \ref{secSM:protocols}). All $6$ independent stabilizers of the toric code with $N=8$ qubits can then be measured sequentially in a total time $\gtrsim 2.4$~$\mu$s. We note that measurements of several stabilizers involving non-overlapping subsets of qubits can be performed \emph{in parallel} using frequency-multiplexing techniques, as the frequency of their respective GUEs can be tuned to be resonant with probe fields with different frequencies. This allows to scale up stabilizer codes without increasing the total measurement time.



\section{Conclusion}

To conclude, we presented the design of a unidirectional artificial atom, and demonstrated its application as an on-chip interface between itinerant photons and stationary qubits. This design can be integrated in a modular architecture of photonic quantum networks, where controllable multi-qubit operations are realized by passively scattering itinerant photons, which we illustrated with the realization of quantum state transfer protocols with high fidelity, as well as the measurement of many-body stabilizer operators, pertinent for topological quantum error correction. 

In contrast to standard strategies for routing quantum information between nodes of a quantum network, our approach does not make use of circulators. In fact, rather than breaking Lorentz reciprocity for the electromagnetic field (i.e., the invariance under the exchange of source and detector) to control and route an itinerant quantum signal, here the propagation of the quantum signal is set by the itinerant photons injected in the network. This allows to achieve an effective non-reciprocal interaction \emph{between stationary qubits} with a rather simple design, and an architecture resilient to noise and perturbations.

\textit{Note added.} We recently became aware of related unpublished work by N. Gheeraert, S. Kono and Y. Nakamura.
 
\begin{acknowledgments}
We thank Hannes Pichler and Nicolas Roch for useful discussions. This work was supported by the Army Research Laboratory Center for Distributed Quantum Information via the project SciNet. AS is funded by the European Union's Horizon 2020 research and innovation program under grant agreement No. 714235. MLJ is funded by the European Union's Horizon 2020 research and innovation program under grant agreement No. 736943.
\end{acknowledgments}

\newpage
\newpage 

\onecolumngrid
\newpage
{
\center \bf \large 
Supplementary Material for: \\
A Unidirectional On-Chip Photonic Interface for Superconducting Circuits \vspace*{0.1cm}\\ 
\vspace*{0.0cm}
}
\begin{center}
P.-O. Guimond,${}^{1,2}$ B. Vermersch,${}^{1,2,3}$ M. L. Juan,${}^{4,2}$ A. Sharafiev,${}^{4,2}$ G. Kirchmair,${}^{4,2}$ and P. Zoller${}^{1,2}$\\
\vspace*{0.15cm}
\small{\textit{${}^1$ Center for Quantum Physics,
University of Innsbruck, Innsbruck A-6020, Austria\\
${}^2$ Institute for Quantum Optics and Quantum Information, Austrian Academy of Sciences, Innsbruck A-6020,
      Austria\\
  ${}^3$ Univ. Grenoble Alpes, CNRS, LPMMC, 38000 Grenoble, France\\
   ${}^4$ Institute for Experimental Physics, University of Innsbruck, A-6020 Innsbruck, Austria  }}\\
\vspace*{0.25cm}
\end{center}

\setcounter{section}{0}
\renewcommand\thesection{\Alph{section}}
\renewcommand\thesubsection{\arabic{subsection}}

\section{Directionality of photon emission}\label{secSM:directionality}
Here we provide details on the derivation of Eqs.~\eqref{eq:Otheisenberg} and \eqref{eq:Gsdef}, which are used in the main text to describe the directionality of photon emission of the GUE. In a frame rotating with a central frequency $\omega_0$, the Hamiltonian for the artificial atoms reads
\begin{equation}
\hat H_a=-\sum_{k=1}^2\Delta_k \hat a^\dagger_k \hat a_k-\left(U_k/2\right) \hat a^\dagger_k\hat a^\dagger_k\hat a_k\hat a_k
  + J\left(\hat a^\dagger_1\hat a_2+\hat a^\dagger_2\hat a_1\right)-\chi \hat a^\dagger_1\hat a_1\hat a^\dagger_2\hat a_2,
\end{equation}
where $\Delta_k=\omega_0-\omega_k$. In an interaction picture with respect to the photonic Hamiltonian $\hat H_\text{ph}$, the interaction between the artificial atoms and the waveguide reads
\begin{equation}
\hat H_\text{int}(t)=\frac{1}{\sqrt{2\pi}}\int d\omega e^{i(\omega-\omega_0)t} \left[\hat b_R^\dagger(\omega)\left(e^{i\omega \overline d/v_g} \hat L_1+ \hat L_2\right) 
+ \hat b_L^\dagger(\omega)\left(\hat L_1+e^{i\omega \overline d/v_g} \hat L_2\right) +\text{h.c.}\right].
\end{equation}
and the total Hamiltonian is given by $\hat H_\text{tot}(t)=\hat H_a+\hat H_\text{int}(t)$. Denoting for the initial time $t_0$, the Heisenberg equations of motion for the field operators then yield
\begin{equation}
\begin{aligned}
\hat b_R(\omega,t)=\hat b_R(&\omega,t_0)-\frac{i}{\sqrt{2\pi}}\int_{t_0}^tdt' e^{i(\omega-\omega_0)t'}\left(e^{i\omega \overline d/v_g}\hat L_1+\hat L_2\right)
\\ \hat b_L(\omega,t)=\hat b_L(&\omega,t_0)-\frac{i}{\sqrt{2\pi}}\int_{t_0}^tdt' e^{i(\omega-\omega_0)t'}\left(\hat L_1+e^{i\omega \overline d/v_g}\hat L_2\right).
\end{aligned}
\end{equation}

Injecting these expression in the Heisenberg equation for an arbitrary atomic operator $\hat O(t)$, which read $(d/dt)\hat O(t)=-i[\hat O(t),\hat H_\text{tot}(t)]$, we obtain the quantum Langevin equation
\begin{equation}\label{eq:heisenbergO}
\begin{aligned}
\frac{d}{dt}\hat O(t)=&-i\left[\hat O,\hat H_a\right]
+\hat L_R^\dagger\left[\hat O,\hat L_2\right]+\hat L_L^\dagger\left[\hat O,\hat L_1\right]
+\left[\hat L_2^\dagger,\hat O\right]\hat L_R+\left[\hat L_1^\dagger,\hat O\right]\hat L_L
\\ & +\sum_{d=R,L}[\hat b_d^\text{in}(t)]^\dagger\left[\hat O(t),\hat L_d\right]+\left[\hat L_d^\dagger,\hat O(t)\right]\hat b_d^\text{in}(t),
\end{aligned}
\end{equation}
where we defined the collective coupling operators as $\hat L_{R}=e^{i\phi}\hat L_1+\hat L_2$ and $\hat L_{L}=\hat L_1+e^{i\phi}\hat L_2$ with $\phi=\omega_0\overline d/v_g$, and the input fields as
\begin{equation}\label{eq:bdintdef}
\hat b_d^\text{in}(t)=\frac{i}{\sqrt{2\pi}}\int d\omega e^{-i(\omega-\omega_0)(t-t_0)}\hat b_d(\omega,t_0),
\end{equation}
(with $d=R,L$) which satisfy $[\hat b_d^\text{in}(t),[\hat b_{d'}^\text{in}(t')]^\dagger]=\delta(t-t')\delta_{d,d'}$. In deriving Eq.~\eqref{eq:heisenbergO}, we used integrals of the form $\int d\omega e^{i\omega t}=2\pi\delta(t)$, and we made use of a markovian approximation where any retardation effects, due e.g. to the finite time-delays in the propagation of photons between the quantum emitters, is set to $0^+$ in the final expression. We then obtain the expression of Eq.~\eqref{eq:Otheisenberg} by rearranging terms and by defining the effective Hamiltonian $\hat H_\text{eff}=\hat H_a+\sin(\phi)\left(\hat L_2^\dagger\hat L_1+\hat L_1^\dagger \hat L_2\right)$. Defining the output fields
\begin{equation}\label{eq:bdoutdef}
\hat b_d^\text{out}(t)=\frac{i}{\sqrt{2\pi}}\int d\omega e^{-i(\omega-\omega_0)(t-t_1)}\hat b_d(\omega,t_1),
\end{equation}
with arbitrary $t_1>t$, we have $[\hat b_d^\text{out}(t),[\hat b_{d'}^\text{out}(t')]^\dagger]=\delta_{d,d'}\delta(t-t')$, and the input-output field relations read \cite{gardiner2004quantum}
\begin{equation}\label{eq:boutdefSM}
\hat b^\text{out}_d(t)=\hat b^\text{in}_d(t)+\hat L_d(t).
\end{equation}

The quantum Langevin equation \eqref{eq:heisenbergO} can be interpreted according to Ito quantum stochastic calculus, and integrated. In the particular case where the waveguide is initially in its vacuum state $\ket{\text{vac}}$, we have $\hat b_d^\text{in}(t)\ket{\text{vac}}=0$. Moving back to the Schr\"odinger picture, we then express the average value of the arbitrary atomic operator $\hat O(t)$ in Eq.~\eqref{eq:heisenbergO} as $\langle \hat O(t)\rangle=\text{Tr}[\hat O\hat\rho(t)]$, where $\hat\rho(t)$ is the density matrix for the atoms, and we obtain from Eq.~\eqref{eq:heisenbergO} the master equation
 \begin{equation}\label{eq:MEatomdef}
 \frac{d}{dt}\hat\rho=-i\big[\hat H_\text{eff},\hat\rho\big]+\mathcal D\big[\hat L_R\big]\hat\rho+\mathcal D\big[\hat L_L\big]\hat\rho,
\end{equation}
where $\mathcal D[\hat a]\hat\rho= \hat a\hat\rho \hat a^\dagger-\frac12\{\hat a^\dagger \hat a,\hat \rho\}$.

In order to obtain Eq.~\eqref{eq:Gsdef}, we consider the situation where the atoms are prepared in state $\ket{R}$ at time $t=0$. Over time, the system will spontaneously emit a photon in the waveguide, with the atoms returning to their ground state $\ket{G}$.
We then make a Wigner-Weisskopf ansatz for the density matrix of the GUE as
\begin{equation}
\hat\rho= P_g(t)\ket{G}\bra{G}+\ket{\Psi(t)}\bra{\Psi(t)},
\end{equation}
where $\ket{\Psi(t)}=c_R(t) \ket{R}+c_L(t)\ket{L}$ with $P_g(t)+|c_R(t)|^2+|c_L(t)|^2=1$, which provides from Eq.~\eqref{eq:MEatomdef}  
\begin{equation}\label{eq:ddtpsit}
i\frac{d}{dt}\ket{\Psi(t)}=\left[\hat H_\text{eff}-\frac{i}2\big(\hat L^\dagger_R\hat L_R+\hat L^\dagger_L\hat L_L\big)\right]\ket{\Psi(t)}.
\end{equation} 
Denoting the Laplace transform of $\ket{\Psi(t)}$ as $\ket{\tilde \Psi(s)}\equiv\mathcal L[\ket{\Psi(t)}](s)$, Eq.~\eqref{eq:ddtpsit} can be solve as $\ket{\tilde \Psi(s)}=\hat F^{-1}(s)\ket{\Psi(t_0)}$, with $\hat F(s)$ defined in Eq.~\eqref{eq:Gsdef} as 
\begin{equation}
\hat F(s)=s+i\hat H_\text{eff}+\frac12 \left(\hat L_R^\dagger \hat L_R+\hat L_L^\dagger \hat L_L\right).
\end{equation}
From Eq.~\eqref{eq:boutdefSM}, this provides for the wavepacket of the emitted photon $f_{R/L}(t)=\bra{G}\hat L_{R/L}\mathcal L^{-1}[\hat F^{-1}(s)\ket{R}](t)$, and we define the emission directionality as $\beta_\text{dir}=\int_{0}^\infty dt|f_R(t)|^2$. 

\section{Superconducting circuit implementation of unidirectional emitters}
\label{secSM:simulations}

The circuit implementing the GUE is represented in Fig.~\ref{fig:fig1}(c), and consists of two transmons interacting via a SQUID and coupled at two points to an open transmission line. Following standard quantization procedures~\cite{Johansson_2006,Lalumiere:2013io,doi:10.1002/cta.2359}, we decompose the transmission line, with inductance and capacitance per unit length $l_0$ and $c_0$, into segments of finite lengths $\Delta x$, and write the Lagrangian of the system as $L= \frac12\dot{\boldsymbol \varphi}^T \bar{\bar C}\dot{\boldsymbol \varphi} - V$, where ${\boldsymbol \varphi}=(\varphi_1,\varphi_2,\varphi_{\text{TL},1},\varphi_{\text{TL},2},\varphi_{\text{TL},3},\ldots)^T$ contains the superconducting phase variables associated to the transmons ($\varphi_{1}$ and $\varphi_2$), and to each segment of the transmission line ($\varphi_{\text{TL},i}$), indexed from left to right. Denoting the indices for the segments coupled to each transmon as $i_1$ and $i_2$, the capacitance matrix reads $\bar{\bar C}=
\begin{pmatrix}
\bar{\bar C}_a & - \bar{\bar C}_{a,\text{TL}}
\\ - \bar{\bar C}_{a,\text{TL}}^T & \bar{\bar C}_\text{TL}
\end{pmatrix}$, with 
\begin{equation}
\bar{\bar C}_a=\begin{pmatrix}
C_1+\overline C+c'_1 & -\overline C
\\ -\overline C & C_2+\overline C+c'_2
\end{pmatrix},
\end{equation}
$\left(\bar{\bar C}_\text{TL}\right)_{j,k}=\delta_{j,k}\left(c_0\Delta x+\delta_{j,i_1}c'_1+\delta_{j,i_2}c'_2\right)$, and $(\bar{\bar C}_{a,\text{TL}})_{j,k}=c'_1\delta_{j,1}\delta_{k,i_1}+c'_2\delta_{j,2}\delta_{k,i_2}$. The potential energy, on the other hand, reads
\begin{equation}
V=\frac{1}{2l_0\Delta x}\sum_i (\varphi_{\text{TL},i+1}-\varphi_{\text{TL},i})^2 - \overline E_J\cos[(\varphi_2-\varphi_1)/\varphi_0]
 - E_J^1\cos(\varphi_1/\varphi_0)-E_J^2\cos(\varphi_2/\varphi_0),
\end{equation}
with $\varphi_0=\hbar/2e$ ($e$ is the elementary charge). 

Defining the conjugate variables ${\boldsymbol Q}=\frac{\partial L}{\partial\dot{\boldsymbol \varphi}}=\bar{\bar C}\dot{\boldsymbol \varphi}$, we obtain the Hamiltonian of the full system 
\begin{equation}
H_\text{tot}={\boldsymbol Q}^T\dot{\boldsymbol \varphi}-L=\frac12{\boldsymbol Q}^T\cdot \left(\bar{\bar C}\right)^{-1}\cdot{\boldsymbol Q}+V,
\end{equation}
which can be decomposed into $H_\text{tot}=H_a+H_\text{ph}+H_\text{int}$, with an atomic term $H_a$, a term for the transmission line $H_\text{ph}$, and an interaction term $H_\text{int}$. For the artificial atoms we obtain
\begin{equation}\label{eq:hacos}
H_a=\frac12\sum_kQ_k^2 \left(\bar{\bar C}\right)^{-1}_{k,k}-E_J^1\cos(\varphi_1/\varphi_0)-E_J^2\cos(\varphi_2/\varphi_0)
 +Q_1Q_2\left(\bar{\bar C}\right)^{-1}_{1,2}-\overline E_J\cos[(\varphi_2-\varphi_1)/\varphi_0].
\end{equation}
We then promote the phase and charge variables to operators satisfying $[\hat \varphi_k,\hat Q_l]=i\delta_{k,l}$, and express the Hamiltonian in terms of bosonic annihilation operators $\hat a_1$ and $\hat a_2$, with
\begin{equation}\label{eq:varphikQkdef}
\begin{aligned}
\hat \varphi_k=\varphi_0\left(\frac{2 E_C^k}{E_J^k}\right)^{1/4}(\hat a^\dagger_k+\hat a_k),
\\ \hat Q_k=2ei\left(\frac{E_J^k}{32E_C^k}\right)^{1/4}(\hat a^\dagger_k-\hat a_k).
\end{aligned}
\end{equation}
Here $E_C^k=e^2/2C_k^\text{eff}$, with $C_k^\text{eff}=1/\left(\bar{\bar C}\right)^{-1}_{k,k}\approx C_k+c'_k+\overline C$ for $(c'_k,\overline C)\ll C_k$. 

The atomic Hamiltonian $\hat H_a$ then takes the expression of Eq.~\eqref{eq:Hadef} by expanding the cosine functions in Eq.~\eqref{eq:hacos} up to fourth order, in the limit $\langle|\hat \varphi_k|\rangle\ll\varphi_0$, which is achieved in the transmon regime $E_C^k\ll E_J^k$, and discarding counter-rotating terms in a rotating wave approximation valid for $\overline C\ll C_k$ and $\overline E_J\ll E_J^k$. To estimate the value of the parameters in Eq.~\eqref{eq:Hadef}, we keep only the leading order terms, and find for the transition frequencies $\omega_k\approx\sqrt{8E_J^kE_C^k}$, while the anharmonicities read $U_k\approx E_C^k$. The linear interaction terms have a capacitive component $J_C$ and an inductive component $J_I$ as expressed in Eq.~\eqref{eq:Jexpr}. We note that the conditions of $\overline C\ll C_k$ and $\overline E_J\ll E_J^k$ are required here in order to be able to neglect counter-rotating terms such as $(J_C+J_I)(\hat a^\dagger_1\hat a_2^\dagger+\hat a_1\hat a_2)$. Similar considerations apply for the non-linear cross-Kerr interaction $\chi$ as expressed in Eq.~\eqref{eq:Uexpr}. 

For the transmission line Hamiltonian on the other hand, in the limit $\Delta x\to0$ the only non-vanishing terms are
$H_\text{ph}=\int dx \left[\partial_x\varphi(x)\right]^2/{2l_0}+{q(x)^2}/{2c_0}$,
where $\varphi(x)$ is the phase variable at position $x$ in the waveguide, and $q(x)$ the charge density. We then express these fields in second quantization in terms of the bosonic operators $\hat b_{R}(\omega)$ and $\hat b_L(\omega)$ as 
\begin{equation}
\begin{aligned}
\!\hat\varphi(x)= & \int d\omega \sqrt{\frac{Z_0}{4\pi\omega}}\left(\hat b_R(\omega)e^{i\omega x/v_g}+\hat b_L(\omega)e^{-i\omega x/v_g}\right)
  + \text{h.c.},
\\ \hat q(x)= & -i\int d\omega \sqrt{\frac{Z_0c_0^2\omega}{4\pi}}\left(\hat b_R(\omega)e^{i\omega x/v_g}+\hat b_L(\omega)e^{-i\omega x/v_g}\right)
  +  \text{h.c.},
\end{aligned}
\end{equation}
where $v_g=1/\sqrt{l_0 c_0}$ is the group velocity of photons in the transmission line and $Z_0=\sqrt{l_0/c_0}$ the transmission line characteristic impedance, and we obtain $\hat H_\text{ph}=\int d\omega \omega [\hat b_R^\dagger(\omega)\hat b_R(\omega)+\hat b_L^\dagger(\omega)\hat b_L(\omega)]$.  

Finally, for the interaction term we obtain
\begin{equation}
\hat H_\text{int}=\frac{1}{c_0}\begin{pmatrix} \hat Q_1 & \hat Q_2\end{pmatrix}
\bar{\bar C}_a^{-1}\begin{pmatrix} c'_1 \hat q(x_1)\\ c'_2 \hat q(x_2)\end{pmatrix},
\end{equation}
where $x_1$ and $x_2$ denote the position of the two coupling points along the waveguide. We then obtain the expression in Eq.~\eqref{eq:Hintdef} by setting $x_1=0$ and $x_2=\overline d$, redefining the phase of the right-propagating modes as $\hat b_R(\omega)\to \hat b_R(\omega)e^{-i\omega \overline d/v_g} $ and approximating the couplings to be constant over the relevant bandwidth. Assuming for simplicity $E^1_J/E_C^1\approx E^2_J/E_C^2$, the coupling rates express as $\gamma_k=(c'_k/C^\text{eff}_k)^2 \omega_0 e^2 Z_0\sqrt{E_J^k/(8E_C^k)}$ and the cross-coupling coefficients as $r_k=\overline C/C_k^\text{eff}$.

We now study numerically the validity of the model of $\hat H_a$ in Eq.~\eqref{eq:Hadef}, for our implementation with superconducting circuits. 
We identify the model parameters (namely $J$, $\chi_k$, $U$ and $\omega_k$) from the full atomic Hamiltonian expressed in Eq.~\eqref{eq:hacos}, with the phase and charge operators operators in Eq.~\eqref{eq:varphikQkdef}. 

Moreover, the effect of counter-rotating terms in the full Hamiltonian, which are the terms that do not preserve the number of excitations $N_\text{exc}=\hat a^\dagger_1\hat a_1+\hat a^\dagger_2\hat a_2$, are accounted for by treating them as perturbation, and applying standard second-order perturbation theory. That is, we decompose the full Hamiltonian into $\hat H_a=\bigoplus_{n_e,n_e'=0}^\infty \hat H_{a}^{n_e,n_e'}$, where the diagonal part $\hat H_{a}^{n_e,n_e}$ is the projection of $\hat H_a$ on the subspace with $n_e$ excitations, while the off-diagonal part $\hat H_{a}^{n_e,n_e'\neq n_e}$ couples subspaces with different excitation numbers $n_e$ and $n_e'$. For small perturbations with respect to the optical frequency $\omega_0$, the renomalized Hamiltonian, which now includes the second-order contribution of these off-diagonal terms, is then obtained as
\begin{equation}
\hat H_a^{(2)}\approx\hat H_a-\sum_{n_e,n_e'\neq n_e}\frac{\hat H_{a}^{n_e,n_e'}\hat H_{a}^{n_e',n_e}}{\omega_0(n_e'-n_e)}.
\end{equation} 

\begin{figure}
\includegraphics[width=0.7\textwidth]{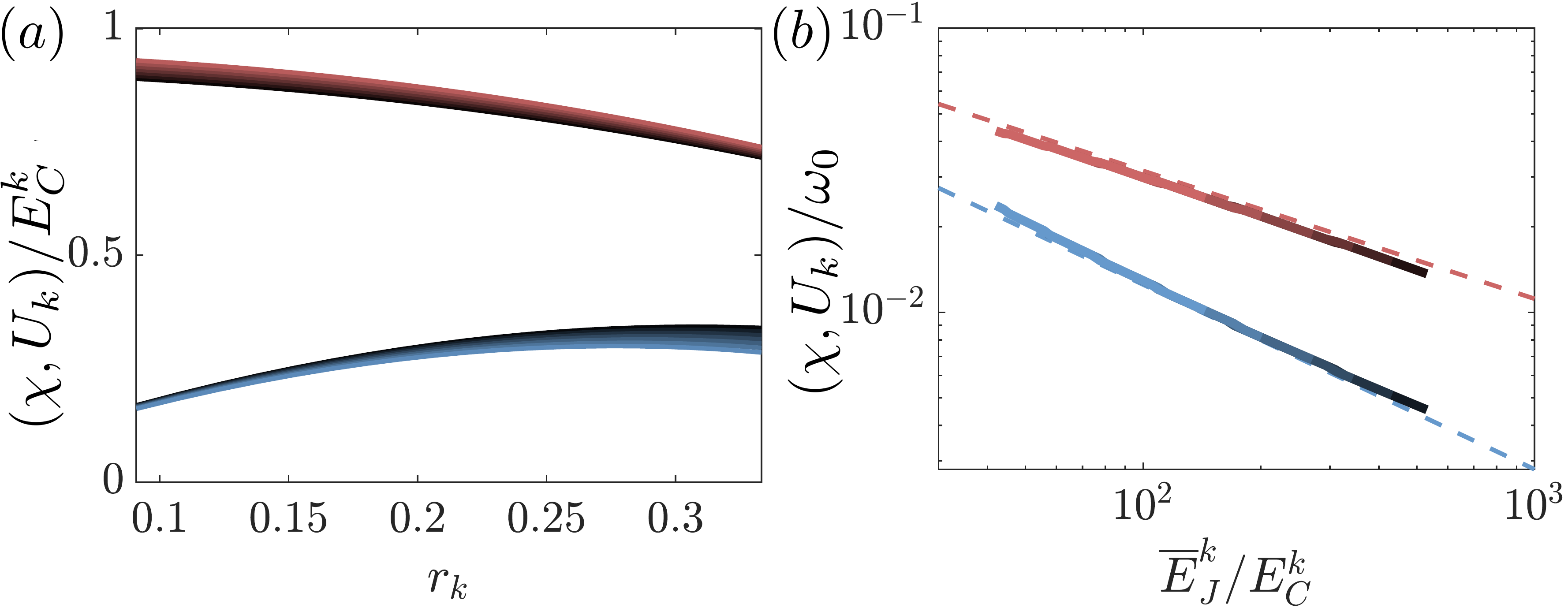}
\caption{\label{fig:fignum}\emph{Superconducting circuit implementation.} Numerical evaluation of the effective anharmonicities $\chi$ (blue) and $U_k$ (red) obtained by optimizing $E_J^k$ and $\overline E_J$ such that $J=J_\text{opt}$ and $\omega_k=\omega_0$. (a)~$\omega_0=2\pi\times8$ GHz and $C_1=C_2\in[200,600]$ fF (dark to light colors), while $r_k=\overline C/C_k^\text{eff}$. (b)~Maximum value of $\chi$, and corresponding $U_k$, obtained by optimizing $r_k$, with $\omega_0\in2\pi\times[5,10]$  GHz (light to dark colors), with the ratio $\overline E_J^k/E_C^k$ varied as a function of $C_1=C_2\in[200,600]$ fF, where $\overline E_J^k=E_J^k+\overline E_J$. Dashed red $\propto (\overline E_J^k/E_C^k)^{-0.45}$. Dashed blue $\propto (\overline E_J^k/E_C^k)^{-0.65}$. }
\end{figure}

 For experimentally realistic parameters, the resulting cross-Kerr interaction $\chi$ and anharmonicities $U_k$ are shown in Fig.~\ref{fig:fignum}, where the Josephson energies $E_J^k$ and $\overline E_J$ are optimized such that $J=J_\text{opt}$ and $\omega_k=\omega_0$.
Fig.~\ref{fig:fignum}(a) shows a linear scaling of $\chi$ for weak cross-coupling coefficients $r_k$, while the anharmonicity $U_k$ decreases, and displays an optimal value for $\chi$ which is achieved with a small but non-negligible $r_k$. This optimal value is represented as a function of the photon frequency $\omega_0$ and the ratio $\overline E_J^k/E_C^k$ in Fig.~\ref{fig:fignum}(b), where $\overline E_J^k=E_J^k+\overline E_J$, which shows that $\chi$ decreases for increasing ratio $\overline E_J^k/E_C^k$. A trade-off must thus be made between working in the transmon regime ($\overline E_J^k/E_C^k\gg 1$) in order for the artificial atoms to have small sensitivity to charge noise, and having large effective anharmonicities. For concreteness, a reasonable such trade-off can be taken as $\omega_0=2\pi\times8$ GHz and $\overline E_J^k/E_C^k=100$, in which case $\chi\approx 2\pi\times80$ MHz and $U_k\approx 2\pi\times240$ MHz.
  
\section{Implementation of unidirectional qubit -- photon interface}
\label{secSM:implementation}

\begin{figure}
\includegraphics[width=0.5\textwidth]{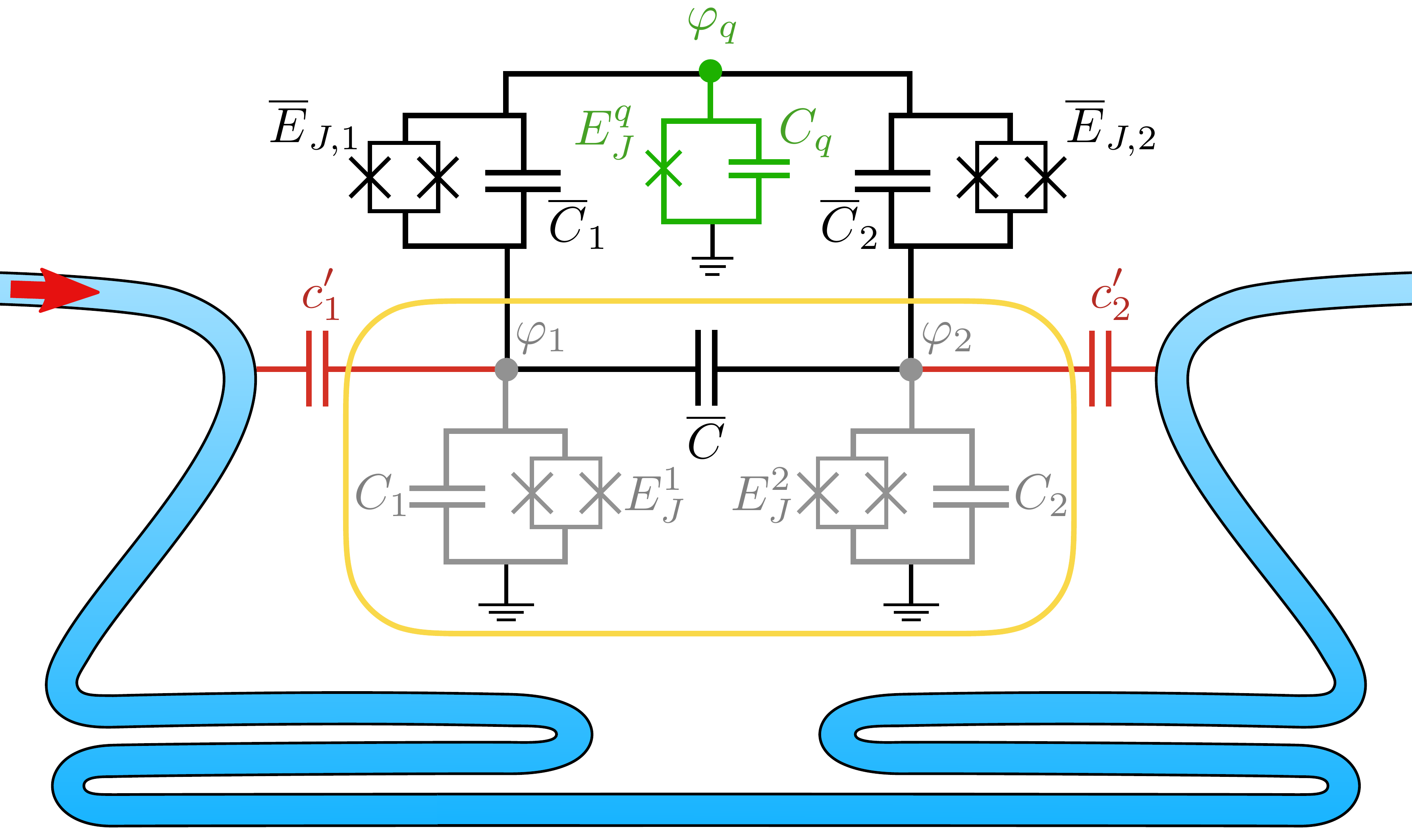}
\caption{\label{fig:figSM2}\emph{Superconducting circuit implementation of unidirectional photonic interface.} A transmon qubit, represented in green, with superconducting phase variable $\varphi_q$, is coupled with a purely non-linear cross-Kerr interaction mediated by two SQUIDs to a GUE, in yellow, which acts as unidirectional photonic interface. }
\end{figure}

Here we discuss the superconducting circuit implementation of the GUE as unidirectional photonic interface for an additional transmon qubit, as represented in Fig.~\ref{fig:figSM2}. Following the quantization procedure as described in Sec.~\ref{secSM:simulations} for the GUE, the full system, including the transmission line, is described by a Hamiltonian $\hat H_\text{tot}=\hat H_\text{ph}+\hat H_{a}+\hat H_\text{int}$, with the transmission line Hamiltonian reading 
\begin{equation}
\hat H_\text{ph}=\int d\omega \omega\left[\hat b_R^\dagger(\omega)\hat b_R(\omega)+\hat b^\dagger_L(\omega) \hat b_L(\omega)\right].
\end{equation}
In the regime of weakly coupled transmons, where $(c'_1,c'_2,\overline C,\overline C_{1},\overline C_{2})\ll(C_q,C_1,C_2)$ and $(\overline E_{J,1},\overline E_{J,2})\ll(E_J^1,E_J^2)$, the Hamiltonian for the artificial atoms, including both the GUE and the additional qubit, reduces to
\begin{equation}\label{eq:Hainterface}
\begin{aligned}
\hat H_a=&\sum_{k=1}^2\omega_k\hat a^\dagger_k \hat a_k -(U_k/2)\hat a^\dagger_k\hat a^\dagger_k\hat a_k\hat a_k
+\omega_q a^\dagger_q a_q -(U_q/2)\hat a^\dagger_q\hat a^\dagger_q\hat a_q\hat a_q+ J_C(\hat a^\dagger_1\hat a_2+\hat a^\dagger_2\hat a_1)\\&+\sum_{k=1}^2(J_{C,k}-J_{I,k})(\hat a^\dagger_q\hat a_k+\hat a^\dagger_k\hat a_q)-V_k\hat a^\dagger_q\hat a_q\hat a^\dagger_k\hat a_k.
\end{aligned}
\end{equation}
In analogy to the quantization of the GUE variables yielding Eq.~\eqref{eq:Hadef}, here we also quantized the variables for the qubit charge ($Q_q$) and phase ($\varphi_q$) as 
\begin{equation}
\begin{aligned}
\hat \varphi_q=\varphi_0\left(\frac{2 E_C^q}{E_J^q}\right)^{1/4}(\hat a^\dagger_q+\hat a_q),
\\ \hat Q_q=2ei\left(\frac{E_J^q}{32E_C^q}\right)^{1/4}(\hat a^\dagger_q-\hat a_q),
\end{aligned}
\end{equation}
where $E_C^q=e^2/2C_q^\text{eff}$, with $C_q^\text{eff}\approx C_q+\sum_kC_{c,k}$ the effective qubit capacitance, and we assumed $E_C^q\ll E_J^q$. In Eq.~\eqref{eq:Hainterface}, the qubit frequency reads $\omega_q\approx \sqrt{8E_C^q E_J^q}$, while the qubit anharmonicity is given by $U_q\approx E_C^q$. For the qubit -- GUE interaction terms, we find a linear exchange interaction term with a capacitive ($J_{C,k}$) and an inductive ($J_{I,k}$) contribution, where
\begin{equation}
J_{C,k}\approx\frac{\sqrt{\omega_0\omega_q}}{2}\frac{\overline C_{k}}{\sqrt{C_q^\text{eff}C_k^\text{eff}}}, \ \ \ \ J_{I,k}\approx \frac{\sqrt{\omega_0\omega_q}}{2}\frac{\overline E_{J,k}}{\sqrt{E_J^qE_J^k}}.
\end{equation}
For the non-linear cross-Kerr terms, we have on the other hand
\begin{equation}
V_k\approx2 \overline E_{J,k}\sqrt{\frac{E_C^kE_C^q}{ E_J^k E_J^q}},
\end{equation}
which is constrained by the rotating wave approximation [which requires $J_{I,k}\ll(\omega_0,\omega_q)$] as $V_k\ll \sqrt{E_C^kE_C^q}$. In practice we consider values up to $V_k\sim2\pi\times50$ MHz, with $E_C^k=E_C^q\sim2\pi\times300$ MHz.

The interaction between qubit and GUE reduces to the expression of Eq.~\eqref{eq:HnVdef}, by setting $J_{C,k}=J_{I,k}$. We note that (i)~the Josephson energies of the coupling SQUIDs $\overline E_{J,k}$ can be independently controlled via flux biases, allowing to fine-tune $V_k$ such that $V_1=V_2=\gamma_r$, as required in the main text, (ii)~the frequency of the qubit ($\omega_q$) and of the GUEs ($\sim\omega_0$) can be far detuned (by several GHz), allowing to relax the condition $J_{C,k}\approx J_{I,k}$, which does not need to be met exactly in order to cancel deleterious excitation exchanges between the GUE and the qubit, and (iii)~as all the applications discussed here are achieved by scattering single photons with the GUE in its ground state $\ket{G}$, the cross-Kerr interaction term within the GUE, arising from the coupling SQUID in Fig.~\ref{fig:fig1}(c), is irrelevant here, and has been removed for simplicity. 

Finally, for the interaction between artificial atoms and the transmission line, we obtain, in analogy to Eq.~\eqref{eq:Hintdef}
\begin{equation}
\begin{aligned}
\hat H_\text{int}=\frac{1}{\sqrt{2\pi}}\int d\omega & \left[\hat b_R^\dagger(\omega)\left(e^{i\omega \overline d/v_g} \hat L_1+ \hat L_2\right) 
+ \hat b_L^\dagger(\omega)\left(\hat L_1+e^{i\omega \overline d/v_g} \hat L_2\right) +\text{h.c.}\right]
\\+\frac{1}{\sqrt{2\pi}}\int d\omega & \left[\hat b_R^\dagger(\omega)\left(e^{i\omega \overline d/v_g} \sqrt{\gamma_{q,1}}+ \sqrt{\gamma_{q,2}}\right)
+ \hat b_L^\dagger(\omega)\left(\sqrt{\gamma_{q,1}}+e^{i\omega \overline d/v_g} \sqrt{\gamma_{q,2}}\right)\right]\hat a_q +\text{h.c.},
\end{aligned}
\end{equation}
where the first component corresponds to the interaction of propagating photons with the GUE, while the second is an additional spurious interaction term coupling directly the qubit atom to the transmission line, with the rates 
\begin{equation}
\gamma_{q,k}\approx\gamma_k\frac{\omega_q}{\omega_0}\sqrt{\frac{E_C^k E_J^q}{E_J^k E_C^q}}\left(\frac{\overline C_{k}}{C_q}\right)^2.
\end{equation} 
Even though we typically have $\gamma_{q,k}\ll \gamma_k$ as ${\overline C_{k}}\ll{C_q}$, the presence of these additional couplings could lower the lifetime of the qubit atom by spontaneous photon emission in the transmission line. This can however be remedied by properly choosing the qubit frequency $\omega_q$. Indeed, let us consider the situation where the qubit is prepared in state $\ket{1}_q$, with the GUE in its ground state $\ket{G}$ and the transmission line in the vacuum state $\ket{\text{vac}}$. The dynamics of this system can be solved by means of a Wigner-Weisskopf ansatz, where the state of the system takes the expression 
\begin{equation}
\ket{\psi(t)}=\Big[c_q(t)e^{-i\omega_q t}\hat a^\dagger_q+c_1(t)e^{-i\omega_0t}\hat a^\dagger_1+c_2(t)e^{-i\omega_0t}\hat a^\dagger_2(t)+\!\!\int\!\! d\omega c_{R}(\omega,t)e^{-i\omega t}\hat b_R^\dagger(\omega)\!+\!c_{L}(\omega,t)e^{-i\omega t}\hat b_L^\dagger(\omega)\Big]\!\ket{0}_q\!\ket{G}\!\ket{\text{vac}},
\end{equation}
with $c_q(0)=1$ and $c_1(0)=c_2(0)=c_R(\omega,0)=c_L(\omega,0)=0$.  For simplicity, let us assume $\omega_k=\omega_0$, $\gamma_k=\gamma$ and $r_k=0$. From $(d/dt)\ket{\psi(t)}=-i\hat H_\text{tot}\ket{\psi(t)}$, we then have 
\begin{equation}
\begin{aligned}
\dot c_q(t)=&-i \sum_{k=1}^2 (J_{C,k}-J_{I,k})e^{-i(\omega_0-\omega_q)t} c_k(t)-\frac{i}{\sqrt{2\pi}}\int d\omega c_R(\omega,t)\left(\sqrt{\gamma_{q,1}}e^{-i\omega \overline d/v_g}+\sqrt{\gamma_{q,2}}\right)e^{-i(\omega-\omega_q)t}
\\ &-\frac{i}{\sqrt{2\pi}}\int d\omega c_L(\omega,t)\left(\sqrt{\gamma_{q,1}}+e^{-i\omega \overline d/v_g}\sqrt{\gamma_{q,2}}\right)e^{-i(\omega-\omega_q)t}
\\ \dot c_1(t)=&-i J_C c_2(t) -i (J_{C,1}-J_{I,1}) e^{i(\omega_0-\omega_q)t} c_q(t)
-i\sqrt{\frac{\gamma}{2\pi}}\int d\omega \left(c_R(\omega,t)e^{-i\omega \overline d/v_g}+c_L(\omega,t)\right)e^{-i(\omega-\omega_0)t}
\\ \dot c_2(t)=&-i J_C c_1(t) -i (J_{C,2}-J_{I,2}) e^{i(\omega_0-\omega_q)t} c_q(t)
-i\sqrt{\frac{\gamma}{2\pi}}\int d\omega \left(c_R(\omega,t)+c_L(\omega,t)e^{-i\omega \overline d/v_g}\right)e^{-i(\omega-\omega_0)t}.
\end{aligned}
\end{equation}
Inserting in these equations the formal solutions for the dynamics of the field variables, obtained from $(d/dt)\ket{\psi(t)}=-i\hat H_\text{tot}\ket{\psi(t)}$ as
\begin{equation}
\begin{aligned}
c_R(\omega,t)=&-\frac{i}{\sqrt{2\pi}}\int_{t_0}^tdt' e^{i(\omega-\omega_q)t'} c_q(t')\left(\sqrt{\gamma_{q,1}}e^{i\omega \overline d/v_g}+\sqrt{\gamma_{q,2}}\right)
-i\sqrt{\frac{\gamma}{2\pi}}\int_{t_0}^tdt' e^{i(\omega-\omega_0)t'} \left(c_1(t')e^{i\omega \overline d/v_g}+c_2(t')\right),
\\ c_L(\omega,t)=&-\frac{i}{\sqrt{2\pi}}\int_{t_0}^tdt' e^{i(\omega-\omega_q)t'} c_q(t')\left(\sqrt{\gamma_{q,1}}+\sqrt{\gamma_{q,2}}e^{i\omega \overline d/v_g}\right)
-i\sqrt{\frac{\gamma}{2\pi}}\int_{t_0}^tdt' e^{i(\omega-\omega_0)t'} \left(c_1(t')+c_2(t')e^{i\omega \overline d/v_g}\right),
\end{aligned}
\end{equation}
we obtain
\begin{equation}
\begin{aligned}
\dot c_q(t)=&-\left(\gamma_{q,1}+\gamma_{q,2}+2\sqrt{\gamma_{q,1}\gamma_{q,2}}e^{i\omega_q\overline d/v_g}\right)c_q(t)
-\left[\sqrt{\gamma\gamma_{q,1}}+\sqrt{\gamma\gamma_{q,2}}e^{i\omega_0\overline d/v_g}-i(J_{C,1}-J_{I,1})\right] e^{i(\omega_q-\omega_0)t}c_1(t)
\\&-\left[\sqrt{\gamma\gamma_{q,1}}e^{i\omega_0\overline d/v_g}+\sqrt{\gamma\gamma_{q,2}}-i(J_{C,2}-J_{I,2})\right] e^{i(\omega_q-\omega_0)t}c_2(t),
\\ \dot c_1(t)=&-\gamma c_1(t)-\left(i J_C+\gamma e^{i\omega_0\overline d/v_g}\right) c_2(t)-\left[\sqrt{\gamma\gamma_{q,1}}+\sqrt{\gamma\gamma_{q,2}}e^{i\omega_q\overline d/v_g}-i(J_{C,1}-J_{I,1})\right]e^{-i(\omega_q-\omega_0)t}c_q(t),
\\ \dot c_2(t)=&-\gamma c_2(t)-\left(i J_C+\gamma e^{i\omega_0\overline d/v_g}\right) c_1(t)-\left[\sqrt{\gamma\gamma_{q,1}}e^{i\omega_q\overline d/v_g}+\sqrt{\gamma\gamma_{q,2}}-i(J_{C,2}-J_{I,2})\right]e^{-i(\omega_q-\omega_0)t}c_q(t).
\end{aligned}
\end{equation}

One can then readily solve these differential equations. In particular, working in a regime with the qubit frequency far detuned with respect to the GUE, i.e., $|\omega_q-\omega_0|\ggg\sqrt{\gamma\gamma_{q,k}}$ and $|\omega_q-\omega_0|\gg |J_{C,k}-J_{I,k}|$, the residual linear exchange rates $(J_{C,k}-J_{I,k})$ between the qubit and the GUE in Eq.~\eqref{eq:Hainterface} adds a contribution to the linear coupling between qubit and waveguide as
\begin{equation}
\gamma_{q,k}^\text{eff}\approx\gamma_{q,k}+\gamma\left(\frac{J_{C,k}-J_{I,k}}{\omega_0-\omega_q}\right)^2.
\end{equation} 
This yields for the qubit amplitude $c_q(t)=e^{-i\Delta_q t}e^{-\gamma_q t}$, where the qubit atom frequency is shifted by $\Delta_q=2\sqrt{\gamma_{q,1}^\text{eff}\gamma_{q,2}^\text{eff}}\sin(\omega_qd/v_g)$, and the qubit undergoes a spontaneous photon emission with rate $\gamma_q=\sum_k\gamma_{q,k}^\text{eff}+2\sqrt{\gamma_{q,1}^\text{eff}\gamma_{q,2}^\text{eff}}\cos(\omega_qd/v_g)$. In particular, this decay rate vanishes if $\gamma_{q,1}^\text{eff}\approx\gamma_{q,2}^\text{eff}$ and if $\omega_q$ is taken such that $\omega_q d/v_g$ is an odd multiple of $\pi$. This is a manifestation of subradiance due to the destructive interference of photons emitted by the qubit atom via the two coupling points.

\section{SLH formalism for modeling input-output photonic quantum networks}  
\label{secSM:SLH}
Here we provide details on the SLH formalism employed for modeling input-output photonic quantum networks, with several GUEs coupled to a waveguide. We start by giving a brief introductory overview of the formalism in Sec.~\ref{sec:SMdefSLH}, including the composition rules for modeling composite photonic systems in a ``bottom-up'' approach. For more details, we refer the reader to the review in Ref.~\cite{doi:10.1080/23746149.2017.1343097}. In Sec.~\ref{sec:SMdrivendissipative} we apply the formalism to the network represented in Fig.~\ref{fig:fig1}(d), with an ensemble of $N$ GUEs coupled to a common waveguide as photonic bath. In Sec.~\ref{sec:SMscatteringoperator}, we derive the expression of the single-photon scattering operator, first for a single qubit coupled to a GUE, as represented in Fig.~\ref{fig:fig6}(b), and then for the more generic setup represented in Fig.~\ref{fig:fig6}(a).

\subsection{Definitions and properties}\label{sec:SMdefSLH}

In the SLH formalism, each element of an open input-output photonic quantum network with $N_c$ input and output photonic channels is represented by a triplet $G=\big({S},\hat {\boldsymbol L},\hat H\big)$. Here ${S}$ is an $N_c\times N_c$ scattering matrix describing the coupling between photonic quantum channels, $\hat {\boldsymbol L}$ is an $N_c\times 1$ vector of coupling operators representing the interaction between the system and the photonic channels, and $\hat H$ is the Hamiltonian of the system. For instance, in the situation represented in Fig.~\ref{fig:fig1}(d) where an ensemble of $N$ GUEs interacts via a common waveguide, each individual GUE couples to $N_c=2$ photonic channels, corresponding to the right- and left-propagating modes of the waveguide. Denoting the various parameters and operators associated with each composite emitter with a corresponding superscript $n$, the SLH triplet for each GUE $n$ is given by $\Big(\mathbb 1,\begin{pmatrix}\hat L_R^n&\hat L_L^n\end{pmatrix}^T,\hat H_\text{eff}^n\Big)$. On the other hand, the propagation of photons between nodes is described by another triplet $\big( e^{i\tilde \phi}, 0,0 \big)$, where $\tilde\phi=\omega_0 l/v_g$ with $l$ the distance between two neighbouring GUEs along the waveguide. 

\begin{figure}
\includegraphics[width=\textwidth]{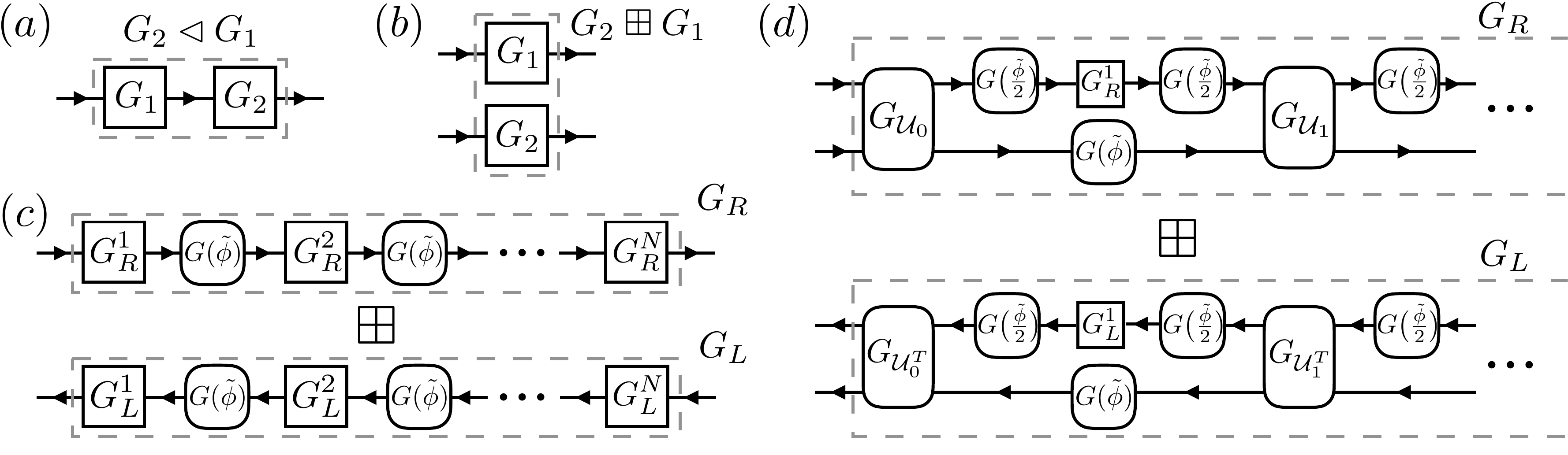}
\caption{\label{fig:figSM3} \emph{Composition rules and applications of the SLH formalism}. (a)~Series product $G_2\triangleleft G_1$ for two cascaded quantum systems, where the output field of the first system becomes the input field of the second one. (b)~Concatenation product $G_2\boxplus G_1$ describing several photonic channels in parallel. (c)~SLH model of the setup of Fig.~\ref{fig:fig1}(d), with $N$ GUEs coupled to a waveguide. The contributions of right- and left-propagating photons (accounted for in $G_R$ and $G_L$) are each decomposed into series products of atomic elements $G_{R/L}^n$ ($n=1,2,\ldots$) and phase shifts $G({\tilde\phi})$. (d)~SLH model for the setup of Fig.~\ref{fig:fig6}(a), containing now two waveguides, each with right- and left-propagating modes contributing to $G_R$ and $G_L$, and linear optics elements $G_{\mathcal U_n}$ coupling the waveguides. }
\end{figure}

To describe larger composite quantum systems, triplets can be combined in a bottom-up approach using different composition rules. In the following we will make use of two composition rules: the \emph{series} product and the \emph{concatenation} product. The series product, represented in Fig.~\ref{fig:figSM3}(a), allows to combine cascaded quantum systems, where the output channel of a first system $G_1$ becomes the input channel of a second one $G_2$, and is denoted $G_2\triangleleft G_1$. The composition rule is 
\begin{equation}
\big({S_2},\hat {\boldsymbol L}_2,\hat H_2\big)\triangleleft \big({S}_1,\hat {\boldsymbol L}_1,\hat H_1\big)
=\left( S_2 S_1, S_2\hat{\boldsymbol L}_1+\hat{\boldsymbol L}_2,\hat H_1+\hat H_2-\frac{i}{2}\big[\hat{\boldsymbol L}_2^\dagger S_2\hat{\boldsymbol L}_1-\hat{\boldsymbol L}_1^\dagger S_2^\dagger \hat{\boldsymbol L}_2\big]\right).
\end{equation}
The concatenation product on the other hand, represented in Fig.~\ref{fig:figSM3}(b), combines different photonic channels in parallel, and is denoted $G_2\boxplus G_1$. The composition rule is 
\begin{equation}
\big({S_2},\hat {\boldsymbol L}_2,\hat H_2\big)\boxplus \big({S}_1,\hat {\boldsymbol L}_1,\hat H_1\big)=\left( \begin{pmatrix}S_2 & 0\\0 & S_1\end{pmatrix},\begin{pmatrix} \hat{\boldsymbol L}_2\\ \hat{\boldsymbol L}_1\end{pmatrix},\hat H_1+\hat H_2\right).
\end{equation}

For a generic (possibly large) quantum system with triplet 
\begin{equation}\label{eq:Gdefgeneric}
G=\left({S},\begin{pmatrix}{\hat L_1}\\\vdots\\\hat L_{N_c}\end{pmatrix},\hat H\right),
\end{equation} with $N_c$ photonic quantum channels, one can straightforwardly access its dynamics. If we assume for simplicity that the scattering matrix $S$ commutes with any arbitrary quantum system operator $\hat O$, which is the case throughout this work, the quantum Langevin equation takes the expression~\cite{doi:10.1080/23746149.2017.1343097}
\begin{equation}\label{eq:HeisenbergSLHdef}
\frac{d}{dt}\hat O(t)=-i[\hat O,\hat H]+\sum_{j=1}^{N_c}\hat{L}_j^\dagger \hat O\hat{L}_j -\frac12\left\{\hat {L}_j^\dagger \hat{L}_j ,\hat O\right\}
+\sum_{i,j}\big[\hat b_i^\text{in}(t)\big]^\dagger(S_{j,i})^*\big[ \hat O,\hat L_j\big]+\big[\hat{L}_j^\dagger,\hat O\big]S_{j,i} \hat b_i^\text{in}(t).
\end{equation}  
Here the operators $\hat b_j^\text{in}(t)$ are the input field operators for each photonic quantum channel ($j=1,2,\ldots,N_c$), and are equivalent to the operators in Eq.~\eqref{eq:bdintdef} defined for the case of right- and left-propagating waveguide modes. Defining analogously the output field operators $\hat b^\text{out}_j(t)$ as in Eq.~\eqref{eq:bdoutdef}, these operators are related to one another via the input-output relation
\begin{equation}\label{eq:SLHinputoutputdefgeneric}
\hat b^\text{out}_j(t)=\sum_{i} S_{j,i} \hat b^\text{in}_i(t)+\hat L_j(t).
\end{equation} 

\subsection{Driven-dissipative dynamics of cascaded GUEs}\label{sec:SMdrivendissipative}

We can now apply the composition rules defined above to model the situation represented in Fig.~\ref{fig:fig1}(d), where an ensemble of $N$ GUEs are coupled to a common waveguide. To this end, the easiest approach consists in decomposing the coupling of each GUE to right- and left-propagating modes of the waveguides, by defining the triplets $G_R^n=\big(1,\hat L_R^n,\hat H_\text{eff}^n\big)$ and  $G_L^n=\big(1,\hat L_L^n,0\big)$. The phase shift $\tilde\phi$ arising in the propagation of the photon between neighbouring GUEs is accounted for with another triplet $G(\phi)=\big( e^{i\tilde \phi}, 0,0 \big)$. We then obtain the triplet $G_R$ representing the coupling of all the GUEs to right-propagating modes [as represented in Fig.~\ref{fig:figSM3}(c)], by recursively applying the series product as
\begin{equation}
G_R=G_R^N\triangleleft G(\tilde\phi)\triangleleft G_R^{N-1}\triangleleft G({\tilde\phi})\triangleleft\ldots\triangleleft G_R^1
 =\left( e^{i\tilde\phi(N-1)}, \hat L_R, \sum_{n=1}^N\hat H_\text{eff}^n+\hat H_R \right),
\end{equation}
where
\begin{equation}
\begin{aligned}
\hat L_R=&\sum_{n=1}^N e^{i\tilde\phi (N-n)}\hat L_R^n,
\\ \hat H_R=&-\frac{i}2\sum_{n,m<n}\left[ \big(\hat L_R^n \big)^\dagger \hat L_R^m e^{i\tilde\phi(n-m)}-\text{h.c.} \right].
\end{aligned}
\end{equation}

Similarly, the collective coupling of the emitters to left-propagating waveguide modes $G_L$ is obtained by recursively applying the series product as represented in Fig.~\ref{fig:figSM3}(c). Note that the ordering of the triplets is however reversed, and we have 
\begin{equation}
G_L=G^1_L\triangleleft G({\tilde\phi})\triangleleft G^2_L\triangleleft G({\tilde\phi})\triangleleft\ldots\triangleleft G_L^N
=\left( e^{i\tilde\phi(N-1)}, \hat L_L, \hat H_L \right),
\end{equation}
where
\begin{equation}
\begin{aligned}
\hat L_L=&\sum_{n=1}^N e^{i\tilde\phi (n-1)}\hat L_L^n,
\\ \hat H_L=&-\frac{i}2\sum_{n,m>n}\left[ \big(\hat L_L^n \big)^\dagger \hat L_L^m e^{i\tilde\phi(m-n)}-\text{h.c.} \right].
\end{aligned}
\end{equation}

The system is then finally described by the concatenation of the right- and left-propagating mode contributions as
\begin{equation}
\begin{aligned}
G=G_R\boxplus G_L=\left(e^{i\tilde\phi(N-1)} \mathbb 1,\begin{pmatrix} \hat{L}_R\\ \hat{L}_L\end{pmatrix},\hat H_\text{eff}\right),
\end{aligned}
\end{equation}
with $\hat H_\text{eff}=\sum_n \hat H_\text{eff}^n + \hat H_R+\hat H_L$.
The dynamics of the system then follows from Eq.~\eqref{eq:HeisenbergSLHdef}, and expresses as in Eq.~\eqref{eq:Otheisenberg}, up to a redefinition of the phase of the input field operators as $\hat b_{R/L}^\text{in}(t) \to e^{-i\tilde\phi(N-1)}\hat b_{R/L}^\text{in}(t)$.

\subsection{Single-photon scattering operator}\label{sec:SMscatteringoperator}

In the following we derive the expression of the single-photon scattering operator for the setup of Fig.~\ref{fig:fig6}(a). We start by defining the single-photon scattering operator, and derive its expression for the situation of a single node (i.e., a single qubit atom interacting with a GUE) coupled to a waveguide, as represented in Fig.~\ref{fig:fig6}(b). We then extend the situation to the more generic setup of Fig.~\ref{fig:fig6}(a).

\subsubsection{Definitions}

Within the SLH formalism, we recall that a (possibly composite) open quantum system with $N_c$ photonic input and output channels, described by the SLH triplet in Eq.~\eqref{eq:Gdefgeneric}, follows the dynamics in Eqs.~\eqref{eq:HeisenbergSLHdef} and \eqref{eq:SLHinputoutputdefgeneric}. In this framework, the single-photon scattering problem consists in solving for the \emph{single-photon scattering operator} $\hat{\mathcal S}_{j,i}(\nu_p,\delta_p)$, where
\begin{equation}\label{eq:Sjinudeltadef}
\hat{\mathcal S}_{j,i}(\nu_p,\delta_p)=\bra{\text{vac}}\hat b^\text{out}_{j}(\nu_p)[\hat b^\text{in}_{i}(\delta_p)]^\dagger\ket{\text{vac}}.
\end{equation}
Here, $\hat{\mathcal S}(\nu_p,\delta_p)$ is a matrix of operators acting on the quantum system, and its elements $\hat{\mathcal S}_{j,i}(\nu_p,\delta_p)$ represent the back-action on the quantum system when a photon with detuning $\delta_p$ (with respect to the central frequency $\omega_0$) scatters on the system from input channel $i$, and leaves the system in output channel $j$ with detuning $\nu_p$, with the the input/output Fourier transform operators defined as
\begin{equation}
\hat b_{j}^\text{in/out}(\delta_p)=\frac{-i}{\sqrt{2\pi}}\int dt \hat b_{j}^\text{in/out}(t) e^{i\delta_p t},
\end{equation}
with $[\hat b_{j}^\text{in/out}(\nu_p),[\hat b_{i}^\text{in/out}(\delta_p)]^\dagger]=\delta_{j,i}\delta(\nu_p-\delta_p)$.
We stress that the single-photon scattering operator in Eq.~\eqref{eq:Sjinudeltadef} is a different quantity from the scattering matrix $S$ in Eq.~\eqref{eq:Gdefgeneric}.
From the input-output relation in Eq.~\eqref{eq:SLHinputoutputdefgeneric} we then have
\begin{equation}\begin{aligned}\label{eq:Sjiexprexpl}
\hat{\mathcal S}_{j,i}(\nu_p,\delta_p)  & =\frac{1}{{2\pi}}\int dt dt' e^{i\nu_p t-i\delta_p t'}\bra{\text{vac}}\hat b_{j}^\text{out}(t)  [\hat b_{i}^\text{in}(t')]^\dagger \ket{\text{vac}}
 \\ & = \delta(\nu_p-\delta_p)S_{j,i}+\frac{1}{{2\pi}}\int dt dt' e^{i\nu_p t-i\delta_p t'}\bra{\text{vac}}\hat L_{j}(t)  [\hat b_{i}^\text{in}(t')]^\dagger, \ket{\text{vac}},
\end{aligned}\end{equation}
where the last term must be evaluated using the quantum Langevin equation~\eqref{eq:HeisenbergSLHdef}. 

\subsubsection{Single node}

We now evaluate the single-photon scattering operator in Eq.~\eqref{eq:Sjiexprexpl} for the case of a single node $n$ as in Fig.~\ref{fig:fig6}(b), where a single qubit atom interacts with a GUE coupled to a waveguide. The GUE is initially in its ground state $\ket{G}_n$, and returns to it after the photon scattering. The SLH triplet describing this system is given by
\begin{equation}
\left(\mathbb 1,\begin{pmatrix}\hat L_R^n\\\hat L_L^n\end{pmatrix},\hat H_\text{eff}^n+\hat H_V^n\right),
\end{equation}
where the non-linear interaction between GUE and qubit is given from Eq.~\eqref{eq:Hainterface} as
\begin{equation}
\hat H_V^n=-(\hat a_q^n)^\dagger \hat a_q^n \left[ V_1^n (\hat a_1^n)^\dagger\hat a_1^n+V_2^n (\hat a_2^n)^\dagger\hat a_2^n\right].
\end{equation}
The single-photon scattering operator is then obtained by integrating Eq.~\eqref{eq:Sjiexprexpl}. In order to express it we first evaluate the matrix elements 
\begin{equation}
v^{s',s}_{R/L}(t)
\equiv \bra{s'}_{q,n}\bra{G}_n\bra{\text{vac}}\hat a_{R/L}^n(t)\int dt' e^{-i\delta_p t'} [\hat b_{R}^\text{in}(t')]^\dagger\ket{\text{vac}}\ket{G}_n\ket{s}_{q,n}, 
\end{equation}
with $\lim_{t\to-\infty}v^{s',s}_{R/L}(t)=0$, where $\ket{s}_{q,n}$ denotes the qubit state (with $s=0,1$). Defining a two-dimensional vector 
\begin{equation}
{\boldsymbol b}^\text{in}_R=\begin{pmatrix}\bra{R}_n(\hat L_R^n)^\dagger\ket{G}_n\\
\bra{L}_n(\hat L_R^n)^\dagger\ket{G}_n\end{pmatrix},
\end{equation}
and a non-hermitian matrix with elements
\begin{equation}
{F}^{s',s}_{d',d}=\bra{s'}_{q,n}\bra{d'}_n\Big[-i\hat H_\text{eff}^n-i\hat H_V^n
-\frac{1}{2}\sum_{d''}(\hat L^n_{d''})^\dagger\hat L^n_{d''}\Big]\ket{d}_n\ket{s}_{q,n},
\end{equation}
we have from the quantum Langevin equation in Eq.~\eqref{eq:HeisenbergSLHdef}
\begin{equation}
\frac{d}{dt}\begin{pmatrix}v^{s',s}_{R}\\v^{s',s}_{L}\end{pmatrix}
= \begin{pmatrix} {F}^{s',s}_{R,R} & {F}^{s',s}_{R,L} \\ {F}^{s',s}_{L,R} & {F}^{s',s}_{L,L}\end{pmatrix}
\begin{pmatrix}v^{s',s}_{R}\\v^{s',s}_{L}\end{pmatrix} - \delta_{s',s} e^{-i\delta_p t} {\boldsymbol b}_R^{\text{in}}.
\end{equation}
The solution of this differential equation reads
\begin{equation}
\begin{pmatrix}v^{s',s}_{R}(t)\\v^{s',s}_{L}(t)\end{pmatrix}
= \delta_{s,s'}e^{-i\delta_p t}\begin{pmatrix} i\delta_p+{F}^{s',s}_{R,R} & {F}^{s',s}_{R,L} \\ {F}^{s',s}_{L,R} & i\delta_p+{F}^{s',s}_{L,L}\end{pmatrix}^{-1}
\cdot {\boldsymbol b}_R^\text{in}.
\end{equation}

Defining two other two-dimensional vectors ${\boldsymbol b}^\text{out}_R$ and ${\boldsymbol b}^\text{out}_L$ as
\begin{equation}
{\boldsymbol b}^\text{out}_d=\begin{pmatrix}\bra{G}_n\hat L_d^n\ket{R}_n\\
\bra{G}_n\hat L_d^n\ket{L}_n\end{pmatrix},
\end{equation}
the single-photon scattering operator from Eq.~\eqref{eq:Sjiexprexpl} reads here, denoting $\hat{\mathcal S}^n_{d',d}(\nu_p,\delta_p)\equiv \bra{G}_n\hat{\mathcal S}_{d',d}(\nu_p,\delta_p)\ket{G}_n$,
\begin{equation}\begin{aligned}
\bra{s'}\hat{\mathcal S}^n_{L,R}(\nu_p,\delta_p)&\ket{s}=\delta(\nu_p-\delta_p)\delta_{s',s}
{\boldsymbol b}_L^\text{out}\cdot\begin{pmatrix} i\delta_p+{F}^{s',s}_{R,R} & {F}^{s',s}_{R,L} \\ {F}^{s',s}_{L,R} & i\delta_p+{F}^{s',s}_{L,L}\end{pmatrix}^{-1}
\cdot {\boldsymbol b}_R^\text{in},
\\ \bra{s'}\hat{\mathcal S}^n_{R,R}(\nu_p,\delta_p)&\ket{s}=\delta(\nu_p-\delta_p)\delta_{s',s}
\left[1+{\boldsymbol b}_R^\text{out}\cdot\begin{pmatrix} i\delta_p+{F}^{s',s}_{R,R} & {F}^{s',s}_{R,L} \\ {F}^{s',s}_{L,R} & i\delta_p+{F}^{s',s}_{L,L}\end{pmatrix}^{-1}
\cdot {\boldsymbol b}_R^\text{in}\right].
\end{aligned}\end{equation}
The explicit expression in terms of the system parameters reads
\begin{equation}\label{eq:SnexprSM}
\begin{aligned}
\hat{\mathcal S}^n_{L,R}(\nu_p,\delta_p)=\delta&(\nu_p-\delta_p)\left[\tilde r_0(\delta_p)\ket{0}_{q,n}\bra{0}+\tilde r_1(\delta_p)\ket{1}_{q,n}\bra{1}\right],
\\ \hat{\mathcal S}^n_{R,R}(\nu_p,\delta_p)=\delta&(\nu_p-\delta_p)\left[\tilde t_0(\delta_p)\ket{0}_{q,n}\bra{0}+\tilde t_1(\delta_p)\ket{1}_{q,n}\bra{1}\right],
\end{aligned}\end{equation}
and we obtain similar expression for $\hat{\mathcal S}^n_{L,L}(\nu_p,\delta_p)$ and $\hat{\mathcal S}^n_{R,L}(\nu_p,\delta_p)$. We stress that in our model the scattering preserves both the number of propagating photons, as well as the qubit atom excitation number, which results in Eq.~\eqref{eq:SnexprSM} with $|\tilde r_0(\delta_p)|^2+|\tilde t_0(\delta_p)|^2=|\tilde r_1(\delta_p)|^2+|\tilde t_1(\delta_p)|^2=1$, where $\tilde r_s$ and $\tilde t_s$ are the reflected and transmitted photon amplitude when the qubit is in state $\ket{s}_{q,n}$. 

Assuming for simplicity that the couplings of the GUE atoms to the waveguide are symmetric, i.e. $\gamma_1=\gamma_2\equiv\gamma$ and $r_1=r_2\equiv r$, these amplitudes take the expression 
\begin{equation}
\begin{aligned}
\tilde r_0(\delta_p)=&-\tfrac{i \gamma  e^{-i \phi } \left[\Delta_1+e^{2 i \phi } \left(\Delta_2+2 J r+r^2 (\Delta_1+\delta_p )+\delta \right)+2 e^{i \phi } \left(J \left(r^2+1\right)+r (\Delta_1+\Delta_2+2 \delta_p )\right)+2 J r+2 \gamma  \left(r^2-1\right)^2 e^{i \phi } \sin (\phi )+\Delta_2 r^2+r^2 \delta_p +\delta_p \right]}{(\Delta_1+\delta_p ) (\Delta_2+\delta_p )-J^2+2 i \gamma  J \left(2 r+\left(r^2+1\right) e^{i \phi }\right)+\gamma ^2 \left(r^2-1\right)^2 \left(-1+e^{2 i \phi }\right)+i \gamma  \left(r^2+2 r e^{i \phi }+1\right) (\Delta_1+\Delta_2+2 \delta_p )},
\\ \tilde t_0(\delta_p)=&\tfrac{\left[(\Delta_1+\delta ) (\Delta_2+\delta_p)+J^2\right]- 2\gamma  \sin(\phi) \left[J \left(r^2+1\right)+r (\Delta_1+\Delta_2+2 \delta_p )\right]}{(\Delta_1+\delta_p ) (\Delta_2+\delta_p )-J^2+2 i \gamma  J \left[2 r+\left(r^2+1\right) e^{i \phi }\right]+\gamma ^2 \left(r^2-1\right)^2 \left(-1+e^{2 i \phi }\right)+i \gamma  \left(r^2+2 r e^{i \phi }+1\right) (\Delta_1+\Delta_2+2 \delta_p )},
\end{aligned}
\end{equation}
while $\tilde r_1(\delta_p)$ and $\tilde t_1(\delta_p)$ are respectively obtained from these expression for $\tilde r_0(\delta_p)$ and $\tilde t_0(\delta_p)$ by replacing $\Delta_{1,2}\to\Delta_{1,2}+V^n_{1,2}$.

In particular, in the regime where the GUEs are unidirectionally coupled to the waveguide, i.e., where $\Delta_1=\Delta_2\equiv \Delta^n+2r\gamma\sin(\phi)$, $J=-\gamma(1+r^2)\sin(\phi)$ and $\phi=\pi/2+2 \text{arctan}(r)$ as discussed in the main text, and assuming a symmetric qubit -- GUE interaction with $V_1^n=V_2^n\equiv V$, the photon scattering becomes unidirectional. We then have $\tilde r_1(\delta_p)=\tilde r_0(\delta_p)=0$, while $\tilde t_0(\delta_p)= t(\Delta^n+\delta_p)$ and $\tilde t_1(\delta_p)= t(\Delta^n+\delta_p+V)$, where
\begin{equation}
t(\delta_p)=\frac{2i\delta_p+\gamma_r}{2i\delta_p-\gamma_r}
\end{equation}
with $\gamma_r\equiv2\gamma\left(1+2r\cos[\phi_\text{opt}]+r^2\right)$, as used in the main text. We write for the single-photon scattering operator in Eq.~\eqref{eq:SnexprSM} $\hat{\mathcal S}^n_{L,R}(\nu_p,\delta_p)=0$ and $\hat{\mathcal S}^n_{R,R}(\nu_p,\delta_p)=\delta(\nu_p-\delta_p)\hat \sigma^n(\delta_p)$, with
\begin{equation}\label{eq:defsigmandeltapSM}
\hat \sigma^n(\delta_p)=\tilde t_0(\delta_p)\ket{0}_{q,n}\!\bra{0}+\tilde t_1(\delta_p)\ket{1}_{q,n}\bra{1}.
\end{equation}

\subsubsection{Photonic quantum network with several nodes}

Here we derive the single-photon scattering operator for the generic setup in Fig.~\ref{fig:fig6}(a), which contains $N$ nodes and two waveguides (labelled ``up'' and ``down''). The waveguides are coupled via linear optical elements, such as beam-splitters, which are described within the SLH framework by triplets of the form 
\begin{equation} 
G_{\mathcal U_n}=\left( \mathcal U_n,0,0\right).
\end{equation}
Here $\mathcal U_n$ are unitary matrices of dimension $2$ representing these couplings between right-propagating modes of the two waveguides. We start by deriving the expression of the SLH triplet describing the system, and derive afterwards the single-photon scattering operator in terms of this triplet.
Similarly to the situation of Sec.~\ref{sec:SMdrivendissipative}, we first evaluate the contribution of right-propagating waveguide modes to the dynamics of the system. Again, this requires applying recursively the series product, as represented in the upper half of Fig.~\ref{fig:figSM3}(d). Notice that, in contrast to the case of Sec.~\ref{sec:SMdrivendissipative}, there are now two right-propagating photonic channels, corresponding to the two waveguides. The corresponding triplet $G_R$ is then obtained recursively, as
\begin{equation}
G_R=\ldots \triangleleft G_{\mathcal U_1} \triangleleft \left[ \left( G({\tilde\phi/2})\triangleleft G_R^1 \triangleleft G({\tilde\phi/2})\right) \boxplus G({\tilde \phi})\right] \triangleleft G_{\mathcal U_0}.
\end{equation}
Here the photon propagation phase $\tilde\phi$ between the elements of the network is accounted for with the triplets $G({\tilde\phi/2})=(e^{i\tilde\phi/2},0,0)$ and $G({\tilde\phi})=(e^{i\tilde\phi},0,0)$, while the triplet for each node $n$ reads $G_R^n=(\mathbb 1,\hat L_R^n,\hat H_\text{eff}^n+\hat H_V^n)$. We then obtain $G_R=(S_R,\hat{\boldsymbol L}_R,\hat H_R+\sum_n\hat H_\text{eff}^n+\hat H_V^n)$, where
\begin{equation}\begin{aligned}\label{eq:SLHRgen}
&S_R=e^{i\tilde\phi N}\prod_{n=0}^N\mathcal U_n,
\\ 
&(\hat{\boldsymbol L}_R)_j=\sum_{n=1}^Ne^{i\tilde\phi(N-n+1/2)}\left(\prod_{m=n}^N\mathcal U_m\right)_{j,\text{up}}\hat L_R^n,
\\
&\hat H_R=-\frac{i}2
\sum_{n,m<n}\left[e^{i\tilde \phi(n-m)}
(\hat L_R^n)^\dagger
\left(\prod_{l=m}^{n-1}\mathcal U_l\right)_{\text{up},\text{up}}\hat L_R^m 
-\text{h.c.}\right],
\end{aligned}
\end{equation}

Similarly, we evaluate the contribution of left-propagating mode as represented in the lower part of Fig.~\ref{fig:figSM3}(d). Again, there are two channels, corresponding to the two waveguides, and their couplings are represented by the transposed matrices $\mathcal U^T_n$. The coupling of the nodes to these modes is given by the triplets $G_L^n=(\mathbb 1,\hat L_L^n,0)$. We then obtain 
\begin{equation}
G_L=G_{\mathcal U_0^T}\triangleleft \left[ \left( G({\tilde\phi/2})\triangleleft G_L^1 \triangleleft G({\tilde\phi/2})\right) \boxplus G({\tilde \phi})\right] \triangleleft G_{\mathcal U_1^T}\triangleleft\ldots
\equiv(S_L,\hat{\boldsymbol L}_L, \hat H_L),
\end{equation}
where
\begin{equation}\begin{aligned}\label{eq:SLHLgen}
\\S_L&=e^{i\tilde\phi (N-1)}\prod_{n=N}^0 \mathcal U_n^T,
\\ (\hat {\boldsymbol L}_L)_j&=\sum_{n=1}^Ne^{i\tilde \phi (n-1/2)}\left(\prod_{m=n-1}^0\mathcal U_m^T\right)_{j,\text{up}}\hat {L}_L^n,
\\ \hat H_L&=-\frac{i}2\sum_{n,m>n}\left[e^{i\tilde \phi(m-n)}(\hat L_L^n)^\dagger
\left(\prod_{l=m-1}^{n}\mathcal U_l^T\right)_{\text{up},\text{up}}\hat L_L^m 
-\text{h.c.}\right],
\end{aligned}
\end{equation}
with the convention $\prod_{n=N}^0A_n\equiv A_0A_1\ldots A_N$. 
We then finally obtain the triplet describing the full system by combining the contribution of the right- and left-propagating modes, as
\begin{equation}
G=G_R\boxplus G_L
=\left( \begin{pmatrix}
S_R &0
\\ 0& S_L
\end{pmatrix}, \begin{pmatrix}
\hat {\boldsymbol L}_R \\ \hat {\boldsymbol L}_L\end{pmatrix}, \hat H_T\right)
\end{equation}
with $\hat H_T\equiv\hat H_R+\hat H_L+\sum_n\hat H_\text{eff}^n+\hat H_V^n$.

Having derived the SLH triplet for the system, one can readily derive the single-photon scattering operator, similarly to the case of a single node above, assuming the GUEs are initially in their ground state $\ket{\mathcal G}=\bigotimes_n\ket{G}_n$ and return to it after the photon scattering. We define the matrix elements 
\begin{equation}v^{\boldsymbol s',\boldsymbol s}_{j,R/L,n}(t)
\equiv \bra{\boldsymbol s}\bra{\mathcal G}\bra{\text{vac}}\!\hat a_{R/L}^n(t) \int dt' e^{-i\delta_p t'} [\hat b_{R,j}^{\text{in}}(t')]^\dagger\ket{\text{vac}}\ket{\mathcal G}\ket{\boldsymbol s}, 
\end{equation}
where $\hat b_{R,j}^{\text{in}}(t')$ is the quantum noise operator for right-propagating photons in line $j$, and $\ket{\boldsymbol s}=\ket{s_1}_{q,1}\ldots\ket{s_N}_{q,N}$ denotes the state of the qubit atoms, with $s_n\in\{0,1\}$. It is convenient to define two $2N$-dimensional vectors ${\boldsymbol b}_j^\text{in}=(b^\text{in}_{j,1}, \ldots, b^\text{in}_{j,2N})^T$ with components 
\begin{equation}\label{eq:binjngendef}
\begin{aligned}
b^\text{in}_{j,n}=e^{i\tilde\phi(n-1/2)}\left(\prod_{m=0}^{n-1}\mathcal U_m\right)_{\text{up},j}\bra{\mathcal G} \hat a_R^n(\hat L_R^n)^\dagger\ket{\mathcal G},
\\ b^\text{in}_{j,N+n}=e^{i\tilde\phi(n-1/2)}\left(\prod_{m=0}^{n-1}\mathcal U_m\right)_{\text{up},j}\bra{\mathcal G} \hat a_L^n(\hat L_R^n)^\dagger\ket{\mathcal G}. 
\end{aligned}
\end{equation}
These vectors represent the absorption amplitudes of a right-propagating photon injected in line $j$, for the two transitions of each GUE $n$. Writing similarly the matrix elements in $2N$-dimensional vectors ${\boldsymbol v}^{\boldsymbol s',\boldsymbol s}_j(t)=(v^{\boldsymbol s',\boldsymbol s}_{j,R,1},\ldots,v^{\boldsymbol s',\boldsymbol s}_{j,R,N},v^{\boldsymbol s',\boldsymbol s}_{j,L,1},\ldots,v^{\boldsymbol s',\boldsymbol s}_{j,L,N})^T$, the quantum Langevin equation in Eq.~\eqref{eq:HeisenbergSLHdef} provides 
\begin{equation}\label{eq:ddtvsspsjt}
\frac{d}{dt}{\boldsymbol v}^{\boldsymbol s',\boldsymbol s}_j(t)= F^{\boldsymbol s',\boldsymbol s}\cdot {\boldsymbol v}^{\boldsymbol s',\boldsymbol s}_j(t)-\delta_{\boldsymbol s',\boldsymbol s}e^{-i\delta_p t}{\boldsymbol b}^\text{in}_j.
\end{equation}
Here the matrix $F^{\boldsymbol s',\boldsymbol s}$ of dimension $2N$ can be decomposed into two submatrices of dimensions $N$ as
\begin{equation}
F^{\boldsymbol s',\boldsymbol s}=\begin{pmatrix}
\bar{\bar A}^{\boldsymbol s',\boldsymbol s} & \left(\bar{\bar B}^{\boldsymbol s',\boldsymbol s}\right)^T\\ \bar{\bar B}^{\boldsymbol s',\boldsymbol s} & \left(\bar{\bar A}^{\boldsymbol s',\boldsymbol s}\right)^T\end{pmatrix}
\end{equation} 
with  
\begin{equation}\label{eq:matricesgendef}
\begin{aligned}
\bar{\bar A}^{\boldsymbol s',\boldsymbol s}_{n,m}=\bra{\boldsymbol s'}\bra{\mathcal G}& \hat a_R^n \Bigg[-i \hat H_T
-\frac12\left(\hat {\boldsymbol L}_R^\dagger\cdot \hat {\boldsymbol L}_R+\hat {\boldsymbol L}_L^\dagger \cdot\hat {\boldsymbol L}_L \right)\Bigg] (\hat a_R^m)^\dagger \ket{\mathcal G} \ket{\boldsymbol s},
\\ \bar{\bar B}^{\boldsymbol s',\boldsymbol s}_{n,m}=\bra{\boldsymbol s'}\bra{\mathcal G}& \hat a_L^n \Bigg[-i \hat H_T
-\frac12\left(\hat {\boldsymbol L}_R^\dagger\cdot \hat {\boldsymbol L}_R+\hat {\boldsymbol L}_L^\dagger \cdot\hat {\boldsymbol L}_L \right)\Bigg] (\hat a_R^m)^\dagger \ket{\mathcal G} \ket{\boldsymbol s}.
\end{aligned}
\end{equation}
The solution of Eq.~\eqref{eq:ddtvsspsjt} reads
\begin{equation}\label{eq:vsspsjtgen}
{\boldsymbol v}^{\boldsymbol s',\boldsymbol s}_j(t)=\delta_{\boldsymbol s',\boldsymbol s} e^{-i\delta_p t}\left(i\delta_p\mathbb 1+ F^{\boldsymbol s',\boldsymbol s}\right)^{-1}\cdot {\boldsymbol b}^\text{in}_j.
\end{equation}
We now define four other $2N$-dimensional vectors ${\boldsymbol b}^\text{out}_{d',j}$ (with $d'=R, L$ and $j\in\{\text{up},\text{down}\}$), with ${\boldsymbol b}^\text{out}_{d',j}=(b^\text{out}_{d',j,1}, \ldots, b^\text{out}_{d',j,2N})$, where
\begin{equation}\label{eq:boudpjngendef}
\begin{aligned}
b^\text{out}_{d',j,n}=\bra{\mathcal G}(\hat {\boldsymbol L}_{d'})_j (\hat a_R^n)^\dagger\ket{\mathcal G},
\\  b^\text{out}_{d',j,N+n}=\bra{\mathcal G}(\hat {\boldsymbol L}_{d'})_j (\hat a_L^n)^\dagger\ket{\mathcal G}.
\end{aligned}
\end{equation}
These vectors represent the amplitudes of photon emission in direction $d'$ and line $j$ for the two excitation modes of each GUE $n$. The single-photon scattering operator from Eq.~\eqref{eq:Sjiexprexpl} is then obtained from Eq.~\eqref{eq:vsspsjtgen}, and reads here, 
\begin{equation}
\begin{aligned}\label{eq:mathcalSnumerical}
\hat {\mathcal S}^{j,i}_{L,R}(\nu_p,\delta_p)=\delta(\nu_p-\delta_p)\sum_{\boldsymbol s}\ket{\boldsymbol s}\bra{\boldsymbol s} & {\boldsymbol b}^\text{out}_{L,j}\cdot
\left(i\delta_p\mathbb 1+ F^{\boldsymbol s,\boldsymbol s}\right)^{-1}\cdot {\boldsymbol b}_i^\text{in},
\\ \hat{\mathcal S}^{j,i}_{R,R}(\nu_p,\delta_p)=\delta(\nu_p-\delta_p)\sum_{\boldsymbol s}\ket{\boldsymbol s}\bra{\boldsymbol s} & \Big[(S_R)_{j,i}
+{\boldsymbol b}^\text{out}_{R,j}\cdot
\left(i\delta_p\mathbb 1+ F^{\boldsymbol s,\boldsymbol s}\right)^{-1}\cdot {\boldsymbol b}_i^\text{in}\Big],
\end{aligned}
\end{equation}
This expression can be evaluated numerically in general.

On the other hand, in the particular case where the coupling of the GUEs to the waveguide is purely unidirectional and where $V^n_1=V^n_2=V$, the coupling operators $\hat L_{R/L}^n$ are proportional to $\hat a_{R/L}^n$, as discussed in the main text. We then obtain from the definitions of Eqs.~\eqref{eq:SLHRgen}, \eqref{eq:SLHLgen}, \eqref{eq:binjngendef}, \eqref{eq:matricesgendef} and \eqref{eq:boudpjngendef} that $\bar {\bar B}^{\boldsymbol s,\boldsymbol s}=0$ and $b^\text{in}_{i,N+n}=b^\text{out}_{L,j,n}=b^\text{out}_{R,j,N+n}=0$, and thus $\hat{\mathcal S}_{L,R}^{j,i}(\nu_p,\delta_p)=0$ in Eq.~\eqref{eq:mathcalSnumerical}.  Moreover, $\bar{\bar A}_{\boldsymbol s}$ becomes block-triangular as $\bar{\bar A}^{\boldsymbol s,\boldsymbol s}_{n,m>n}=0$ and $\bar{\bar A}^{\boldsymbol s,\boldsymbol s}_{n,m<n}=-e^{i\tilde\phi(n-m)}\gamma_r\left(\prod_{l=m}^{n-1}\mathcal U_l\right)_{\text{up},\text{up}}$. This allows to perform the inversion in Eq.~\eqref{eq:mathcalSnumerical} by using recursively the property
\begin{equation}
\begin{pmatrix}
M_1&0
\\M_2&M_3
\end{pmatrix}^{-1}
=
\begin{pmatrix}
M_1^{-1}&0
\\-M_3^{-1}M_2M_1^{-1}&M_3^{-1}
\end{pmatrix}.
\end{equation} 
One then obtains that Eq.~\eqref{eq:mathcalSnumerical} can be factorized as
\begin{equation}
\begin{aligned}
 \hat{\mathcal S}^{j,i}_{R,R}&(\nu_p,\delta_p) 
\\= &  e^{i\tilde\phi N}\delta(\nu_p-\delta_p)\sum_{\boldsymbol s}\ket{\boldsymbol s}\bra{\boldsymbol s}
 \Bigg[  \left(\prod_{n=0}^N\mathcal U_n\right)_{j,i} \!\!+\gamma_r\!\!\sum_{n_2\geq n_1\geq1}\!\!e^{-i\tilde\phi (n_2-n_1)}\left(\prod_{n=n_2}^N\mathcal U_n\right)_{j,\text{up}}
\!\!\left(i\delta_p\mathbb 1+ \bar{\bar A}^{\boldsymbol s,\boldsymbol s}\right)^{-1}_{n_2,n_1}\!\!\left(\prod_{n=0}^{n_1-1}\mathcal U_n\right)_{\text{up},i}\Bigg]
\\ =  & \sum_{i_1} e^{i\tilde\phi N}\delta(\nu_p-\delta_p)\sum_{\boldsymbol s}\ket{\boldsymbol s}\bra{\boldsymbol s}
 \Bigg[  \left(\prod_{n=1}^N\mathcal U_n\right)_{j,i_1} \!\!\!\!+\gamma_r\!\!\!\!\sum_{n_2\geq n_1\geq2}^N\!\!\!e^{-i\tilde\phi(n_2-n_1)}\left(\prod_{n=n_2}^N\mathcal U_n\right)_{j,\text{up}}
\!\!\!\!\!\!\left(i\delta_p\mathbb 1+ \bar{\bar A}^{\boldsymbol s,\boldsymbol s}\right)^{-1}_{n_2,n_1}\left(\prod_{n=1}^{n_1-1}\mathcal U_n\right)_{\text{up},i_1}\Bigg]
\\ & \left(1+\gamma_r\left(i\delta_p\mathbb 1+ \bar{\bar A}^{\boldsymbol s,\boldsymbol s}\right)^{-1}_{1,1}\delta_{i_1,\text{up}}\right)\left(\mathcal U_0\right)_{i_1,i}
\\ =  &\sum_{i_1,i_2} e^{i\tilde\phi N}\delta(\nu_p-\delta_p)\sum_{\boldsymbol s}\ket{\boldsymbol s}\bra{\boldsymbol s}
 \Bigg[  \left(\prod_{n=2}^N\mathcal U_n\right)_{j,i_2} \!\!\!\!+\gamma_r\!\!\!\!\sum_{n_2\geq n_1\geq3}^N\!\!\!e^{-i\tilde\phi(n_2-n_1)}\left(\prod_{n=n_2}^N\mathcal U_n\right)_{j,\text{up}}
\!\!\!\!\!\!\left(i\delta_p\mathbb 1+ \bar{\bar A}^{\boldsymbol s,\boldsymbol s}\right)^{-1}_{n_2,n_1}\left(\prod_{n=2}^{n_1-1}\mathcal U_n\right)_{\text{up},i_2}\Bigg]
\\ & \left(1+\gamma_r\left(i\delta_p\mathbb 1+ \bar{\bar A}^{\boldsymbol s,\boldsymbol s}\right)^{-1}_{2,2}\delta_{i_2,\text{up}}\right)\left(\mathcal U_1\right)_{i_2,i_1} \left(1+\gamma_r\left(i\delta_p\mathbb 1+ \bar{\bar A}^{\boldsymbol s,\boldsymbol s}\right)^{-1}_{1,1}\delta_{i_1,\text{up}}\right)\left(\mathcal U_0\right)_{i_1,i}
\\ & = \ldots = \sum_{i_1,i_2,\ldots,i_N} e^{i\tilde\phi N}\delta(\nu_p-\delta_p)\sum_{\boldsymbol s}\ket{\boldsymbol s}\bra{\boldsymbol s}\left(\mathcal U_N\right)_{j,i_N}\prod_{n=1}^N\left[\left(1+\gamma_r\left(i\delta_p\mathbb 1+ \bar{\bar A}^{\boldsymbol s,\boldsymbol s}\right)^{-1}_{n,n}\delta_{i_n,\text{up}}\right)\left(\mathcal U_{n-1}\right)_{i_n,i_{n-1}}\right],
\end{aligned}
\end{equation}
with $i_0\equiv i$. This expression for $\hat{\mathcal S}^{j,i}_{R,R}(\nu_p,\delta_p)$ can finally be reduced to Eq.~\eqref{eq:SiRjRdef} by defining the two-dimensional diagonal matrix of qubit atom operators $\hat S_n(\delta_p)$, with $[\hat S_n(\delta_p)]_{\text{down},\text{down}}=\mathbb 1$ and $[\hat S_n(\delta_p)]_{\text{up},\text{up}}=\hat \sigma^n(\delta_p)$, where we identified the operator in Eq.~\eqref{eq:defsigmandeltapSM} as
\begin{equation}
\hat \sigma^n(\delta_p)=\sum_{s=0,1}\ket{s}_{q,n}\!\bra{s}\left[1+\gamma_r\left(i\delta_p\mathbb 1+ \bar{\bar A}^{\boldsymbol s,\boldsymbol s}\right)^{-1}_{n,n}\right].
\end{equation} 

\section{Application for single-photon detectors}
\label{secSM:detector}
Here we discuss how the setup of a single qubit atom coupled to a GUE can be used to detect individual itinerant photons. Considering the setup as in Fig.~\ref{fig:fig6}(b), the scattering of a photon on qubit atom $n$, assuming a unidirectional coupling, is described by the operator in Eq.~\eqref{eq:defsigmandeltapSM}, which, with $\Delta^n=-\gamma_r/2$ and $V=\gamma_r$, yields
\begin{equation}
\hat\sigma^n(\delta_p)=\frac{2i(\delta_p-\gamma_r/2)+\gamma_r}{2i(\delta_p-\gamma_r/2)-\gamma_r}\ket{0}_{q,n}\!\bra{0}
+\frac{2i(\delta_p+\gamma_r/2)+\gamma_r}{2i(\delta_p+\gamma_r/2)-\gamma_r}\ket{1}_{q,n}\!\bra{1},
\end{equation} 
which becomes the Pauli operator $\hat \sigma_z^n=\ket{0}_{q,n}\!\bra{0}-\ket{1}_{q,n}\!\bra{1}$, up to an irrelevant global phase, for resonant photons (i.e., with $\delta_p=0$). Resonant photons can thus be detected by preparing qubit $n$ in state $\ket{+}_{q,n}$, which will be flipped to the orthogonal state $\ket{-}_{q,n}$ upon photon scattering. The photon is then effectively detected by measuring qubit $n$, after applying a Ramsey $\pi/2$-pulse on the qubit. For off-resonant photons, the detection probability is given by 
\begin{equation}
P_\text{det}(\delta_p)=\left| \bra{-}_{q,n}\hat\sigma^n(\delta_p)\ket{+}_{q,n}\right|^2=\frac{\gamma_r^4}{(\gamma_r^2-2\delta_p^2)^4+4\gamma_r^2\delta_p^2} 
 =1-4(\delta_p/\gamma_r)^4+\mathcal O(\delta_p/\gamma_r)^8,
\end{equation}
which represents a detection bandwidth of $\sim\gamma_r$.

For photons with finite wavepacket bandwidth, the back-action of this detection will also alter the shape of the wavepacket. Considering a photon with frequency distribution $f(\delta_p)$, i.e., an input state $\int d\delta_p  f(\delta_p) [\hat b_R^\text{in}(\delta_p)]^\dagger\ket{\text{vac}, G_n}\ket{+}_{q,n}$, the photon will leave the system as 
\begin{equation}
\int d\delta_p d\nu_p f(\delta_p) [\hat b_R^\text{out}(\nu_p)]^\dagger\ket{\text{vac}, G_n}\hat{\mathcal S}_{R,R}^n(\nu_p,\delta_p)\ket{+}_{q,n}.
\end{equation} 
After subsequently measuring the qubit, the frequency distribution of the wavepacket will be deformed, up to normalization constants, as
\begin{equation}\label{eq:fdeltadet}
 f(\delta_p)\to  f(\delta_p)\bra{-}_{q,n}\hat\sigma^n(\delta_p)\ket{+}_{q,n}
=-\frac{\gamma_r^2}{\gamma_r^2-2i\gamma_r\delta_p-2\delta_p^2} f(\delta_p)
\end{equation}
upon successful detection with probability $\int d\delta_p P_\text{det}(\delta_p) | f(\delta_p)|^2$, and 
\begin{equation}\label{eq:fdeltanodet}
 f(\delta_p)\to  f(\delta_p)\bra{+}_{q,n}\hat\sigma^n(\delta_p)\ket{+}_{q,n}
=\frac{2i\delta_p^2}{\gamma_r^2-2i\gamma_r\delta_p-2\delta_p^2} f(\delta_p),
\end{equation}
if the detection fails. Note that the fact that the {phases} of two factors in Eqs.~\eqref{eq:fdeltadet} and \eqref{eq:fdeltanodet} depend on $\delta_p/\gamma_r$ is due to the temporal deformation of the wavepacket in the dynamics of the photon absorption and reemission. On the other hand, the fact that their \emph{norms} depend on $\delta_p/\gamma_r$ is a consequence of the frequency filtering due to the measurement back-action.  


\section{Application for preparation of matrix product states}\label{secSM:MPS}
In this section we provide examples of generation of matrix product states with a single photon scattering, namely, GHZ and 1D cluster states. 
\subsection{GHZ state}
\label{sec:GHZ}
We wish to prepare $N$ qubits in the state
\begin{equation}\label{eq:GHZdef}
\ket{\text{GHZ}}=\frac{1}{\sqrt{2}}\left(\bigotimes_{n=1}^N\ket{+}_{q,n}+\bigotimes_{n=1}^N\ket{-}_{q,n} \right).
\end{equation}
This is achieved with the same setup as represented in Fig.~\ref{fig:fig8}(c). With the qubits initialized in state $\bigotimes_n\ket{+}_{q,n}$, taking $\mathcal U_0=\mathcal U_N=\hat{\mathcal H}$ with $\hat{\mathcal H}$ the Hadamard gate as defined in the main text, for a resonant photon the scattering operator in Eq.~\eqref{eq:SiRjRdef} realizes the projection
\begin{equation}
\begin{aligned}
\hat{\mathcal S}^{\text{down},\text{down}}_{R,R}(\nu_p,\delta_p=0)\bigotimes_n\ket{+}_{q,n}=&\delta(\nu_p)e^{i\tilde\phi N}\frac{1}{2}\left[\mathbb 1-\prod_{m=1}^N\hat \sigma_z^m\right]\bigotimes_n\ket{+}_{q,n},
\\\hat{\mathcal S}^{\text{up},\text{down}}_{R,R}(\nu_p,\delta_p=0)\bigotimes_n\ket{+}_{q,n}=&\delta(\nu_p)e^{i\tilde\phi N}\frac{1}{2}\left[\mathbb 1+\prod_{m=1}^N\hat \sigma_z^m\right]\bigotimes_n\ket{+}_{q,n},
\end{aligned}
\end{equation}
and we obtain the state in Eq.~\eqref{eq:GHZdef}, up to a sign depending on the output waveguide which can be corrected by applying a single $\hat\sigma_x^n$ gate on one of the qubits. 

\subsection{1D cluster state}\label{sec:1dcluster}
The 1D cluster state on $N$ qubits is defined as
\begin{equation}\label{eq:C1Ddef}
\ket{\mathcal C_\text{1D}}=\left(\prod_{m=1}^{N-1}\hat {\mathcal Z}_{m,m+1}\right)\bigotimes_{n=1}^N\ket{+}_{q,n}.
\end{equation}
Here we defined the two-qubit controlled-Z gate 
\begin{equation}
\label{eq:Zlmdef}
\hat {\mathcal Z}_{n,m}=\mathbb 1-2 \hat P_n\hat P_m,
\end{equation} 
with the projectors $\hat P_n=\ket{1}_{q,n}\!\bra{1}$, $n=1,\ldots,N$.
The setup and protocol generating this state are obtained by using $\mathcal U_n=\hat{\mathcal H}$ for all $0\leq n\leq N$, with the qubits initialized again in state $\bigotimes_n\ket{+}_{q,n}$. One then obtains the following property \cite{Lindner2009}:
\begin{equation}\label{eq:S_cluster_singlequbit}
\begin{aligned}
\left(\mathcal U_n\hat {S}_n(\delta_p=0)\right)_{\text{down},\text{down}}\ket{+}_{q,n}=&\frac{1}{\sqrt{2}}\hat {\mathcal H}_n\hat\sigma_z^n\ket{0}_{q,n},
\\ 
\left(\mathcal U_n\hat {S}_n(\delta_p=0)\right)_{\text{down},\text{up}}\ket{+}_{q,n}=&\frac{1}{\sqrt{2}}\hat {\mathcal H}_n\hat\sigma_z^n\ket{1}_{q,n},
\\
\left(\mathcal U_n\hat S_n(\delta_p=0)\right)_{\text{up},\text{down}}\ket{+}_{q,n}=&\frac{1}{\sqrt{2}}\hat{\mathcal H}_n\ket{0}_n,
\\
\left(\mathcal U_n\hat S_n(\delta_p=0)\right)_{\text{up},\text{up}}\ket{+}_{q,n}=&\frac{1}{\sqrt{2}}\hat{\mathcal H}_n\ket{1}_n,
\end{aligned}
\end{equation}
with $\hat {\mathcal H}_n=\ket{+}_{q,n}\bra{0}+\ket{-}_{q,n}\bra{1}$ the Hadamard gate on qubit $n$. From Eq.~\eqref{eq:SiRjRdef} and Eq.~\eqref{eq:S_cluster_singlequbit}, one can then show recursively that 
\begin{equation}\label{eq:Soperatorcluster1D}
\begin{aligned}
\hat{\mathcal S}^{\text{down},\text{down}}_{R,R}(\nu_p,\delta_p=0)\bigotimes_n\ket{+}_{q,n}
= & \delta(\nu_p)e^{i\tilde\phi N}\frac{1}{\sqrt{2}}\prod_{n=1}^N\left(\hat{\mathcal H}_n\hat \sigma_z^n \right)
\ket{\mathcal C_\text{1D}},
\\ \hat{\mathcal S}^{\text{up},\text{down}}_{R,R}(\nu_p,\delta_p=0)\bigotimes_n\ket{+}_{q,n}
= & \delta(\nu_p)e^{i\tilde\phi N}\frac{1}{\sqrt{2}}\hat{\mathcal H}_N\prod_{n=1}^{N-1}\left(\hat{\mathcal H}_n\hat \sigma_z^n \right)
\ket{\mathcal C_\text{1D}}.
\end{aligned}
\end{equation}
We thus obtain the 1D cluster state $\ket{\mathcal C_\text{1D}}$ after scattering a single right-propagating photon injected in line ``down'', by applying the inverse of the single-qubit gates in the right-hand side of Eq.~\eqref{eq:Soperatorcluster1D}, conditional on the output waveguide of the photon.

\section{Quantum state transfer protocol}\label{sec:QST}

The average fidelity $\overline{\mathcal F_\text{QST}}$ for the quantum state transfer protocol is evaluated by applying the protocol on an initially maximally entangled state  between qubit $1$ and a virtual ancilla qubit (denoted $a$) \cite{Nielsen:2002ks} as $\ket{\Psi_i}=(\ket{0}_{q,1}\ket{0}_a+\ket{1}_{q,1}\ket{1}_a)\ket{+}_{q,N}/\sqrt{2}$. After performing the protocol as represented in Fig.~\ref{fig:fig7}, we obtain the state of the system of qubit $N$ and ancilla as a density matrix $\hat \rho_f$. Ideally, the state of qubit $N$ and the ancilla should be pure and entangled as $\ket{\Psi_\text{ideal}}=(\ket{0}_{q,N}\ket{0}_a+\ket{1}_{q,N}\ket{1}_a)/\sqrt{2}$. The average fidelity is then defined as $\overline{\mathcal F_\text{QST}}=\text{Tr}\left[\bra{\Psi_\text{ideal}}\hat \rho_f\ket{\Psi_\text{ideal}}\right]$. 

 Here, the state of the system before the scattering of the photon, injected in the system from line ``down'', reads {$\ket{\text{in}}=\int d\delta_p f(\delta_p)[\hat b_{R,\text{down}}^\text{in}(\delta_p)]^\dagger\ket{\text{vac},\mathcal G}\ket{\Psi}_{i}$}, with frequency distribution $ f(\delta_p)$. Assuming unidirectional photon -- GUE interactions, from the expression in Eq.~\eqref{eq:SiRjRdef} the state after the scattering reads $\ket{\text{out}}=\sum_j \int d\delta_p f(\delta_p)[\hat b_{R,j}^\text{out}(\delta_p)]^\dagger\ket{\text{vac},\mathcal G}\ket{\Psi_j(\delta_p)}$, where $\ket{\Psi_j(\delta_p)}\!=\!\left[{\mathcal H}\hat S_N(\delta_p)\mathcal H \hat S_1(\delta_p)\mathcal H\right]_{j,\text{down}}\!\ket{\Psi}_i$ represents the state of the qubits when the photon is scattered to line $j$. The (unnormalized) qubit density matrix $\hat\rho_j$, conditioned on the detection of the photon at the output of line $j$, is then obtained as $\hat\rho_j=\int d\delta_p|f(\delta_p)|^2 \ket{\Psi_j(\delta_p)}\bra{\Psi_j(\delta_p)}$. 
 
 Denoting all the other operations performed in the circuit of Fig.~\ref{fig:fig7} after the photon scattering and subsequent detection at the output of line $j$, including the projective measurements of qubit atom $1$, as superoperators $\hat P_j$, we then obtain the reduced density matrix as $\hat\rho_f=\text{Tr}_{q,1}\left(\hat P_\text{up}[\hat \rho_\text{up}]+\hat P_\text{down}[\hat \rho_\text{down}]\right)$, where $\text{Tr}_{q,1}$ denotes the trace over qubit $1$. The average fidelity finally expresses as $\overline {\mathcal F_\text{QST}}=\int d\delta_p|f(\delta_p)|^2\mathcal F_\text{QST}(\delta_p)$, with
\begin{equation}
\mathcal F_\text{QST}(\delta_p)= \frac{\gamma_r^8-2\gamma_r^6\delta_p^2 -2\gamma_r^5\delta_p^3+3\gamma_r^4\delta_p^4+2\gamma_r^3\delta_p^5+4\delta_p^8}{\left(\gamma^4_r+4\delta_p^4\right)^2}
\\=1-2(\delta_p/\gamma_r)^2+\mathcal O(\delta_p/\gamma_r)^3.
\end{equation}

Including in the description the finite probability $P_d$ of losing the photon in the process, due for instance to amplitude attenuation in the waveguides or to a faulty photon detection, the overall transfer fidelity is $(1-P_d)\overline{\mathcal F_\text{QST}}$. Standard strategies for quantum error correction can however be applied to correct for such photon losses. For example, following Ref.~\cite{PhysRevLetters78.4293}, we can add an ancillary \emph{backup} stationary qubit $b$ to node $1$ and, before performing the state transfer protocol, entangle it with qubit $1$ as 
\begin{equation}\label{eq:entangledqubit}
\begin{pmatrix}
\ket{0}_{q,1}
\\ \ket{1}_{q,1}
\end{pmatrix}
\ket{0}_{b}\ket{+}_{q,N}\to 
\begin{pmatrix}
\ket{0}_{q,1}\ket{1}_b+\ket{1}_{q,1}\ket{0}_b
\\ \ket{1}_{q,1}\ket{1}_b+\ket{0}_{q,1}\ket{0}_b
\end{pmatrix}
\ket{+}_{q,N}.
\end{equation}
In case the photon is \emph{not} detected after the scattering, the initial superposition can then be retrieved by measuring qubit $1$, as the photon scattering operator in Eq.~\eqref{eq:mathcalSnRR} is diagonal in the computational basis of the qubits. From Eq.~\eqref{eq:entangledqubit}, for the measurement outcome $\ket{0}_{q,1}$, the state of the backup qubit is projected to $\begin{pmatrix}\ket{1}_b\\\ket{0}_b\end{pmatrix}$, while the outcome $\ket{1}_{q,1}$ yields $\begin{pmatrix}\ket{0}_b\\\ket{1}_b\end{pmatrix}$. This allows to prepare the system back to the entangled state \eqref{eq:entangledqubit}, and repeat the procedure until the photon is successfully detected at the output, which requires on average $1/(1-P_d)$ trials. At this stage, the state transfer protocol can resume normally, which transfers the entanglement with the backup qubit $b$ from qubit $1$ to qubit $N$, yielding
\begin{equation}
\begin{pmatrix}
\ket{1}_b\ket{0}_{q,N}+\ket{0}_b\ket{1}_{q,N}
\\ \ket{1}_b\ket{1}_{q,N}+\ket{0}_b\ket{0}_{q,N}
\end{pmatrix}.
\end{equation}
The qubit superposition is then finally transferred to qubit $N$ by measuring the backup qubit $b$ and, depending on the outcome, performing a local $\hat\sigma^N_x$ gate on qubit $N$.

\section{Protocols for toric code generation and manipulation}\label{secSM:protocols}
\subsection{Toric code generation}

The toric code is a stabilizer code where physical qubits are located on the edges of a 2D lattice with periodic boundary conditions \cite{Kitaev:2003jw}. The code has two types of stabilizers: as represented in Fig.~\ref{fig:fig8}(a), for each plaquette $p$ of the lattice we define an operator $\hat A_p=\prod_{n\in p}\hat \sigma_z^n$, and, similarly, for each vertex $v$ we define $\hat B_v=\prod_{n\in v}\hat \sigma_x^n$. With an $N_l\times N_l$ lattice [e.g. $N_l=2$ in Fig.~\ref{fig:fig8}(a)], the number of physical qubits is $N=2N_l^2$, while the number of independent stabilizers is $2(N_l^2-1)$. Thus, the manifold of states $\ket{\Phi}$ satisfying the constraints $\hat A_p\ket{\Phi}=\ket{\Phi}$ and $\hat B_v\ket{\Phi}=\ket{\Phi}$ for all $p$ and $v$ is four-dimensional. One such state can be expressed as
\begin{equation}
\ket{\Phi_1}=\frac1{\sqrt{2}}\prod_p\left(\mathbb 1+\hat A_p\right)\bigotimes_{n=1}^N\ket{+}_{q,n},
\end{equation}
which projects $\bigotimes_{n=1}^N\ket{+}_{q,n}$ on the eigenmanifold of each $\hat A_p$ with eigenvalue $+1$. In the protocol of Fig.~\ref{fig:fig8}, we prepare state $\ket{\Phi_1}$ by performing sequential measurements of the operators $\hat A_p$. From the list of the measurement outcomes, one can always perform single-qubit $\hat \sigma_x^n$ gates such that the state becomes eigenstate of all $\hat A_p$ with eigenvalue $+1$. Since $\hat B_v\bigotimes_{n=1}^N\ket{+}_{q,n}=\bigotimes_{n=1}^N\ket{+}_{q,n}$ for all $v$, and as all the stabilizers commute, the state remains an eigenstate of all $\hat B_v$ with eigenvalue $+1$ throughout the protocol, and we prepared the system in state $\ket{\Phi_1}$.

The other three code states can be obtained from $\ket{\Phi_1}$ by applying products of $\hat \sigma_z^n$ gates on all qubits along horizontal or vertical cyclic paths passing through lattice sites. 
We denote such horizontal and vertical cyclic paths as $\boldsymbol \gamma_1$ and $\boldsymbol \gamma_2$ respectively. For the minimal instance represented in Fig.~\ref{fig:fig8}(a), one can take for instance $\boldsymbol \gamma_1=\{5,6\}$ and $\boldsymbol \gamma_2=\{4,8\}$. Conversely, we define vertical and horizontal cyclic paths passing through plaquette centers, which we denote respectively as $\boldsymbol \gamma_1^*$ and $\boldsymbol \gamma_2^*$. In the example of Fig.~\ref{fig:fig8}(a) one can use for instance $\boldsymbol \gamma_1^*=\{1,5\}$ and $\boldsymbol \gamma_2^*=\{3,4\}$. We then define logical operators on the code manifold as the string operators $\hat Z_1=\prod_{n\in\boldsymbol \gamma_1}\hat \sigma_z^n$,$\hat X_1=\prod_{n\in\boldsymbol \gamma_1^*}\hat \sigma_x^n$, $\hat Z_2=\prod_{n\in\boldsymbol \gamma_2}\hat \sigma_z^n$ and  $\hat X_2=\prod_{n\in\boldsymbol \gamma_2^*}\hat \sigma_x^n$. This allows to define the other three code states as $\ket{\Phi_2}=\hat Z_1\ket{\Phi_1}$, $\ket{\Phi_3}=\hat Z_2\ket{\Phi_1}$ and $\ket{\Phi_4}=\hat Z_2\hat Z_1\ket{\Phi_1}$, which satisfy $\hat X_\alpha\ket{\Phi_{\beta}}=(-1)^{s_{\alpha,\beta}}\ket{\Phi_\beta}$ with $s_{1,1}=s_{1,3}=s_{2,1}=s_{2,2}=1$ and $s_{1,2}=s_{1,4}=s_{2,3}=s_{2,4}=-1$. We note that in our setup, the logical operators $\hat{Z}_{1,2}$ and $\hat{X}_{1,2}$ are measurable in the very same way as the stabilizers, without having to measure individual physical qubits which would project the state out of the code subspace.

\subsection{Fidelity of stabilizer measurements with finite photon pulse}
We consider here the scattering of a photon wavepacket with a temporal distribution given by  
\begin{equation}
\tilde f(t)=A e^{-t^2/(4\sigma_t^2)} \Theta(T/2-|t|),
\end{equation}
where $T$ is the temporal extent of the wavepacket, $\sigma_t$ its temporal width, $\Theta$ the Heaviside step function, and $A$ a normalization constant such that $\int_{-T/2}^{T/2}|\tilde f(t)|^2dt=1$. With the qubits initialized in state $\ket{\Psi_+}$ as expressed above, this realizes a scattering with the input state $\ket{\text{in}}=\int d\delta_p  f(\delta_p)[\hat b_{R,\text{down}}^{\text{in}}(\delta_p)]^\dagger\ket{\text{vac}}\ket{\Psi_+}$, where $ f(\delta_p)=(1/\sqrt{2\pi})\int dt \tilde f(t) e^{i\delta_p t}$ is the Fourier transform of $\tilde f(t)$. Without necessarily assuming purely unidirectional photon -- GUE interactions, the system is then left after the scattering in state $\ket{\text{out}}=\sum_{d',j}\int d\delta_p f(\delta_p) [\hat b_{d',j}^{\text{out}}(\delta_p)]^\dagger\ket{\text{vac},\mathcal G} \ket{\Psi_{d',j}(\delta_p)}$, where $\ket{\Psi_{d',j}(\delta_p)}=\int d\nu_p\hat{\mathcal S}_{d',R}^{j,\text{down}}(\nu_p,\delta_p)\ket{\Psi_+}$, and we made use of the fact that the single-photon scattering operator is proportional to $\delta(\nu_p-\delta_p)$ (with its general expression provided in the Supplementary Section \ref{secSM:SLH}). 

Detecting the photon at the right output of waveguide $j$ yields for the qubits the (unnormalized) density matrix $\hat\rho_j=\int d\delta_p|f(\delta_p)|^2\ket{\Psi_{R,j}(\delta_p)}\bra{\Psi_{R,j}(\delta_p)}$. The measurement fidelity is then defined as $\overline{\mathcal F_{\mathcal Z}}=\overline{\mathcal F_\text{down}}+\overline{\mathcal F_\text{up}}$ with 
\begin{equation}
\overline{\mathcal F_j}=\bra{\Psi_{j}^\text{ideal}}\hat\rho_j\ket{\Psi_j^\text{ideal}}=\int d\delta_p|f(\delta_p)|^2\mathcal F_{\mathcal Z}(\delta_p),
\end{equation}
with $\mathcal F_{\mathcal Z}(\delta_p)$ as expressed in Eq.~\eqref{eq:FmathcalZdeltapdef}.

\begin{figure}
\includegraphics[width=\textwidth]{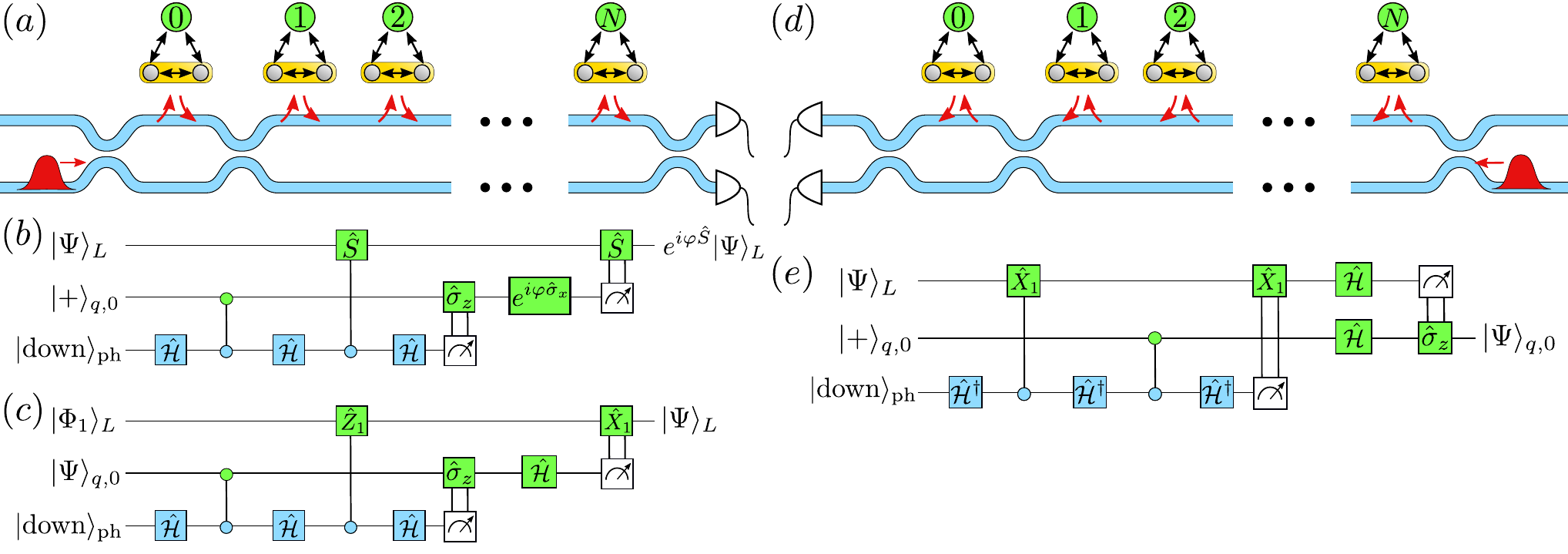}
\caption{\label{fig:figSM1} \emph{Operations on the toric code and exponentiated string operators with an additional qubit.} (a)~Setup with an additional qubit atom $n=0$, located to the left of the other qubit atoms as topological quantum memory. (b,c)~Protocols for (b)~applying an exponentiated gate $e^{i\varphi \hat S}$ on the quantum memory (denoted $L$), and (c)~for transferring a qubit state superposition from qubit $0$ to the quantum memory, with the setup depicted in (a). (d)~Inverted setup, with a photon injected from the right and detected at the left output. (e)~Protocol for transferring back a state superposition from the quantum memory to qubit $0$, with the setup depicted in (d).}
\end{figure}

\subsection{Logical qubit gates and exponentiated string operators}

We now provide details on the protocols for manipulating the toric code with our setup, namely for applying arbitrary gates on logical qubit states, as well as for the write-in and read-out of logical superposition states. 
We first note that the application of any string operator $\hat S=\prod_{n\in\mathcal I}\hat{\boldsymbol \sigma}^n$, with $\mathcal I$ a subset of qubit atoms and $\hat{\boldsymbol \sigma}^n$ an arbitrary rotation of the Pauli operator $\hat \sigma_z^n$ on the Bloch sphere, requires only single-qubit gates. The logical qubit operators $\hat Z_{1,2}$ and $\hat X_{1,2}$, as defined above, are examples of such operators. As represented in Fig.~\ref{fig:figSM1}(a,b), this allows the application of \emph{exponentiated} string operators $e^{i\varphi \hat S}$ with $\varphi$ an arbitrary phase, by using an additional ancilla node $(n=0)$ located to the left of the other ``topological quantum memory'' qubits with $n=1,\ldots N$. There, the controlled-$\hat S$ gate is realized by (i)~performing single qubit rotations on the quantum memory (denoted $L$) before and after the photon scattering, and (ii)~engineering the photon scattering in Eq.~\eqref{eq:mathcalSnRR} such that $\hat \sigma^n(\delta_p=0)=\sigma_z^n$ if $n\in\mathcal I$, and $\hat \sigma^n(\delta_p=0)=\mathbb 1$ otherwise. The exponentiated gate is performed on the ancilla qubit which, after measurement, is transferred to the quantum memory. Assuming a logical qubit in the memory $\ket{\Psi}_L$ is encoded in a superposition of states $\ket{\Phi_1}$ and $\ket{\Phi_2}=\hat Z_1\ket{\Phi_1}$, any logical single-qubit gate can be decomposed into a product $e^{i\varphi_1 \hat X_1}e^{i\varphi_2 \hat Z_1}e^{i\varphi_3 \hat X_1}$, which is thus performed with our protocol in three steps.

 \subsection{Quantum state write-in and read-out}
 The protocol described above can also be used to write a qubit superposition state in the quantum memory, with the same setup. This ``write-in'' protocol, adapted from the quantum state transfer protocol of Fig.~\ref{fig:fig7}, is represented in Fig.~\ref{fig:figSM1}(c), where the ancilla qubit is initialized in a superposition state $\ket{\Psi}_{q,0}=c_0\ket{0}_{q,0}+c_1\ket{1}_{q,0}$ (with $|c_0|^2+|c_1|^2=1$). After the protocol, the superposition state is transferred to the quantum memory as $\ket{\Psi}_L=c_0\ket{\Phi_1}_L+c_1\ket{\Phi_2}_L$. The inverse protocol, consisting in reading-out the quantum memory by mapping the superposition state back to the ancilla, is represented in Fig.~\ref{fig:figSM1}(d,e). This requires to invert the setup, and use left-propagating photons to carry the quantum information from the quantum memory to the ancilla.
 
\end{document}